\documentclass[acmsmall]{acmart}

\AtBeginDocument{%
  \providecommand\BibTeX{{%
    \normalfont B\kern-0.5em{\scshape i\kern-0.25em b}\kern-0.8em\TeX}}}

\setcopyright{acmcopyright}
\copyrightyear{2021}
\acmYear{2021}
\acmDOI{10.1145/0000XXX.0000XXX}

\acmJournal{TOIS}
\acmVolume{1}
\acmNumber{1}
\acmArticle{1}
\acmMonth{4}

\usepackage[utf8]{inputenc} % allow utf-8 input
\usepackage[T1]{fontenc}    % use 8-bit T1 fonts
\usepackage{url}            % simple URL typesetting
\usepackage{booktabs}       % professional-quality tables
\usepackage{amsfonts}       % blackboard math symbols
\usepackage{nicefrac}       % compact symbols for 1/2, etc.
\usepackage{microtype}      % microtypography
\usepackage{amsmath, amsthm}
\usepackage{graphicx, subcaption, float}
\usepackage{siunitx, multirow}
\usepackage{algorithm}  
\usepackage{algorithmicx}  
\usepackage{algpseudocode}

\newtheorem{definition}{Definition}

\usepackage{bbm}
\usepackage{enumitem}
\usepackage{hyperref}
\usepackage{wrapfig}
\usepackage{placeins}
\usepackage{dsfont}
\usepackage{xcolor}
\usepackage{cancel}
\usepackage{wrapfig}
\usepackage{tikz}
\newcommand*{\circled}[1]{\lower.7ex\hbox{\tikz\draw (0pt, 0pt)%
		circle (.5em) node {\makebox[1em][c]{\small #1}};}}

\newcommand{\minisection}[1]{\vspace{5pt}\noindent\textbf{#1.}}

\begin{document}

%%
%% The "title" command has an optional parameter,
%% allowing the author to define a "short title" to be used in page headers.
\title[GraphHINGE]{GraphHINGE: Learning Interaction Models of Structured Neighborhood on Heterogeneous Information Network}
% \title[GraphHINGE]{Structured Neighborhood Interaction Models for Personalized Recommendation on Heterogeneous Information Network}
% \title[GraphHINGE]{Learning Interaction Models of Structured Neighborhood\\for Personalized Recommendation on Heterogeneous Information Network}

%%
%% The "author" command and its associated commands are used to define
%% the authors and their affiliations.
%% Of note is the shared affiliation of the first two authors, and the
%% "authornote" and "authornotemark" commands
%% used to denote shared contribution to the research.
\author{Jiarui Jin}
\email{jinjiarui97@sjtu.edu.cn}
\author{Kounianhua Du}
\email{kounianhuadu@gmail.com}
\author{Weinan Zhang}
\email{wnzhang@sjtu.edu.cn}
\affiliation{%
  \institution{Shanghai Jiao Tong University}
  \city{Shanghai}
  \country{China}
}
\author{Jiarui Qin}
\email{qjr1996@sjtu.edu.cn}
\author{Yuchen Fang}
\email{arthur\_fyc@sjtu.edu.cn}
\author{Yong Yu}
\email{yyu@sjtu.edu.cn}
\affiliation{%
  \institution{Shanghai Jiao Tong University}
  \city{Shanghai}
  \country{China}
}

\author{Zheng Zhang}
\email{zhaz@amazon.com}
\author{Alexander J. Smola}
\email{alex@smola.org}
\affiliation{%
	\institution{Amazon Web Services}
	\country{United States}
}
%\author{Lars Th{\o}rv{\"a}ld}
%\affiliation{%
%  \institution{The Th{\o}rv{\"a}ld Group}
%  \streetaddress{1 Th{\o}rv{\"a}ld Circle}
%  \city{Hekla}
%  \country{Iceland}}
%\email{larst@affiliation.org}

%%
%% By default, the full list of authors will be used in the page
%% headers. Often, this list is too long, and will overlap
%% other information printed in the page headers. This command allows
%% the author to define a more concise list
%% of authors' names for this purpose.
\renewcommand{\shortauthors}{Jin, et al.}

%%
%% The abstract is a short summary of the work to be presented in the
%% article.
\begin{abstract}
Heterogeneous information network (HIN) has been widely used to characterize entities of various types and their complex relations.
% which is promising to carry key semantics in recommender systems.
Recent attempts either rely on explicit path reachability to leverage path-based semantic relatedness
% between users and items
%; e.g., metapath-based similarities
or graph neighborhood to learn heterogeneous network representations 
% (i.e., a single embedding) 
before predictions. 
These weakly coupled manners overlook the rich interactions among neighbor nodes, which introduces an early summarization issue. 
In this paper, we propose GraphHINGE (\textbf{\underline{H}}eterogeneous \textbf{\underline{IN}}teract and aggre\textbf{\underline{G}}at\textbf{\underline{E}}), which captures and aggregates the interactive patterns between each pair of nodes through their structured neighborhoods.
Specifically,
% we analyze the significance of learning interactions in HINs and then propose a novel formulation to capture the interactive patterns between each pair of nodes through their metapath-guided neighborhoods.
% Given that each metapath represent a specific semantic, 
we first introduce Neighborhood-based Interaction (NI) module to model the interactive patterns under the same metapaths, and then extend it to Cross Neighborhood-based Interaction (CNI) module to deal with different metapaths.
Next, in order to address the complexity issue on large-scale networks, we formulate the interaction modules via a convolutional framework and learn the parameters efficiently with fast Fourier transform. 
Furthermore, we design a novel neighborhood-based selection (NS) mechanism, a sampling strategy, to filter high-order neighborhood information based on their low-order performance. 
The extensive experiments on six different types of heterogeneous graphs demonstrate the performance gains by comparing with state-of-the-arts in both click-through rate prediction and top-N recommendation tasks. 
\end{abstract}

%%
%% The code below is generated by the tool at http://dl.acm.org/ccs.cfm.
%% Please copy and paste the code instead of the example below.
%%


%%
%% Keywords. The author(s) should pick words that accurately describe
%% the work being presented. Separate the keywords with commas.

%%
%% This command processes the author and affiliation and title
%% information and builds the first part of the formatted document.
\maketitle

\section{Introduction}
Recommender systems have been playing an increasingly important role in various online services for helping users find items of interest.
However, the existing methods are challenged by problems of data sparsity and cold start, i.e., most items receive only a few feedbacks (e.g., ratings and clicks) or no feedback at all (e.g., for new items).
Previous approaches usually leverage side information to obtain better user/item representations to tackle these issues \citep{wang2018ripplenet,huang2018improving,qu2019end}, which facilitate the learning of user preference and finally promote the recommendation quality. 
Although this auxiliary data is likely to contain useful information \citep{shi2018heterogeneous,hu2018local,jin2020efficient}, it is difficult to model and use the heterogeneous and complex information in recommender systems.

\begin{figure}[h]
\centering
% \vspace{-2mm}
\includegraphics[width=0.95\textwidth]{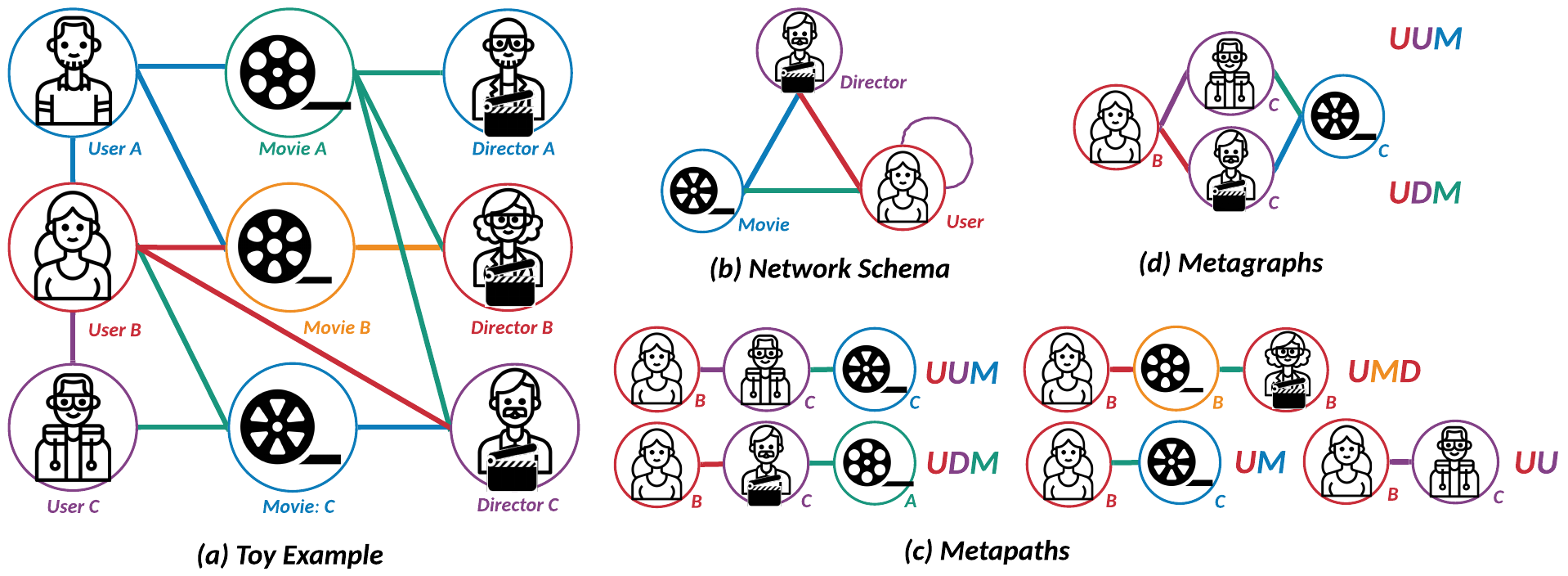}
\vspace{-2mm}
\caption{The network schema and meta relations of Heterogeneous Information Network (HIN). (a) an instance of HIN where we can model user $u_\text{A}$ and define its metapath-guided neighborhoods (e.g, $\mathcal{N}_\text{UMD}(u_\text{A}) = \{(u_\text{A}, m_\text{A}, d_\text{A}), (u_\text{A}, m_\text{A}, d_\text{B}), (u_\text{A}, m_\text{A}, d_\text{C}), (u_\text{A}, m_\text{B}, d_\text{B})\}$). (b) three types of nodes (user, movie, director) and four types of relations (user-user, user-movie, user-director, movie-director) in solid lines. (c) metapaths involved (e.g., User-User (UU), User-Movie (UM), User-User-Movie (UUM), User-Movie-Director (UMD) and User-Director-Movie (UDM)). (d) metagraphs invloved which can be presented by combination of metapaths (e.g., User-Director-Movie (UDM), User-User-Movie (UUM)). 
We will further illustrate the difference among several definitions in Fig.~\ref{fig:def}.}
\label{fig:instance}
% \vspace{-2mm}
\end{figure}

Recently, the heterogeneous information network (HIN)\footnote{The terms ``heterogeneous information network'' and ``heterogeneous graph'' are used interchangeably in the related literature \cite{shi2018heterogeneous,wang2019heterogeneous}. 
In this paper, we mainly use ``heterogeneous information network'' (HIN).}, consisting of multiple types of nodes and/or links, has been used as a powerful modeling method to fuse complex information and successfully applied to many recommender system tasks, which are called HIN-based recommendation methods \cite{shi2018heterogeneous,shi2016integrating}.
In Fig.~\ref{fig:instance}, we present an instance of movie data characterized by an HIN.
HIN is built from historical feedback and side information. 
The nodes can be users, movies, or entities of side information (e.g., directors, movie genres), and two nodes are linked together based on relevance or co-occurrence.
We can easily see that the HIN contains multiple types of entities connected by different types of relations.
Hence, HIN is able to integrate various recommendation tasks, e.g., user-user relations for content-based recommendations and user-movie-user relations (i.e., users who have viewed the same movie may also share the same preference) for collaborative recommendation \citep{shi2015semantic}.
A variety of graph mining methods have been proposed on HINs to capture the rich semantic information, which roughly fall into the following two categories.
One school is graph-based methodologies, which promotes expressive ability by taking the local structures into consideration.
For example, graph convolution networks \citep{zhang2019heterogeneous,kipf2016semi} can integrate high-order neighborhood information in an end-to-end way, and graph attention networks \citep{velivckovic2017graph,wang2019heterogeneous} can capture key information to simulate user preferences.
However, these technologies usually compress the information of a node and its neighborhood into a single embedding vector before making the prediction \cite{fu2017hin2vec,shi2018heterogeneous}.
In such a case, only two nodes and one edge are activated, yet other nodes and their connections are mixed and relayed, which introduces an \textbf{early summarization} issue \cite{qu2019end}.
As an example illustrated in Fig.~\ref{fig:instance},
a system is recommending a movie to user $u_\text{B}$ based on an HIN, where $u_\text{B}$ has shown her preference for $m_\text{B}$ and $m_\text{C}$.
We know that $m_\text{B}$ is directed by $d_\text{B}$, and $m_\text{C}$ is directed by $d_\text{C}$.
Hence, the connections between $m_\text{B}$ and $d_\text{B}$, $m_\text{C}$ and $d_\text{C}$, could be helpful for recommendation, meanwhile, the connections between $m_\text{B}$ and $d_\text{C}$, $m_\text{C}$ and $d_\text{B}$, are nonsense and not expected.
We argue that these meticulous local structures are valuable, and a good recommender system should be able to capture these valuable patterns and filter out other noise. 

The other school is metapath-based approaches.
A metapath means composite relation connecting two objects at the network schema level. 
It has been adopted to capture semantic information \citep{liu2018interactive,wang2019heterogeneous}.
Taking the movie data in Fig.~\ref{fig:instance} as an example, the relations between user and item can be revealed by the metapath User-User-Movie (UUM) for the co-user relation and User-Director-Movie (UDM) for the co-director relation.
However, as stated in \citep{shi2018heterogeneous}, metapath-based methods, heavily relying on explicit path reachability, may obtain bad performance when path connections are sparse or noisy.
Also, rich structured information of nodes outside metapaths (e.g., their neighborhoods) is omitted in these approaches.

Based on the above analysis, when designing HIN-based recommendation, state-of-the-art methods have not well solved, even may not be aware of, the following challenges,
% faced by HIN-based recommendation, 
which we address in this paper:
\begin{itemize}[topsep = 3pt,leftmargin =10pt]
	\item (\textbf{C1}) How to tackle the early summarization issue?
	Due to the complex structures and large scales of the graphs, it is hard to make predictions directly. 
	Hence, we consider the interactive\footnote{In this paper, we specifically use ``interactive'' patterns or ``interactions'' to denote the interactive patterns or interactions between features, nodes, paths, and neighborhoods of nodes, as illustrated in Fig.~\ref{fig:interaction}.} local structures are valuable and helpful.
	% and a good system should be able to capture useful patterns, and filter out other noise. 
	For example, a system is recommending a movie to user $u_\text{B}$ based on an HIN as Fig.~\ref{fig:instance} shows.
	When we consider a candidate movie such as $m_\text{C}$, it is natural to consider their co-users and co-directors; namely there may exist a relationship between $u_\text{B}$ and $m_\text{C}$ since they share the same neighbor user $u_\text{C}$.
	Actually, this is an example of ``\textbf{AND}'' operation between users' neighborhoods.
% 	This ``AND'' operation also can mine the hidden connections between ($m_B$, $d_B$) and ($m_C$, $d_C$). 
	This interactive structured information characterized by ``AND'' can be categorized as the similarity between the same type of nodes (e.g., user and user) and the co-rating between different types of nodes (e.g., user and movie). 
	We argue that these local structures are hidden and not fully utilized in previous methods.
	
	\item \textbf{(C2)} How to mine cross semantic information? 
	The intuitive motivation behind cross semantic mining is natural.
	As aforementioned, an HIN shown in Fig.~\ref{fig:instance} can integrate various kinds of recommendations according to semantic relations such as user-movie-user relation for collaborative recommendation and user-user relation for content-based recommendation \citep{shi2015semantic}.
	Similar to hybrid recommendation \citep{yu2013recommendation}, we aim to build combinations of various semantics and interactions between them.
	In HIN, considering that each kind of metapath represents a specific type of semantic information, we aim to mine interactive patterns between different metapaths to form cross semantic features. 
	A recent attempt \citep{liu2018interactive} focused on measuring the semantic proximity by interactive-paths connecting source and target nodes.
	However, it suffered from high computations and overlooked rich information hidden in these node neighborhoods. 
	
	\item (\textbf{C3}) How to design an end-to-end framework to capture and aggregate the interactive patterns between neighborhoods?
	Recent works \citep{qu2018product,qu2019end} investigated the methods of mining interactive patterns through inner product and aggregate with deep neural network (DNN).
	However, HIN contains diverse semantic information reflected by metapaths \cite{sun2011pathsim}.	
	Also, there are usually various nodes in different types involved in one path.
	Different paths/nodes may contribute differently to the final performance. 
	Hence, besides a powerful interaction module, a well-designed aggregation module to distinguish the subtle difference of these paths/nodes and select the informative ones is required.
	
	\item (\textbf{C4}) How to learn the whole system efficiently?
	Learning interactive information on HINs is always time-consuming, especially when faced with paths in different types and lengths for metapath-based approaches \cite{liu2018interactive} and large-scale high-order information for graph-based approaches \cite{qu2019end}.
	A methodology to efficiently and effectively learn the rich interactive information on HINs is always expected.
	
	\item (\textbf{C5})
	 How to integrate high-order neighborhood information?
	 Introducing high-order neighborhood information has shown to be effective but time-consuming and intractable for complex operations on large graphs \citep{qu2019end,ying2018graph,hamilton2017inductive}.
	 One solution to facilitate high computation is to design a sampling strategy such as neighbor sampling \citep{hamilton2017inductive}, walk-based sampling \citep{ying2018graph}, and importance sampling \citep{chen2018fastgcn}.
	 Hence, when modeling high-order neighborhood information, it is a severe challenge to design a mechanism to select useful information while filtering out the noise.
\end{itemize}

To tackle these challenges, we propose GraphHINGE (\textbf{\underline{H}}eterogeneous \textbf{\underline{IN}}teract and aggre\textbf{\underline{G}}at\textbf{\underline{E}}) to capture and aggregate the interactive patterns between each pair of nodes through their structured neighborhoods.
First, we extend the definition of the neighborhood in homogeneous graphs to the metapath-guided neighborhood in heterogeneous graphs.
Next, we design a heterogeneous graph neural network architecture with two modules to capture and aggregate key information of sampled neighborhoods in the previous step.
The first module, namely the interaction module, constructs an interactive neighborhood and captures latent information with ``AND'' operation.
Specifically, we introduce the Neighborhood-based Interaction (NI) module to model interactive patterns of source and target nodes' structured neighborhoods guided by the same metapaths (i.e., the same semantics).
We also extend to the Cross Neighborhood-based Interaction (CNI) module to deal with different metapaths. 
The second module is the aggregation module, which mainly consists of two components: (i) an element-level attention mechanism to measure the impacts of different interaction elements inside each path and (ii) a path-level attention mechanism to aggregate the content embeddings of different paths.
Furthermore, we formulate the interaction operation via a convolutional framework and learn efficiently with fast Fourier transform (FFT). 
In order to efficiently integrate high-order neighborhood information, we propose to utilize the performances of low-order neighborhoods to guide the selection of their high-order neighborhoods.
% The interaction module here serves as the interactive pattern extractor, which restricts the interactions within the same type of metapath from source and target sides. 
% To further generalize the interaction from the same semantic to cross architecture, we extend the convolutional operation, leading to cross NI (CNI) to make up for the deficiency of NI in modeling cross semantic interactions.

To summarize, the main contributions of our work are:
\begin{itemize}[topsep = 3pt,leftmargin =10pt]
	\item We analyze an important, but seldom exploited, early summarization issue on HINs. 	
	We structure the node neighborhoods with metapaths,
% 	as metapath-guided neighborhodds 
	and propose an end-to-end framework named GraphHINGE to capture and aggregate the interactive patterns on HINs. 
	\item We develop the neighborhood-based interaction (NI) module to model interactive patterns in neighborhoods guided by the same metapaths and extend to the cross neighborhood-based interaction (CNI) module for different metapaths.
	\item We formulate the interaction module via a convolutional framework and propose an efficient end-to-end learning algorithm incorporated with FFT.
	\item We design a novel neighborhood-based selection (NS) mechanism to efficiently select and integrate useful high-order neighborhood information.  
\end{itemize}
We conduct extensive experiments on six benchmark datasets of different graph types, including two with large-scale graphs. 
Our results demonstrate the superior performance of GraphHINGE compared with state-of-the-art methods in both click-through rate prediction and top-N recommendation tasks.

\section{Related Work}
\label{sec:related-work}
\subsection{Heterogeneous Information Network based Recommendation}
As an emerging direction, HIN \citep{shi2016survey} can naturally characterize complex objects and rich relations in recommender systems.
There is a surge of works 
% learning to capture this auxiliary information,
on learning representation in heterogeneous networks, e.g., metapath2vec \citep{dong2017metapath2vec}, HetGNN \citep{zhang2019heterogeneous}, HIN2vec \citep{fu2017hin2vec}, eoe \citep{xu2017embedding}; and their applications, e.g., relation inference \citep{sun2012will}, classification \citep{zhang2018deep}, clustering \citep{ren2014cluscite}, author identification \citep{chen2017task}, user profiling \citep{chen2019semi}, message receiver recommendation \citep{yi2020heterogeneous}.
%% attention is increasingly shifting towards heterogeneous networks and their applications in recent years.
Among them, HIN-based recommendation has been increasingly attracting researchers' attention,
% in both academic and industry fields, 
which roughly falls into two folds.  
% Researchers have began to be aware of evaluate the importance of HIN-based recommendation.
One direction is to leverage structured information of HIN to enhance the representation of user and item in recommendation \citep{feng2012incorporating,hu2018local,zhang2020graph}.
For instance,
\citet{feng2012incorporating} incorporated social networkl, tag semantics and item profiles to alleviate the cold start issue with the heterogeneous information network contained in the social tagged system.
\citet{hu2018local} proposed to integrate both local and global information from HIN to enhance the recommendation performance. \citet{wang2019kgat} (KGAT) modeled the high-order connectivities in collaborative knowledge graphs in an end-to-end manner to enhance the recommendation.
The other direction is to define specific metapaths in HIN to capture the corresponding semantic information \citep{yu2014personalized,hu2018leveraging}.
For example, 
\citet{yu2013recommendation} introduced metapath-based methods to represent the connectivity between users and items along different paths into a hybrid recommender system.
\citet{yu2014personalized} leveraged personalized recommendation framework via taking advantage of different types of entity relationships in HIN.
\citet{luo2014hete} proposed a collaborative filtering based social recommendation to effectively utilize multiple types of relations.
\citet{shi2015semantic} proposed to integrate heterogeneous information and obtain prioritized and personalized weights representing user preferences on paths by a weighted heterogeneous information network.
\citet{shi2016integrating} introduced a matrix factorization based dual regularization framework to flexibly integrate different types of information through adopting the similarity of users and items as regularization on latent factors of users and items.
Recently, \citet{bi2020heterogeneous} designed a cross-domain recommendation method on HIN to address cold start problem. 
\citet{hu2018leveraging} proposed a novel deep neural network with the co-attention mechanism for leveraging rich meta-path
based context for top-N recommendation.
As stated in \citep{shi2018heterogeneous}, most existing HIN-based recommendation methods rely on the path-based similarity, which may not fully mine latent features of users and items.
Instead, we provide the generic definition of metapath-guided neighborhood, and propose a novel framework to capture rich structured interactive patterns hidden in the neighborhoods from both user and item sides.

\subsection{Graph Representation}
Graph representation learning is mainly leveraged to learn latent, low dimensional representations of graph vertices, while preserving graph structure, e.g.,~topology structure and node content.
In general, graph representation algorithms can be categorized in two types.
% , \emph{i.e.}, unsupervised methods and semi-supervised methods.
One school is unsupervised graph representation algorithm, which aims at preserving graph structure for learning node representations \citep{perozzi2014deepwalk,grover2016node2vec,ribeiro2017struc2vec,tang2015line,wang2016structural}.
For instance, \citet{perozzi2014deepwalk} (DeepWalk) utilized random walks to generate node sequences and learn node representations with methods borrowed from word2vec learning.
\citet{grover2016node2vec} (Node2vec) further exploited a biased random walk strategy to capture more flexible contextual structures.
\citet{ribeiro2017struc2vec} (Struc2vec) constructed a multilayer graph to encode structured similarities and generate a structured context for nodes.
\citet{tang2015line} (LINE) proposed an edge-sampling algorithm to
%addressesing the limitation of the classical stochastic gradient descent and 
improve both the effectiveness and efficiency of the inference.
\citet{wang2016structural} (SDNE) built multiple layers of non-linear functions to capture highly non-linear network structures.
% for both the local and global network
Another school is semi-supervised model \citep{huang2017label,kipf2016semi,velivckovic2017graph}, where there exist some labeled vertices for representation learning.
For example, 
\citet{bruna2013spectral} first designed the graph convolution operation in Fourier domain by the graph Laplacian.
Then, \citet{defferrard2016convolutional} further employed the Chebyshev expansion of the graph Laplacian to improve the efficiency.
\citet{huang2017label} (LANE) incorporated label information into the attributed network embedding while preserving their correlations.
\citet{kipf2016semi} (GCN) proposed a localized graph convolutions to improve the performance in a classification task.
\citet{velivckovic2017graph} (GAT) used self-attention network for information propagation, which leverages a multi-head attention mechanism.
\citet{hamilton2017inductive} (GraphSAGE) proposed to sample and aggregate features from local neighborhoods of nodes with mean/max/LSTM pooling. 
\citet{wu2019net} (DEMO-Net) designed a degree-aware feature aggregation process to explicitly express in graph convolution for distinguishing structure-aware node neighborhoods. 
\citet{abu2019mixhop} (MixHop) proposed to aggregate feature information from both first-order and higher-order neighbors in each layer of network, simultaneously.
Most of the current graph neural networks (GNNs) essentially focus on fusing network topology and node features to learn node embedding, which can be regarded as plug-in graph representation modules in heterogeneous graph, such as HetGNN \citep{zhang2019heterogeneous}, HAN \citep{wang2019heterogeneous} and HERec \citep{shi2018heterogeneous}.
Recently, \citet{he2020lightgcn} (LightGCN) largely simplified the design of GCN to
make it more concise and effective for graph-based recommendation. \citet{zhu2020bgcn} (BGCN) inspired by \citet{he2017Neufm} proposed a bilinear aggregator to express the local nodes interactions and enhance representations. 
In this paper, our purpose is not only to deliver graph representation learning techniques on HIN, but also provides a novel efficient structured neighborhood-based interaction module to capture interactive information on HIN, which to our knowledge should be the first work in the literature.

\begin{figure}[t]
	\centering
	\vspace{-2mm}
	\includegraphics[width=0.65\textwidth]{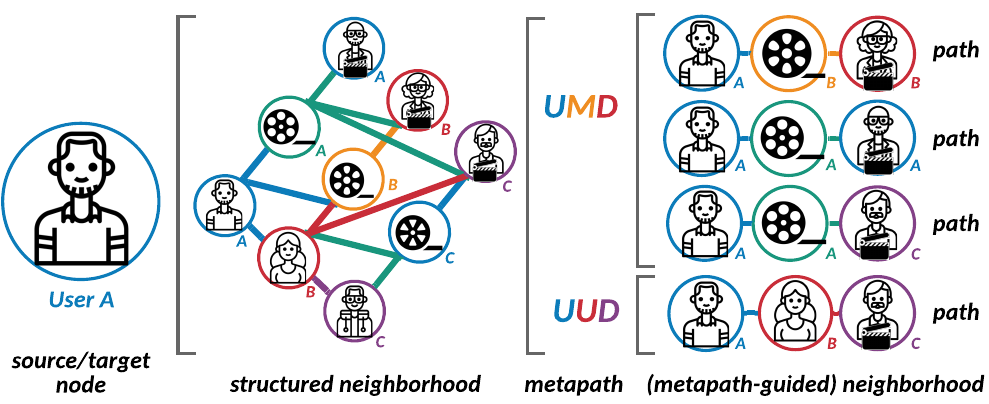}
% 	\vspace{-2mm}
	\caption{
		An illustrated example for several definitions in this paper. 
		We use structured neighborhood to denote the original neighborhood in an HIN.
		We show the generic definition of metapath-guided neighborhood in \textsc{Definition~\ref{def:neighbor}}, which includes several paths (e.g.,~($u_\text{A}, m_\text{B}, d_\text{B}$)) guided by the metapath (e.g.,~UMD).
		For convenience, we use `neighborhood' to denote the metapath-guided neighborhood, unless otherwise stated.
	}
	\label{fig:def}
	\vspace{-2mm}
% 	\vspace{-5mm}
\end{figure}

\subsection{Cross Feature Modeling}
Cross feature modeling aims to mine interactive information among features, especially for multi-field categorical data, to address the data sparsity and improve the performance.
For example, 
% factorization machine  
\citet{rendle2010factorization} (FM) projected each feature into a low-dimensional vector and modeled feature interactions by inner product.
% Field-aware factorization machine  
\citet{juan2016field} (FFM) further enabled each feature to have multiple vector representations to interact with features from other fields. 
An efficient algorithm for training arbitrary-order
higher-order FM (HOFM) was proposed in \citep{blondel2016higher} by introducing the
ANOVA kernel. 
% As shown in \citep{xiao2017attentional}, HOFM achieves only marginal improvement over FM whereas using many more parameters.
Recent attempts, such as Attention FM \citep{xiao2017attentional} and Neural FM \citep{he2017neural}, investigated to improve FM by incorporating a multi-layer perceptron (MLP).
Moreover, \citet{cheng2016wide} (Wide \& Deep) jointly trained a wide model for artificial features and a deep model for raw features. 
\citet{guo2017deepfm} (DeepFM) incorporated an FM layer to replace the wide component in Wide \& Deep to explicitly model the pairwise feature interactions. 
\citet{qu2016product} (PNN) proposed to leverage MLP to model the interaction of a product layer and feature embedding while \citet{qu2018product} (PIN) generalized such product operators in PNN and DeepFM to a network-in-network architecture to model pairwise feature interactions with sub-networks.
DeepFM and PIN cannot explicitly model higher-order feature interactions, which could further improve
model performance. 
\citet{wang2017deep} (CrossNet) introduces a cross operation to learn explicit feature interactions. 
% However, as stated in \citep{lian2018xdeepfm}, such a
% cross operation can not learn effective interactions since the output of cross networks is a scalar multiplication of raw embedding.
\citet{lian2018xdeepfm} (xDeepFM) used a Compressed Interaction Network (CIN) to enumerate and compress all feature interactions, for modeling an explicit order of interactions.
% However, it uses so many parameters that great challenges are posed to identify important feature interactions in the huge combination space. 
\citet{liu2019feature} (FGCNN) combined a convolutional neural network (CNN) and MLP to generate new features for feature augmentation.
Instead of operating on tabular data, in this paper, we investigate to incorporate with structured information and enhance the representation with graph neural network to model the cross features in an HIN, which should be the first time to our knowledge.
% Furthermore, to deal with the higher computation cost issue, we introduce FFT to accelerate the interaction procedure.

\begin{table}[t]
    \caption{A summary of main notations in this paper.}  
    \label{tab:notation}
    \small
    \vspace{-3mm}
    \begin{center}  
    \begin{tabular}{lc}
    \toprule
    \textbf{Notation} & \textbf{Explanation} \\
    \midrule
    $s; t; o$ & Source (node); target (node); object (node)\\
    $I;E$ & Metapath length; dimension of embedding \\
    $L;P;K$ & Number of paths; number of metapaths; number of metapath combinations\\
    $\rho_p$; $\rho^l_p$ & metapath $\rho_p$; $l$-th path guided by metapath $\rho_p$ \\
    $\langle\rho_s,\rho_t\rangle_k$ & $k$-th combination of metapaths $\rho_s$ for source $s$, $\rho_t$ for target $t$\\
    $\mathcal{N}^i_{\rho_o}(o)$ & $i$-th step metapath $\rho_o$ guided neighborhood of object $n_o$ according to \textsc{Definition~\ref{def:neighbor}} \\
    $\mathbf{H}[\mathcal{N}_{\rho_o}(o)]_l$ & Embedding matrix of $l$-th path neighborhood of $n_o$ according to Eq.~(\ref{eqn:embedding})\\
    $\mathbf{H}[\mathcal{N}_{\rho_s}(s),\mathcal{N}_{\rho_t}(t)]_l$ & Interaction matrix of $l$-th path of neighborhoods of $s$, $t$ according to Eq.~(\ref{eqn:crossinteract})\\
    $\cdot; \odot; \oplus$ & Product; convolution; concatenation/stack operation\\
    \bottomrule
    \end{tabular}  
    \end{center} 
    \small
    \vspace{-1mm}
\end{table}

\section{Preliminaries}
\label{sec:pre}	
In this section, we introduce the concept of the content-associated heterogeneous information network that will be used in the paper.

\vspace{-1mm}
\begin{definition}
    \rm
    \label{def:rec}
	\textbf{Neighborhood-based Interaction-enhanced Recommendation.}
	The HIN-based recommendation\footnote{Different from previous works \citep{he2020lightgcn,wang2019neural} built on user-item bipartite graph, HINs here not only include historical data between users and items, but also contain heterogeneous attributes related to users and/or items.} task can be represented as a tuple $\langle\mathcal{U}, \mathcal{I}, \mathcal{A}, \mathcal{R}\rangle$, where $\mathcal{U} = \{u_1, \ldots, u_p\}$ denotes the set of $p$ users; $\mathcal{I} = \{i_1, \ldots, i_q\}$ means the set of $q$ items; $a \in \mathcal{A}$ denotes the attributes associated with objects, and $r \in \mathcal{R}$ represents the relations (e.g.,~ratings and/or clicks) between different types of objects.
	The purpose of recommendation is to predict the relation/connection $r(u_s, i_t)$ (e.g.,~click-through rate and link) between two objects, namely source user $u_s$ and target item $i_t$.
	In our setting, interactions between source and target neighborhoods are leveraged to enhance the performance.
	\end{definition}
\vspace{-1mm}
	
	To clarify the definition of the term `neighborhood', we model our recommendation task in the setting of \textbf{Heterogeneous Information Network} (HIN).
	An HIN is defined as a graph $\mathcal{G}=(\mathcal{V}, \mathcal{E})$, which consists of more than one node type or link type.
	In HINs, \textbf{network schema} is proposed to present the meta structure of a network, including the object types and their relations.

	Fig.~\ref{fig:instance}(a) shows a toy example of HINs and the corresponding network schema is presented in Fig.~\ref{fig:instance}(b).
	We can see that the HIN consists of multiple types of objects and rich relations.
	The set of objects (e.g.,~$\mathcal{U}$ (user), $\mathcal{I}$ (movie), $\mathcal{A}$ (director)) and  relations (e.g.,~$\mathcal{R}$ (relation)) constitute $\mathcal{V}$ and $\mathcal{E}$ in the HIN. 
	The relations here can also be explained as preferences between different types of nodes (e.g.,~$r(u_\text{A}, d_\text{C})$) and similarities between the same type of nodes (e.g.,~$r(u_\text{A}, u_\text{C})$).
% 	In this case, user-item rating history, user-director preference, and movie-director knowledge constitute the interaction set $\mathcal{R}$.
    Hence, one example of \textsc{Definition~\ref{def:rec}} is that when predicting relation $r(u_\text{A}, m_\text{C})$ between source user $u_\text{A}$ and target movie $m_\text{C}$, we consider existing relations between their neighborhoods, namely $r(m_\text{A}, d_\text{C})$, $r(m_\text{B}, u_\text{C})$, to help the final prediction.
	
	In order to capture the structured and semantic relation, the metapath \cite{sun2011pathsim} is proposed as a relation sequence connecting two objects.
	As illustrated in Fig.~\ref{fig:instance}(c), the User-Movie-Director (UMD) metapath indicates that users favor movies and that these movies are guided by some directors \citep{liu2018interactive,wang2019heterogeneous}.
	Besides the metapath, there are some other structures such as metagraph \citep{zhao2017meta,huang2016meta} (as illustrated in Fig.~\ref{fig:instance}(d)) on HINs to capture complicated semantics, which can be presented by a set of metapaths.
% 	As stated in \citep{,fang2016semantic,}, the metagraphs, as illustrated in Figure~\ref{fig:instance}(d), are better at capturing complicated semantics than metapaths.
	In order to build a unified structure to model these semantics, we introduce the definition of metapath-guided neighborhood as follows.

\begin{figure}[t]
	\centering
	\vspace{-2mm}
	\includegraphics[width=0.95\textwidth]{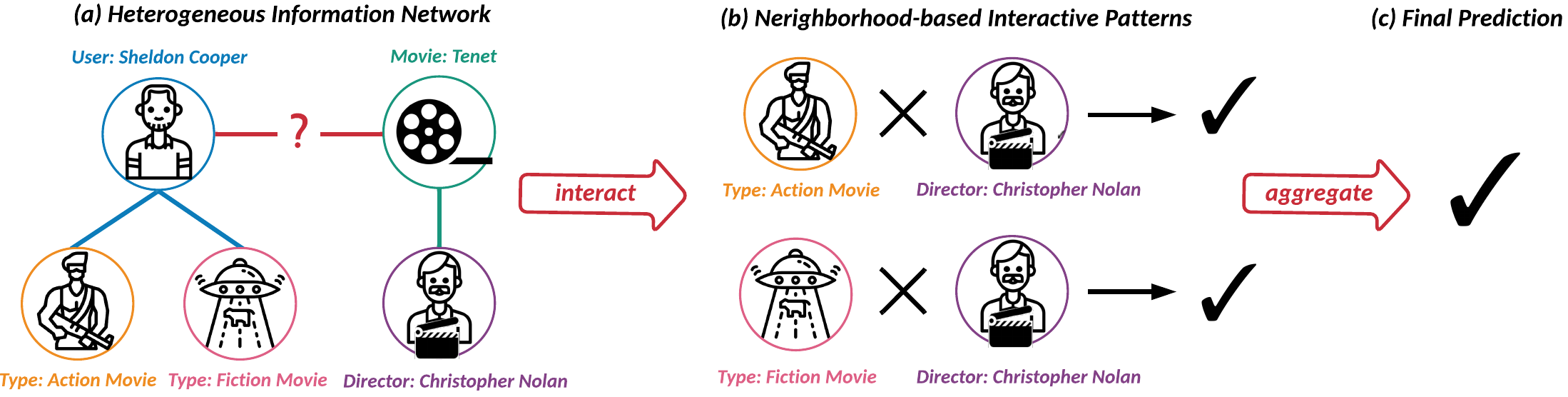}
	\vspace{-2mm}
	\caption{
		An illustrated example of the motivation of GraphHINGE, where we adopt ``interact'' and ``aggregate'' to formulate ``AND'' operation, capture and fuse the interactive patterns.
		}
	\label{fig:example}
% 	\vspace{-5mm}
\end{figure}

	\vspace{-1mm}
	\begin{definition}
		\label{def:neighbor}
		\rm
		\textbf{Metapath-guided Neighborhood.}
		Given an object $o$ and a metapath $\rho$ (start from $o$) in an HIN, the metapath-guided neighborhood is defined as the set of all visited objects when the object $o$ walks along the given metapath $\rho$.
		In addition, we denote the neighborhoods of object $o$ after $i$-th steps sampling as $\mathcal{N}_\rho^i(o)$.
		Specifically, $\mathcal{N}_\rho^0(o)$ is $o$ itself.
		For convenience, let $\mathcal{N}_\rho(o)$ denote $\mathcal{N}^{I-1}_\rho(o)$, where $I$ means the length of metapath.
		Note that different from the similar concept proposed in \cite{wang2019heterogeneous}, which is used to derive homogeneous neighbors on heterogeneous graphs, the metapath-guided neighborhood in this paper can preserve semantic contents since it consists of heterogeneous information.
	\end{definition}	
	\vspace{-1mm}

	Taking Fig.~\ref{fig:instance} as an instance, given the metapath ``User-Movie-Director (UMD)'' and a user $u_\text{A}$, we can get the corresponding metapath-guided neighborhoods as $\mathcal{N}_\rho^1(u_\text{A})$ = $\{(u_\text{A}, m_\text{A})$, $(u_\text{A}, m_\text{B})\}$, $\mathcal{N}_\rho(u_\text{A})$ = $\mathcal{N}_\rho^2(u_\text{A})$ = $\{(u_\text{A}, m_\text{A}, d_\text{A})$, $(u_\text{A}, m_\text{A}, d_\text{B})$, $(u_\text{A}, m_\text{A}, d_\text{C})$, $(u_\text{A}, m_\text{B}, d_\text{B})\}$.
	
	\minisection{Remark}
	Note that different from the metapath proposed in previous works \citep{liu2018interactive,wang2019heterogeneous}, where a path following the specific schema connects the source and target nodes; the metapath adopted here is to guide the neighborhood sampling.
	Hence, there only exists a requirement about where a path starts and no requirement about where a path ends.
% 	As shown in \citep{qiu2020gcc}, this sampling strategy is able to capture both graph-based and path-based structure in neighborhood.
	
	Many efforts have been made for HIN-based recommendations.
	However, most of these works focus on leveraging graph neural networks to aggregate structured message but overlook the effect of early summarization and rich interactive information.
	Given the above preliminaries, we are ready to introduce our GraphHINGE model as a solution.
	% especially in real-world scenario, often noisy and complex.

\section{Methodology}
\label{sec:method}
\subsection{Overview}
The basic idea of our methodology is to design an end-to-end GraphHINGE framework that involves an interaction module to capture the rich interactive patterns hidden in HINs and an aggregation module to integrate the information for the final prediction. 
Fig.~\ref{fig:example} illustrates an example.
Assume that user $u_\text{Sheldon Cooper}$ favors action movies (i.e., $t_\text{Action Movie}$) and fiction movies (i.e., $t_\text{Fiction Movie}$), our task is to predict his preference on movie $m_\text{Tenet}$ directed by director $d_\text{Christopher Nolan}$.
This task can be formulated via link prediction between $u_\text{Sheldon Cooper}$ and $m_\text{Tenet}$ nodes based on the HIN shown in Fig.~\ref{fig:example}(a).
In this HIN, movie types $t_\text{Action Movie}$ and $t_\text{Fiction Movie}$ are neighbors of user $u_\text{Sheldon Cooper}$, and director $d_\text{Christopher Nolan}$ is a neighbor of $m_\text{Tenet}$.
In this paper, we propose to find the co-relations between neighbors to help with the final prediction.
As Fig.~\ref{fig:example}(b) illustrates, director $d_\text{Christopher Nolan}$ is good at action movies (i.e., $t_\text{Action Movie}$) and fiction movies (i.e., $t_\text{Fiction Movie}$).
Therefore, we can infer that there is a link between user $u_\text{Sheldon Cooper}$ and movie $m_\text{Tenet}$ (i.e., $u_\text{Sheldon Cooper}$ favors $m_\text{Tenet}$) in Fig.~\ref{fig:example}(c).
As stated in \citep{qu2018product,rendle2020neural}, these interactive patterns cannot simply captured by neural networks and require explicitly modelling ``AND'' operations in interaction modules.

\begin{figure}[t]
	\centering
	\vspace{-2mm}
	\includegraphics[width=1.0\textwidth]{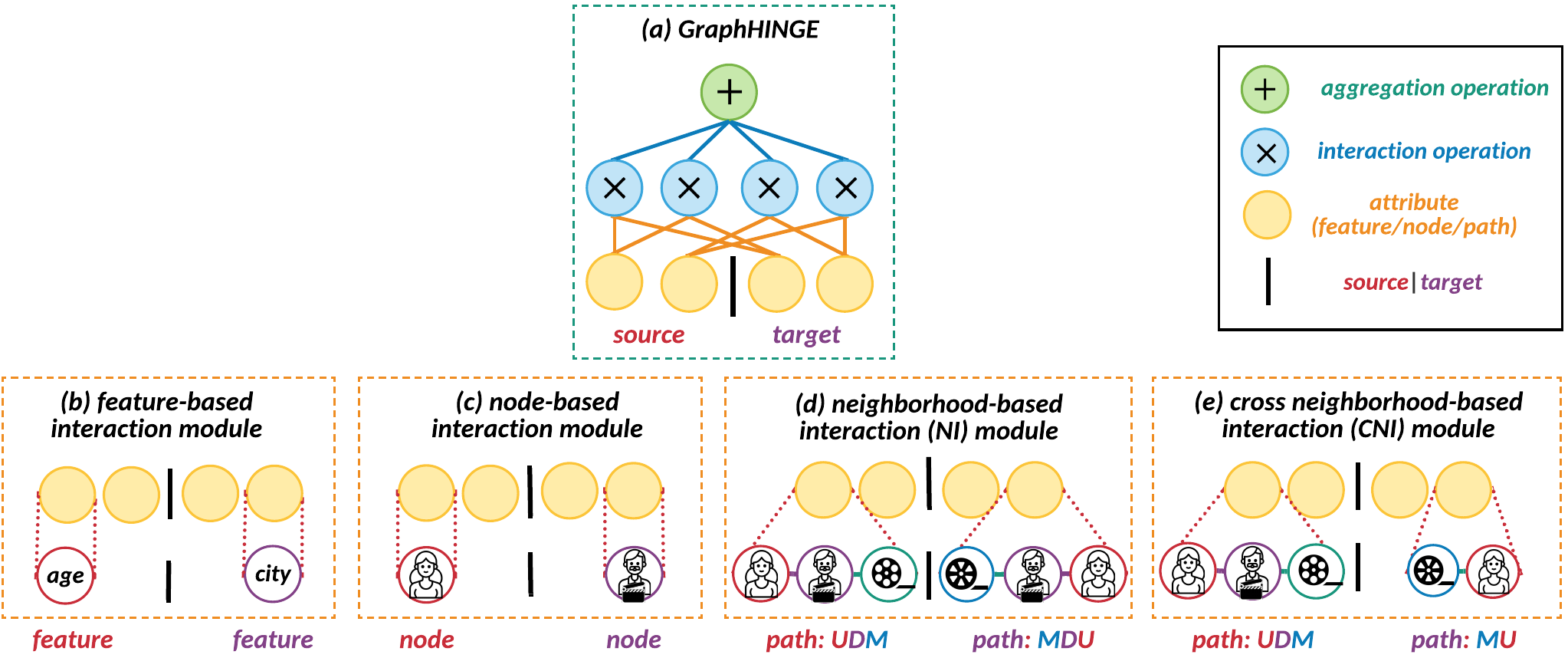}
	\vspace{-5mm}
	\caption{
		Illustration of Interaction Modules. 
		(a) The overview of the GraphHINGE model, where yellow nodes mean one-hot inputs of fields; green ``$+$'' nodes mean aggregation operations; blue ``$\times$'' nodes mean interaction operations.
		(b) In the feature-based interaction module, the yellow nodes represent features in multiple fields (e.g.,~age, location, date).
		(c) In the node-based interaction module, the yellow nodes represent nodes in multiple types (e.g.,~user, movie, director).
		(d) In the neighborhood-based interaction (NI) module, the yellow nodes represent paths guided by the single metapath 
		%(and metapth in reverse order) 
		(e.g.,~UMD and DMU).
		(e) In the cross neighborhood-based (CNI) module, the yellow nodes represent paths guided by multiple metapaths (e.g.,~UMD and UM).
		}
	\label{fig:interaction}
% 	\vspace{-5mm}
\end{figure}

As illustrated in Fig.~\ref{fig:interaction}(a), a naive solution is to extend previous feature-based interaction modules into the HIN case.
As Fig.~\ref{fig:interaction}(b) shows, feature-based interaction modules such as FM \citep{rendle2010factorization} are designed to capture interactive patterns from data in the multi-field categorical form.
% For example, a user with `\textsc{location=China}' would perform several specific purchases and browse logs at `\textsc{date=Chinese New Year}', while another user with `\textsc{location=USA}' has similar performance at `\textsc{date=Thanksgiving}'.
% Hence, the connections between `\textsc{location=China}' and `\textsc{date=Chinese New Year}' are useful information for model training while the connections between `\textsc{location=China}' and `\textsc{date=Thanksgiving}' should be regarded as noise.
If we directly extend these interaction modules into the HIN case by treating nodes as features and types as fields as illustrated in Fig.~\ref{fig:interaction}(c), nodes will be regarded equally and interact with each other freely.
However, one should be noted that nodes located at different topological positions would contribute differently to the final prediction.
For instance, nodes close to the source/target node can be more informative than those far away from source/target node when predicting whether there is a link between the source and target nodes.

Therefore, as Fig.~\ref{fig:interaction}(d) shows, instead of nodes, we define interaction as an operation between paths guided by the metapaths.
According to Fig.~\ref{fig:def}, different from traditional paths connecting source and target nodes, paths here represent neighborhoods. 
Thus this interaction module is called the neighborhood-based interaction (NI) module.
Note that the metapaths on HINs always show the semantics.
In the NI module, we generate the interactive patterns with neighborhoods of users and items guided by the same metapaths (i.e.,~the same semantics).
% we consider the interactive patterns captured by NI are under the same semantic.
For instance, Fig.~\ref{fig:interaction}(d) illustrates the interactive patterns between user $u_\text{B}$ and movie $m_\text{C}$ guided by UMD, which denotes that user $u_\text{B}$ likes movie $m_\text{C}$ directed by director $d_\text{C}$.
Following the semantics, NI investigates whether there exists common/similar neighbor nodes (e.g.,~director $d_\text{C}$, movie $m_\text{A}$) of user $u_\text{B}$ and movie $m_\text{C}$. 
It is natural to consider extracting rich structured information of user $u_\text{B}$ and movie $m_\text{C}$ under different semantics even with different lengths.
As illustrated in Fig.~\ref{fig:interaction}(e), we further extend NI to CNI module, where we operate to seek for correlations between two neighborhoods guided by different semantics (e.g.,~source $s=u_\text{B}$ and target $t=m_\text{C}$'s neighbor $u_\text{B}$). 
% which extracts directors and movies in neighborhood of user $u_B$ to obtain highly related movies, and users in neighborhood of movie $m_C$ to obtain related users.
% Then we can present relationship between $u_B$ and $m_C$ according the interaction between related movies of $u_B$ and related users of $m_C$.

\begin{figure}[t]
	\centering
	\vspace{-2mm}
	\includegraphics[width=1.0\textwidth]{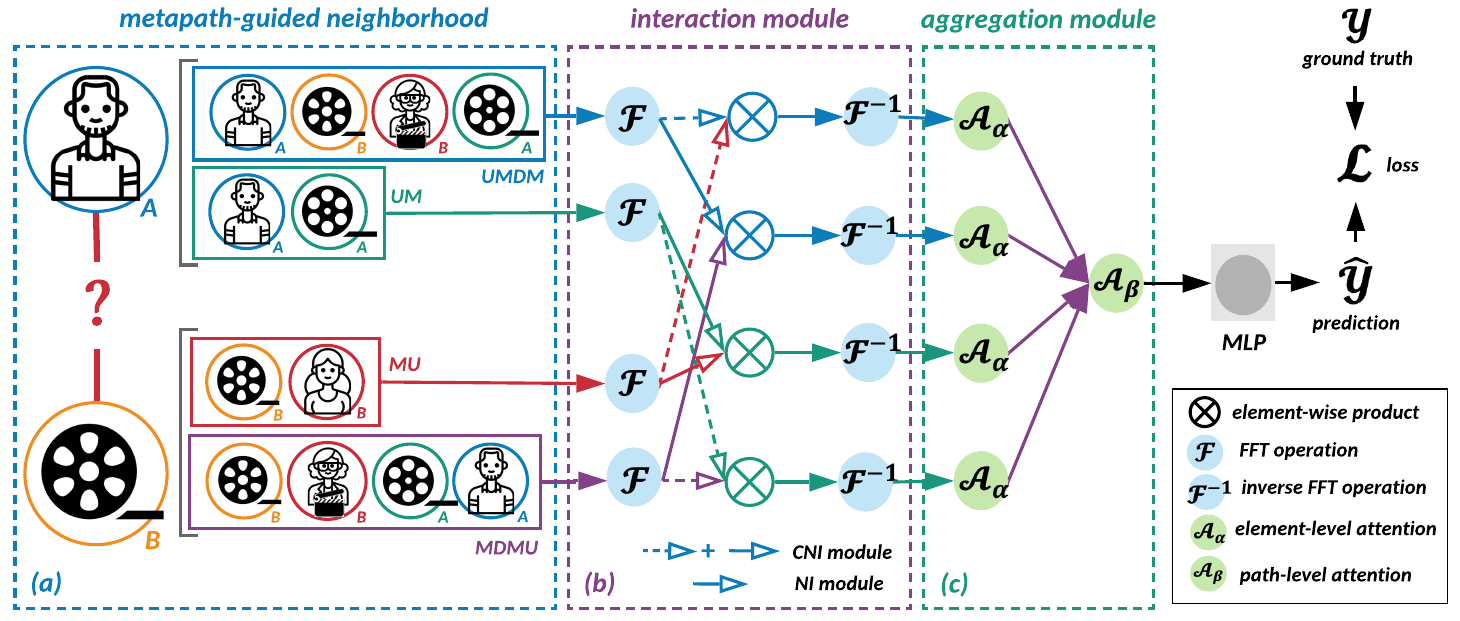}
	\vspace{-2mm}
	\caption{
		The overall architecture of GraphHINGE: (a) it first samples metapath-guided neighborhood (Section~\ref{subsec:sample}); (b) next constructs interactive information via interaction layer, where the interaction operations in NI module (Section~\ref{subsec:interaction}) are represented in solid lines and those in CNI module are represented in solid and dotted lines (Section~\ref{subsec:crossinteraction}); (c) finally combines rich information through aggregation layer (Section~\ref{subsec:aggregation}).
	}
	\label{fig:overview}
% 	\vspace{-5mm}
\end{figure}

With the help of interaction modules, our approach is able to leverage metapaths to guide the selection of different-step and -type neighbors, capture and aggregate the abundant interactive patterns.
Moreover, we represent different types of metapaths with a unified learning procedure and design an efficient learning algorithm with fast Fourier transform (FFT).  
We provide the overall framework of our model in Fig.~\ref{fig:overview}.
First, we use the multiple-object HIN containing $\langle$user, item, attribute, relation$\rangle$ as the input.
Second, we select metapath-guided neighborhoods for source and target nodes via neighbor samplings (Fig.~\ref{fig:overview}(a)).
Third, we introduce the interaction module and formulate it via a convolutional framework to generate interaction information among their neighborhoods (Fig.~\ref{fig:overview}(b)).
Specially, the NI module is designed to model the interactive patterns between source and target nodes under the same semantics, which is illustrated in solid lines, while the CNI module can implement interactions under different semantics and provide additional cross semantic information besides interaction information generated by the NI module, which is shown in dotted and solid lines.
% The inverse operation in Figure~\ref{fig:overview}(a) is designed to represent both similarity and co-rating interactive patterns in Figure~\ref{fig:overview}(b) (see Figure~\ref{fig:fft} and Section~\ref{subsec:interaction} for details).
After that, we capture the key interactions and aggregate information through attention mechanisms in both node and path levels (Fig.~\ref{fig:overview}(c)).
Finally, our model provides the final prediction.
% after interaction and end-to-end optimization.
We illustrate the architecture in detail in the following subsections.

\subsection{Neighborhood Sampling}
\label{subsec:sample}
To generate meaningful node sequences, the key technique is to design an effective random walk strategy that is able to capture the complex semantics reflected in HINs.
Hence, we propose to use the metapath-guided random
walk method with restart.
Given an HIN $\mathcal{G} = \{\mathcal{V}, \mathcal{E}\}$ and a metapath $\rho: A_0, \ldots, A_i, \ldots, A_{I-1}$, where $A_i \in \mathcal{A}_i$ denotes the $i$-th node guided by metapath $\rho$ (note that we include user set $\mathcal{U}$ and item set $\mathcal{I}$ in attribute set $\mathcal{A}$ for convenience), the walk path is generated according to the following distribution:
\begin{equation}
\label{eqn:neighbor}
P(n_{k+1}=x|n_k=v) = \left\{
\begin{aligned}
\frac{1}{|\mathcal{N}^1_\rho(v)|}, \ & \ (v, x) \in \mathcal{E} \ \text{and} \ v, x \in \mathcal{A}_k, \mathcal{A}_{k+1} \\
0, \ & \ \text{otherwise};
\end{aligned}
\right.
\end{equation}
where $n_k$ is the $k$-th node in the walk, $\mathcal{N}^1_\rho(v)$ means the first-order neighborhood of node $v$ guided by metapath $\rho$.
A walk will follow the pattern of the metapath repetitively until it reaches the predefined length and always start from the source/target node.
It is worth noting that, according to \textsc{Definition}~\ref{def:neighbor}, there is no need to sample a complete path from source to target nodes.
% Also, metagraph structure can be obtained by this sampling strategy.

\subsection{Neighborhood-based Interaction Module}
\label{subsec:interaction}
In previous HIN-based recommendations, most approaches leverage graph representation techniques to find key nodes or metapaths \cite{wang2019heterogeneous} and capture the complex structure \cite{zhang2019heterogeneous}.
To further mine interaction information and deal with the early summarization issue, we propose an interaction module based on metapath-guided neighborhoods.

Due to the heterogeneity of nodes, different types of nodes have different feature spaces.
Hence, for each type of nodes (e.g.,~node with type $\phi_i$), we design a type-specific transformation matrix $\mathbf{M}_{\phi_i}$ to project the features of different types of nodes into a unified feature space.
The project process can be shown as follows:
\begin{equation} 
\label{projection}
e_i = \mathbf{M}_{\phi_i} \cdot e'_i,
\end{equation}
where $e'_i$ and $e_i$ are the original and projected features of node $i$, respectively.
By the type-specific projection operation, our model is able to handle arbitrary types of nodes.

\begin{figure}[t]
	\centering
	\vspace{-2mm}
	\includegraphics[width=0.7\textwidth]{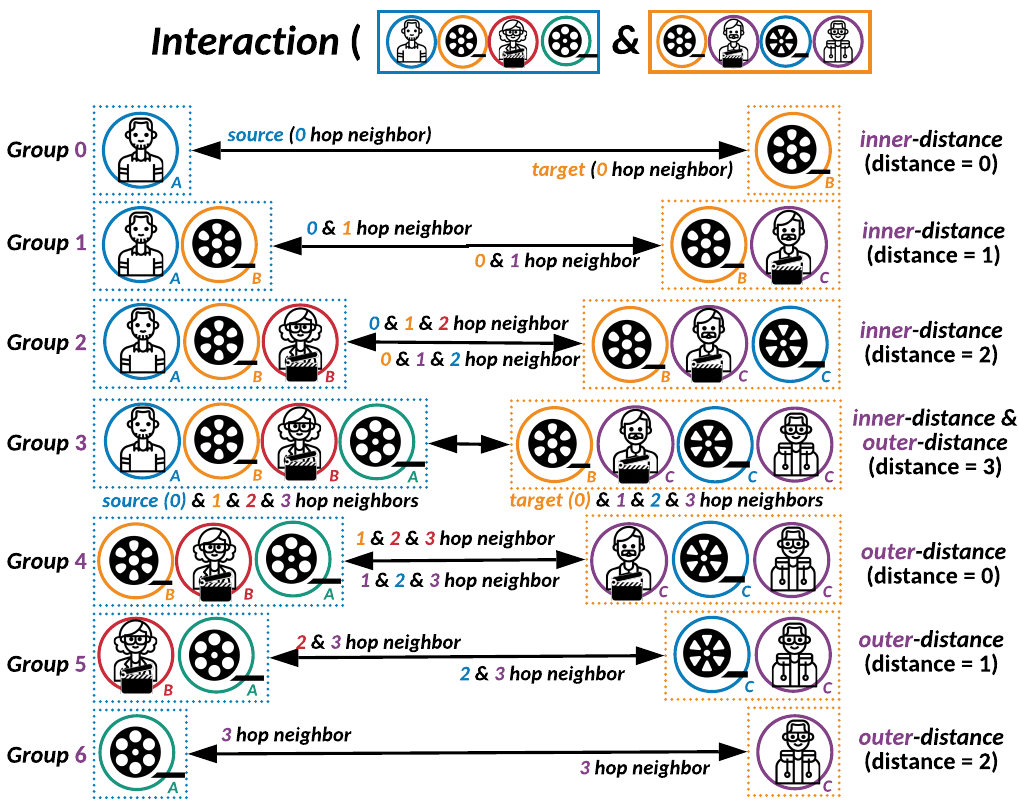}
	\vspace{-2mm}
	\caption{
		An illustrated example of the motivation in the interaction module design. Neighborhoods are grouped according
		to the distance to the source/target node. Interaction is only
		employed between corresponding neighborhoods in each group.
	}
	\label{fig:motivation}
	\vspace{-2mm}
\end{figure}

Considering that neighbors in different distances to the source/target node usually contribute differently to the final prediction, we divide the sampled metapath-guided neighborhood into several inner-distance and outer-distance neighborhood groups.
As illustrated in Fig.~\ref{fig:motivation}, when we set distance as 1, we can get inner-distance \emph{Group} 1 and outer-distance \emph{Group} 5.
In a similar way, we can obtain $2I-1$ groups, where $I$ denotes the metapath length.
We argue that interactions should only be employed in corresponding groups.
% Note that each inner-distance neighborhood (\emph{i.e.}, \emph{Group} 0-0) together with correspond outer-distance neighborhood  (\emph{i.e.}, \emph{Group} 0-1) composes the complete metapath-guided neighborhood.
% For example, as illustrated in Figure~\ref{fig:mov}, 
% We then conduct interaction between inner-distance neighbors (itself as 0 hop neighbor in this case) in \emph{Group} 0-0; and outer-distance neighbors (1 \& 2 \& 3 hop neighbors in this case) in \emph{Group} 0-1.
In order to perform interaction in each neighborhood group, we need to face two situations.
If there is only one node in \emph{Group}, we adopt an element-wise product (i.e.,~``AND'') operation to measure their similarity or co-ratings (e.g., $r(u_\text{A}, m_\text{B})$ in \emph{Group} 0 case).
When there are more than one nodes, we first do interaction by production and then aggregate by summarization, (e.g.,~$r(u_\text{A}, d_\text{C}) + r(m_\text{B}, m_\text{B})$ in \emph{Group} 1 case).
% \weinan{to explain why production operation can capture the `AND' pattern, suggest to refer the analysis of PIN paper here.}

\begin{figure}[t]
	\centering
	\vspace{-2mm}
	\includegraphics[width=1\textwidth]{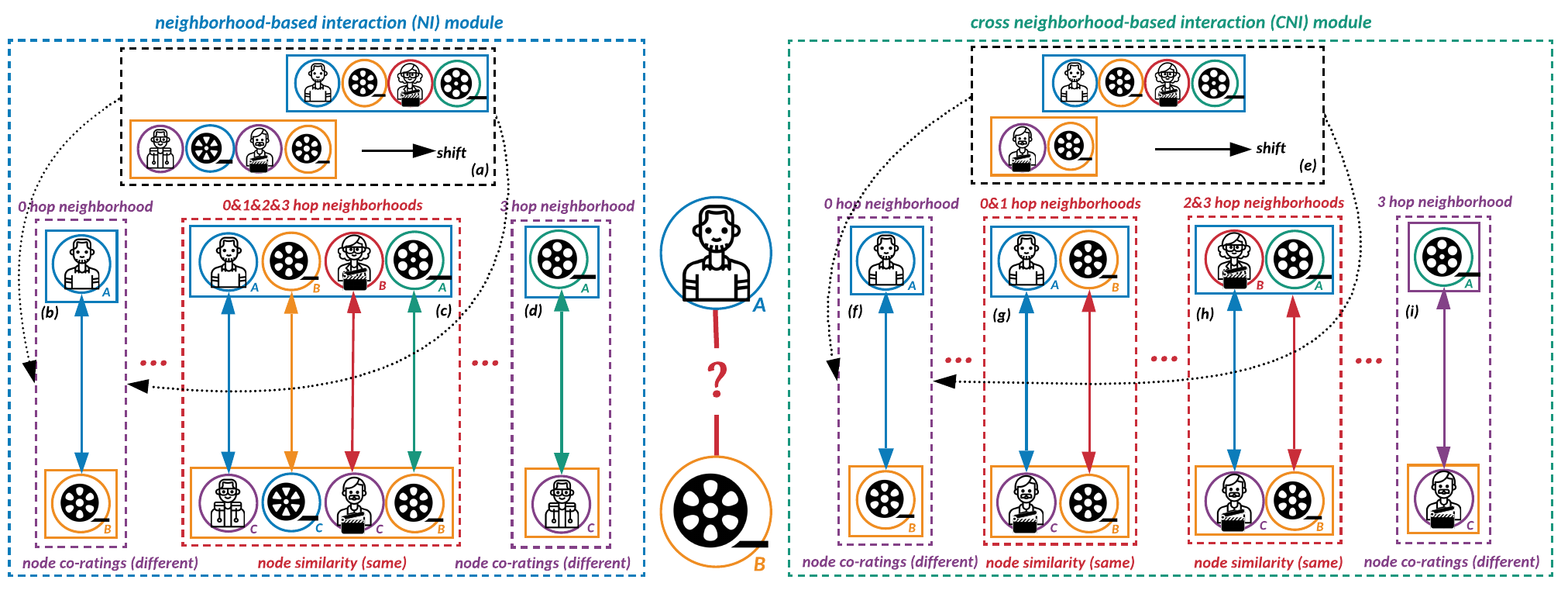}
	\vspace{-6mm}
	\caption{
		An illustrated example of interaction operations with the left part for the NI module and the right part for the CNI module.
		(a) and (b) are the illustrations of interaction operation of NI and CNI modules respectively via a convolutional framework.
		The other parts show the results which contain information of both nodes similarities ((c) in NI module and (g) \& (h) in CNI module) and node co-ratings ((b) \& (d) in NI module and (f) \& (i) in CNI module).
	}
	\label{fig:fft}
	\vspace{-2mm}
\end{figure}

Inspired by work in signal processing field \citep{oppenheim1999discrete}, this neighborhood division and interaction can be formulated as a unified operation named convolution.
Roughly speaking, the convolution mainly contains three kinds of operations on paths, namely shift, product, and sum, which are employed repeatedly until they meet the end of the path.
Take Fig.~\ref{fig:fft}(a) for example.
Our task is to calculate the interaction between source user neighborhood path $(u_\text{A}, m_\text{B}, d_\text{B}, m_\text{A})$ and target movie neighborhood path $(m_\text{B}, d_\text{C}, m_\text{C}, u_\text{C})$.
First, we inverse the order of target movie path and obtain $(u_\text{C}, m_\text{C}, d_\text{C}, m_\text{B})$.
We shift it from left to right and product the overlapping nodes during the shift.
As shown in Fig.~\ref{fig:fft}(a), the first overlapping happens between source and target nodes, namely the 0-hop neighborhood.
We conduct the product operation and obtain the co-ratings between different types of nodes $r(u_\text{A}, m_\text{B})$ (as Fig.~\ref{fig:fft}(b) shows).
Then, we repeatedly shift, product, and sum, and then reach the situation where all nodes are overlapped.
The result in this situation is the similarity between the same type of nodes $r(u_\text{A}, u_\text{C}) + r(m_\text{B}, m_\text{C}) + r(d_\text{B}, d_\text{C}) + r(m_\text{A}, m_\text{B})$ (as shown in Fig.~\ref{fig:fft}(c)).
In a similar way, the last interaction happens between different types of nodes $r(m_\text{A}, u_\text{C})$ (as shown in Fig.~\ref{fig:fft}(d)).

Let $\mathbf{H}[\mathcal{N}_\rho(o)]$ denote the embedding matrix of metapath $\rho$ guided neighborhood of object $o$.
$\mathbf{H}[\mathcal{N}_\rho(o)]_l$ represents the embedding matrix of the $l$-th path, which can be formulated as
\begin{equation}
\label{eqn:embedding}
\mathbf{H}[\mathcal{N}_\rho(o)]_l = [e^{\rho^l}_0 \oplus e^{\rho^l}_1 \oplus \cdots \oplus e^{\rho^l}_{I-1}],
\end{equation}
where $l=0,1,\ldots,L-1$, $e_i^{\rho^l}$ means the embedding of the node in the $i$-th step of the $l$-th path guided by metapath $\rho$, $\oplus$ denotes the stack operation\footnote{We regard concatenation and stack operations as the same operation, and denote it with ``$\oplus$''. Namely, the reshape operation is omitted here.}, $I$ means the metapath length.
Hence, as illustrated in Fig.~\ref{fig:cube}(a), $\mathbf{H}[\mathcal{N}_\rho(o)]$ is a $\mathbb{R}^{L \times I \times E}$ matrix, where $L$ is the number of paths guided by the metapath, $I$ is the length of the metapath, $E$ means the dimension of the node embedding.
Based on convolutional operations, we further define the interaction between neighborhoods of source and target objects as
\begin{equation}
\mathbf{H}[\mathcal{N}_\rho(s), \mathcal{N}_\rho(t)]_{l} = \mathbf{H}[\mathcal{N}_\rho(s)]_{l} \odot \mathbf{H}[\mathcal{N}_\rho(t)]_{l},
\end{equation}
where $\odot$ denotes the convolutional operation.
According to the definition of convolution, one can write the formulation as
\begin{equation}
\label{eqn:interact}
\mathbf{H}[\mathcal{N}_\rho(s), \mathcal{N}_\rho(t)]_{l, n} = \sum_{\substack{a, b\\ a+b ~ \text{mod} ~ N = n}} \mathbf{H}[\mathcal{N}_\rho(s)]_{l, a} \cdot \mathbf{H}[\mathcal{N}_\rho(t)]_{l, b}.
\end{equation}

One can find that $\mathbf{H}[\mathcal{N}_\rho(s), \mathcal{N}_\rho(t)] \in \mathbb{R}^{L \times N \times E}$ (as shown in Fig.~\ref{fig:cube}(c)), where $N$ is the length of convolution outputs and it equals $2I-1$ where $I$ denotes the metapath length.
% of source and that of target nodes respectively.
% In this case, the metapaths of source and target nodes are equal with each other.
% For convenience, we denote $I=I_s=I_t$ and derive $N=2I-1$.

The well-known convolution theorem states that circular convolutions in the spatial domain are equivalent to pointwise products in the Fourier domain. 
Let $\mathcal{F}$ denote the fast Fourier transform (FFT) and $\mathcal{F}^{-1}$ its inverse, and we can compute convolution as
\begin{equation}
\label{eqn:interaction}
\mathbf{H}[\mathcal{N}_\rho(s), \mathcal{N}_\rho(t)] = \mathcal{F}^{-1}(\mathcal{F}(\mathbf{H}[\mathcal{N}_\rho(s)]) \cdot \mathcal{F}(\mathbf{H}[\mathcal{N}_\rho(t)])).
\end{equation}

Let $\mathbf{H}[\mathcal{N}_\rho]$ denote $\mathbf{H}[\mathcal{N}_\rho(s), \mathcal{N}_\rho(t)]$ in the following sections for convenience.
As stated in \cite{mathieu2013fast}, the time complexity of plain convolution is $\mathcal{O}(I^2)$, and it is reduced to $\mathcal{O}(I \log (I))$ when using FFT.	
According to the analysis above, we can conclude that not only can this structure capture both nodes similarities and co-ratings in grouped neighborhoods, but also it can be implemented with high efficiency.
% We provide efficiency analysis of FFT method in Section~\ref{sec:analysis}.

\subsection{Cross Neighborhood-based Interaction Module}
\label{subsec:crossinteraction}
In the previous section, we introduce the neighborhood-based interaction module built on HIN with the user's and item's neighborhoods guided by the same metapaths.
In HINs, each metapath represents one corresponding semantic information \citep{liu2018interactive,wang2019heterogeneous}.
For instance, as illustrated in Fig.~\ref{fig:instance}(c), UM metapath shows that user $u_\text{B}$ likes movie $m_\text{C}$, while UU shows that users $u_\text{B}$ and $u_\text{C}$ are friends.
It's natural to consider the hidden relation between user $u_\text{C}$ and movie $m_\text{C}$ under the combination of these two different semantics.
To this end, a well-designed interaction module should also take the interactions between different metapaths into consideration.
As shown in Fig.~\ref{fig:overview}(b), the interaction operations in the NI module are illustrated in solid lines.
These interactions implemented between neighborhoods of user $u_\text{A}$ and movie $m_\text{B}$ guided by the same metapaths are designed to capture the interactive information under the specific semantics.
In order to enhance the interaction module with interactive patterns between different semantics, we further propose to model the interaction operations with different metapaths even in different lengths, as illustrated with the dotted lines in Fig.~\ref{fig:overview}(b).
We call it the Cross Neighborhood-based Interaction (CNI) module.
%, which aims to capture the interactive patterns under the cross semantics.
Note that the NI module is actually a special case of the CNI module, where the interaction operation is limited between neighborhoods guided by the same metapaths. 
The solid and dotted lines in Fig.~\ref{fig:overview}(b) together represent the interaction operations in our CNI module.

Let $\mathbf{H}[\mathcal{N}_{\rho_s}(s)]$ and $\mathbf{H}[\mathcal{N}_{\rho_t}(t)]$ denote the embedding matrices of neighborhood guided by metapath $\rho_s$ of source node $s$ and neighborhood guided by $\rho_t$ of target node $t$, respectively.
$\mathbf{H}[\mathcal{N}_{\rho_o}(o)]_l$ represents the embedding matrix of the $l$-th path, which can be formulated as
\begin{equation}
\label{eqn:stembedding}
\mathbf{H}[\mathcal{N}_{\rho_s}(s)]_l = [e^{\rho_s^l}_0 \oplus e^{\rho_s^l}_1 \oplus \cdots \oplus e^{\rho_s^l}_{I_s-1}], \quad \mathbf{H}[\mathcal{N}_{\rho_t}(t)]_l = [e^{\rho_t^l}_0 \oplus e^{\rho_t^l}_1 \oplus \cdots \oplus e^{\rho_t^l}_{I_t-1}];
\end{equation}
where $l=0,1,2,\ldots,L-1$, $e^{\rho_o^l}_i$ means the embedding of the node in the $i$-th step of the $l$-th path guided by metapath $\rho_o$, $\oplus$ denotes the stack operation, $I_o$ is the path length. 
Here, $o$ can denote both $s$ for source node and $t$ for target node.

\begin{figure}[t]
	\centering
	\vspace{-2mm}
	\includegraphics[width=1\textwidth]{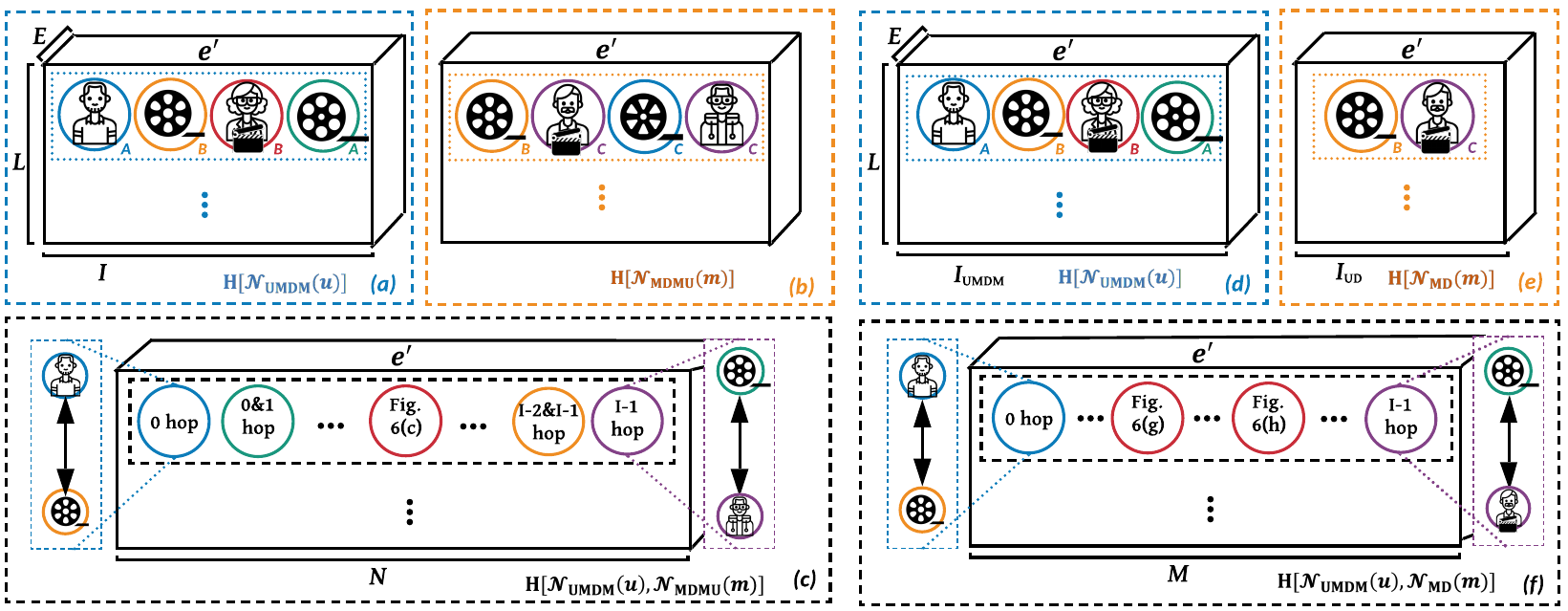}
	\vspace{-6mm}
	\caption{
		An illustrated example of the embedding matrix of metapath guided neighborhoods of source node $s$ $\mathbf{H}[\mathcal{N}_{\rho_s}(s)]$ and target node $t$ $\mathbf{H}[\mathcal{N}_{\rho_t}(t)]$ generated according to Eq.~(\ref{eqn:embedding}).
		In the NI module, the neighborhoods of both source and target nodes are guided by the same metapath $\rho=\rho_s=\rho_t$ ((a) \& (b)), while the CNI module allows neighborhoods guided by different metapaths ((d) \& (e)).
		The interaction matrix $\mathbf{H}[\mathcal{N}_{\rho_{s,t}}]$ is calculated according to Eq.~(\ref{eqn:interaction}) in the NI module (c) and Eq.~(\ref{eqn:crossinteraction}) in the CNI module (f).
	}
	\label{fig:cube}
% 	\vspace{-2mm}
\end{figure}

As illustrated in Fig.~\ref{fig:cube}(e) and (f), $\mathbf{H}[\mathcal{N}_{\rho_o}(o)] \in \mathbb{R}^{L \times I_o \times E}$, where $L$ is the number of paths sampled by each metapath, $E$ means the dimension of the node embedding.
Based on convolutional operations, we further define the interaction between neighborhoods of source and target objects as
\begin{equation}
\mathbf{H}[\mathcal{N}_{\rho_s}(s), \mathcal{N}_{\rho_t}(t)]_{l} = \mathbf{H}[\mathcal{N}_{\rho_s}(s)]_{l} \odot \mathbf{H}[\mathcal{N}_{\rho_t}(t)]_{l},
\end{equation}
where $\odot$ denotes the convolutional operation.
Here, we define the convolution as the sum of the product operations, which can be formulated as
\begin{equation}
\label{eqn:crossinteract}
\mathbf{H}[\mathcal{N}_{\rho_s}(s), \mathcal{N}_{\rho_t}(t)]_{l, m} = \sum_{\substack{a, b\\ a+b ~ \text{mod} ~ M = m}} \mathbf{H}[\mathcal{N}_{\rho_s}(s)]_{l, a} \cdot \mathbf{H}[\mathcal{N}_{\rho_t}(t)]_{l, b},
\end{equation}
where $\mathbf{H}[\mathcal{N}_{\rho_s}(s), \mathcal{N}_{\rho_t}(t)] \in \mathbb{R}^{L \times M \times E}$ (as shown in Fig.~\ref{fig:cube}(h)), $M$ is the length of convolution outputs and it is equal to $I_s + I_t - 1$ where $I_s$, $I_t$ denote the metapath length of source and that of target node, respectively.
Note that when $I=I_s=I_t$, CNI and NI modules share the same formulation.
In fact, the NI module is a special case of the CNI module when the lengths of metapaths of the source and target are the same.
% \begin{equation}
% \label{eqn:crossinteract}
% \mathbf{H}[\mathcal{N}_{\rho_s}(s), \mathcal{N}_{\rho_t}(t)]_{l, m} = \sum_{\substack{a, b, \kappa\\ a+b ~ \text{mod} ~ M = m}} \mathbf{H}[\mathcal{N}_{\rho_s}(s)]_{l, a\sim a+\kappa} \times \mathbf{H}[\mathcal{N}_{\rho_t}(t)]_{l, b-\kappa\sim b},
% \end{equation}
% where $\times$ presents Hadamard product operation, $\kappa$ is the size of kernel, $\sum$ here means the sum over all elements in the matrix, and $\mathbf{H}[\mathcal{N}_{\rho_o}(o)]_{l, x\sim x+\kappa}=[e^{\rho_o}_x \oplus e^{\rho_o}_{x+1} \oplus \cdots \oplus e^{\rho_{o}}_{x+\kappa}]$.
% One can find that $\mathbf{H}[\mathcal{N}_{\rho_s}(s), \mathcal{N}_{\rho_t}(t)]_{l, m} \in \mathbb{R}^{L \times M \times E}$ (as shown in Fig.~\ref{fig:cube}(h)), $M$ is the length of convolution outputs and it equals to $I_s + I_t -2\kappa + 1$ where $I_s$, $I_t$ denote the metapath length of source and that of target nodes respectively.
% One should be noted that $\kappa$ is restricted from $1$ to $I_\text{min}$.
% When $\kappa=1$ and $I=I_s=I_t$, then CNI and NI share the same formulation.
% Namely, NI is the special case of CNI.
% Similarly, this interaction operation can be regarded as the sliding window approach.
% An illustration of interaction 
We propose to replace the convolutional operation with the fast Fourier transform (FFT) using the discrete analogue of the convolution theorem as
\begin{equation}
\label{eqn:crossinteraction}
\mathbf{H}[\mathcal{N}_{\rho_s}(s), \mathcal{N}_{\rho_t}(t)] = \mathcal{F}^{-1}(\mathcal{F}(\mathbf{H}[\mathcal{N}_{\rho_s}(s)]) \cdot \mathcal{F}(\mathbf{H}[\mathcal{N}_{\rho_t}(t)])).
\end{equation}
We use $\mathbf{H}[\mathcal{N}_{\langle\rho_s,\rho_t\rangle}]$ to denote $\mathbf{H}[\mathcal{N}_{\rho_s}(s), \mathcal{N}_{\rho_t}(t)]$ for convenience.
As stated in \cite{pratt2017fcnn}, the time complexity of plain convolution is $O(I^2)$, and it is reduced to $O(I \log (I))$ when using FFT, where $I=\max\{I_s,I_t\}$.
According to the analysis above, we can see that not only can this structure mine the interactions between different metapaths, but also it can be regarded as the extension of NI and accelerated with FFT. 

In the CNI module, let $\langle\rho_s,\rho_t\rangle$ denote the combinations of metapaths $\rho_s$ and $\rho_t$ for convenience, where $\rho_s$ guides the neighborhood $\mathcal{N}_{\rho_s}(s)$ on source node $s$ side and $\rho_t$ guides the neighborhood $\mathcal{N}_{\rho_t}(t)$ on target node $t$ side.
We further use $\langle\rho_s,\rho_t\rangle_k$ to denote the $k$-th metapath combination, where $k=0,1,2,\ldots,K-1$.

\subsection{Aggregation Module}
\label{subsec:aggregation}
Considering that the NI module in Section~\ref{subsec:interaction} is a special case of the CNI module in Section~\ref{subsec:crossinteraction}, without loss of generality, we introduce an aggregation module based on the interaction result of the CNI module in this section.
Specifically, we consider the aggregation module in two sides.
On the first side, from Fig.~\ref{fig:cube}, we can see that elements in the interaction matrix $\mathbf{H}[\mathcal{N}_{\langle\rho_s,\rho_t\rangle}]$ contain interactions between various types of nodes.
Hence, it is natural to capture the key interaction at the element-level during the aggregation procedure.
For instance, the interaction between $u_\text{A}$ and $m_\text{A}$ is important when predicting the relation between $u_\text{A}$ and $m_\text{B}$ in Fig.~\ref{fig:instance}.
On the second side, for each type of metapath $\rho$ (or metapath combination $\langle\rho_s,\rho_t\rangle$), we sample $L$ paths according to Eq.~(\ref{eqn:neighbor}).
Therefore, a good aggregation module is required to distinguish useful paths and filter out the noise at the path-level.
For example, following the UMD schema, we sample $L$ paths to predict the relation between $u_\text{A}$ and $m_\text{B}$, among which path $(u_\text{A},m_\text{A},d_\text{C})$ may be more helpful than path $(u_\text{A},m_\text{A},d_\text{A})$ in Fig.~\ref{fig:instance}.
On the third side, for every source-target pair $(s,t)$, HIN contains multiple types of (cross) semantic information represented with different metapaths $\rho_0, \rho_1, \cdots, \rho_{P-1}$ where $P$ is the number of metapaths (or metapath combinations $\langle\rho_s,\rho_t\rangle_0,\langle\rho_s,\rho_t\rangle_1,\ldots,\langle\rho_s,\rho_t\rangle_{K-1}$ where $K$ is the number of metapath combinations).
This further causes various interaction matrices $\mathbf{H}[\mathcal{N}_{\rho_0}]$, $\mathbf{H}[\mathcal{N}_{\rho_1}]$, $\cdots$, $\mathbf{H}[\mathcal{N}_{\rho_{P-1}}]$ (or $\mathbf{H}[\mathcal{N}_{\langle\rho_s,\rho_t\rangle_0}],\mathbf{H}[\mathcal{N}_{\langle\rho_s,\rho_t\rangle_1}],\ldots,\mathbf{H}[\mathcal{N}_{\langle\rho_s,\rho_t\rangle_{K-1}}]$).
To capture the key message in a complex graph, we need to fuse multiple semantics revealed by different metapaths. 
Take Fig.~\ref{fig:instance} as an example, UM should be paid more attention than UMD.

From the statements above, one can see that metapath $\rho_p$ and metapath combination $\langle\rho_s,\rho_t\rangle_k$ are used interchangeably in NI and CNI modules.
For convenience, we mainly use metapath $\rho_p$ in the following sections and one can easily extend it to the metapath combination $\langle\rho_s,\rho_t\rangle_k$ case.
In light of the analysis above, we adopt an attention mechanism at two levels, i.e., element- and path-level.
As illustrated in Fig.~\ref{fig:agg}, at element-level, we seek to capture the key elements of each path of the interaction matrix (i.e., ~$\mathbf{H}[\mathcal{N}_{\rho_p}]_l$, as shown in (a)) and aggregate interactive information of each path into an embedding vector (as shown in (b)).
We then design a path-level attention mechanism to further aggregate the information from different paths including all paths guided by the involved metapaths.
Following element- and path-level attention mechanisms, we are able to get the final prediction (as shown in (c)).
The detailed descriptions are listed as follows:

\begin{figure}[t]
	\centering
	\vspace{-2mm}
	\includegraphics[width=1.0\textwidth]{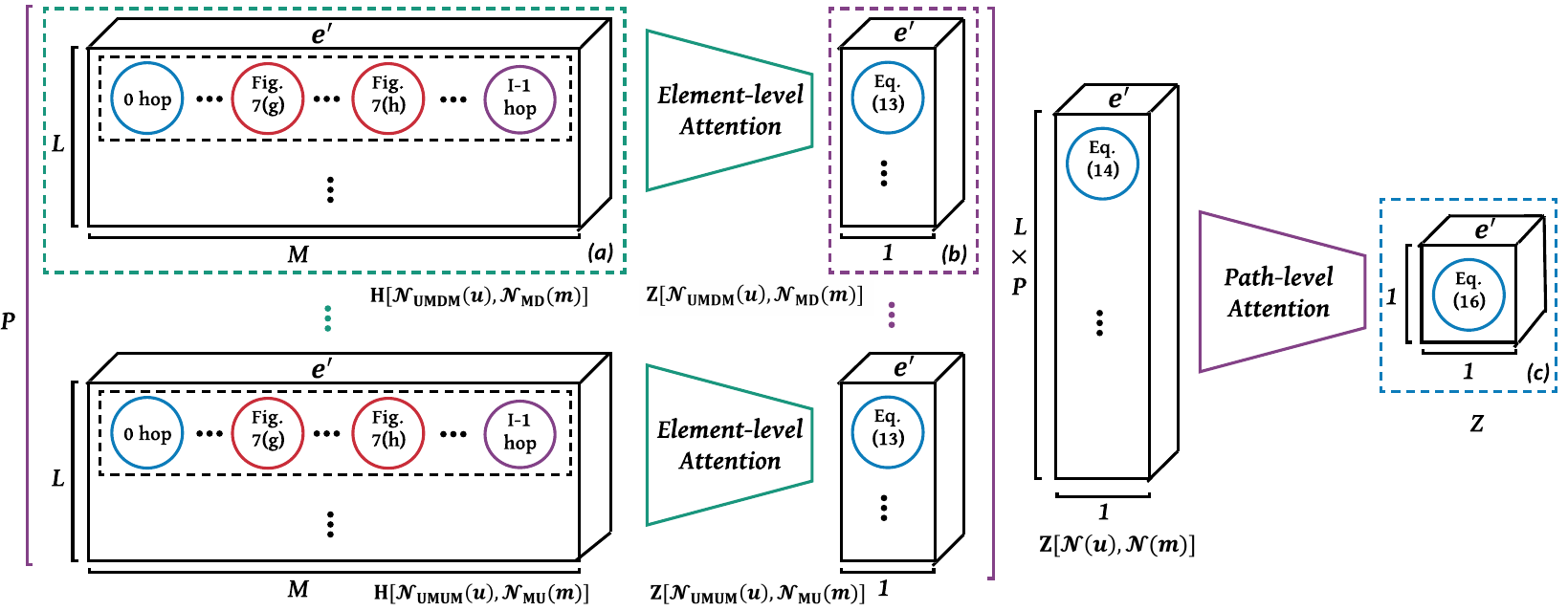}
	\vspace{-2mm}
	\caption{
		An illustrated example of the aggregation module to aggregate the interaction matrix $\mathbf{H}[\mathcal{N}_{\rho_s}(s),\mathcal{N}_{\rho_t}(t)]$ (a) (i.e.,~$\mathbf{H}[\mathcal{N}_{\langle\rho_s,\rho_t\rangle}]\in \mathbb{R}^{L\times M \times E}$) according to Eq.~(\ref{eqn:crossinteract}). We first adopt element-level attention mechanism to obtain $\mathbf{Z}[\mathcal{N}_{\rho_s}(s),\mathcal{N}_{\rho_t}(t)]$ (b) (i.e.,~$\mathbf{Z}[\mathcal{N}_{\langle\rho_s,\rho_t\rangle}]\in \mathbb{R}^{L \times E}$) according to Eq.~(\ref{eqn:nodeattention}), and then employ path-level attention mechanism to obtain $Z\in \mathbb{R}^{E}$ (c) according to Eq.~(\ref{eqn:pathattention}).
	}
	\label{fig:agg}
	\vspace{-2mm}
\end{figure}

\minisection{Element-level Attention} 
Let $\rho^l_p$ denote the $l$-th path guided by metapath $\rho_p$, then $\mathbf{H}[\mathcal{N}_{\rho^l_p}]$ and $\mathbf{H}[\mathcal{N}_{\rho_p}]_l$ are the same.
Similar to \cite{wang2019heterogeneous}, we leverage an attention mechanism to learn the weights among various kinds of interaction elements in path $\rho^l_p$\footnote{In this section, we omit the superscript $\rho^l_p$ for simplification; e.g., we use $h_{0j}$, $h_0$, $h_{j}$, $\alpha_{j}$, and $z$ to denote $h_{0j}^{\rho^l_p}$, $h_{0}^{\rho^l_p}$, $h_j^{\rho^l_p}$, $\alpha_{j}^{\rho^l_p}$, and $z^{\rho^l_p}$, respectively.} as
\begin{equation}
\label{eqn:nodeh}
% h^{\rho^l_p}_{0j} = (h^{\rho^l_p}_0W_T)^T \cdot (h^{\rho^l_p}_{j}W_S),
h_{0j} = (h_0W_T)^T \cdot (h_{j}W_S),
\end{equation}
where $W_T, W_S$ are trainable weights and $h_{j}$ is an element of interaction matrix $\mathbf{H}[\mathcal{N}_{\rho^l_p}]$ (i.e.,~$\mathbf{H}[\mathcal{N}_{\rho_p}]_l$), where $j=0,1,\ldots,M-1$ and $l=0,1,\ldots,L-1$.
$h_{0j}$ can show how important interaction element $h_j$ will be for interaction element $h_0$. 
% (i.e., the interaction result between source and target nodes).
% We set 
$h_0$ is the first element of $\mathbf{H}[\mathcal{N}_{\rho^l_p}]$, which denotes the interaction result of source node $s$ and target node $t$.
% Also, it should be noted that this attention mechanism preserve the asymmetry which is a critical property of heterogeneous graph, since the importance of interaction $h_i$ to $h_j$ is quit different from that of interaction $h_j$ to $h_i$.
To retrieve a general attention value in path ${\rho^l_p}$, we further normalize this value in path scope as
\begin{equation}
\label{eqn:nodealpha}
\alpha_{j} = \text{softmax}(h_{0j}/\iota) = \frac{\text{exp}(h_{0j}/\iota)}{\sum^{M-1}_{i=0} \text{exp}(h_{0i}/\iota)},
\end{equation}
where $\iota$ denotes the temperature factor.
% and $\mathcal{N}_{\rho^l}$ here covers all the interaction elements guided by path $\rho^l$, namely $\mathbf{H}[\mathcal{N}_\rho]_l$.
To jointly attend to the paths from different representation subspaces and learn stably, we leverage the multi-head attention as in previous works \cite{wang2019heterogeneous} to extend the observation as
\begin{equation}
\label{eqn:nodeattention}
z = \sigma(W_q \cdot \frac{1}{H} \sum_{n=1}^H \sum^{M-1}_{j=0} \alpha_{jn}(h_j W_{Cn}) + b_q),
\end{equation}
where $H$ is the number of attention heads and $W_C, W_q, b_q$ are trainable parameters.
Hence, the metapath-based embedding $z$ is aggregated based on the metapath-guided neighborhood with single path (i.e., $\mathbf{H}[\mathcal{N}_{\rho_p}]_l$).
In other words, for each path $\rho^l_p$, we have a specific embedding vector $z^{\rho_p^l}$ (i.e., $z$ in Eq.~(\ref{eqn:nodeattention})).
Note that in our setting, there are $P$ metapaths in an HIN and $L$ paths guided by each metapath.
So, the embedding obtained from element-level attention is $\mathbf{Z}[\mathcal{N}]\in\mathbb{R}^{(L\times P)\times E}$, which is the concatenation of the set of $\{\mathbf{Z}[\mathcal{N}_{\rho_0}],\mathbf{Z}[\mathcal{N}_{\rho_1}],\ldots,\mathbf{Z}[\mathcal{N}_{\rho_{P-1}}]\}$:
\begin{equation}
\label{eqn:z}
\mathbf{Z}[\mathcal{N}] = [\mathbf{Z}[\mathcal{N}_{\rho_0}] \oplus \mathbf{Z}[\mathcal{N}_{\rho_1}] \oplus \cdots \oplus \mathbf{Z}[\mathcal{N}_{\rho_{P-1}}]],
\end{equation}
where $\mathbf{Z}[\mathcal{N}_{\rho_p}]\in\mathbb{R}^{L\times E}$ and $p=0,1,2,\ldots,P-1$.

\minisection{Path-level Attention}
Similarly, we adopt a self-attention mechanism to learn the attention vector among various path embeddings in $\mathbf{Z}[\mathcal{N}]$.
For each path $z_i \in \mathbf{Z}[\mathcal{N}]$, after an MLP, it is fed to a softmax layer:
\begin{equation}
\label{eqn:pathbeta}
\beta_{i} = \text{softmax}(z_{i}/\upsilon) = \frac{\text{exp}(z_{i}/\upsilon)}{\sum_{j=0}^{L\times P -1} \text{exp}(z_{j}/\upsilon)},
\end{equation}
where $\upsilon$ denotes the temperature factor.
With the learned weights as coefficients, we can fuse these semantics-specific embeddings to obtain the final embedding $Z$ via
\begin{equation}
\label{eqn:pathattention}
Z=\sum^{L \times P-1}_{j=0}\beta_j\cdot z_j.
\end{equation}
Hence, we have obtained the final aggregation result, which involves interaction information at both element- and path-level.

% \minisection{Metapath-level Attention}
% To obtain the importance of each metapath, we first transform the semantic-specific embedding through a non-linear transformation.
% We then measure the path-level attention value as the average of the importance of all the semantic-specific path-level embeddings.
% The importance of each metapath $\rho_j$, denoted as $\omega^{\rho_j}$, is shown as follows:
% \begin{equation}
% \label{eqn:metapathomega}
% \omega^{\rho_j} = \frac{1}{|\mathcal{V}|} \sum_{i \in \mathcal{V}} w^T \cdot \text{tanh} (W_k \cdot x^{\rho_j}_i + b_k),
% \end{equation}
% where $w$ is a semantic-level attention vector, and $W_k,b_k$ are trainable parameters.
% We then normalize the above importance of all metapaths via a softmax function
% \begin{equation}
% \label{eqn:metapathgamma}
% \gamma^{\rho_j} = \text{softmax}(\omega^{\rho_j}) = \frac{\text{exp}(\omega^{\rho_j}/\tau)}{\sum_{j=0}^{P-1}\text{exp}(\omega^{\rho_j}/\tau)},
% \end{equation}
% where $\tau$ is the temperature factor.
% It can be explained as the contribution of the metapath $\rho_j$ in a specific task.
% With the learned weights as coefficients, we can fuse these semantic-specific embeddings to obtain the final embedding $X$ via
% \begin{equation}
% \label{eqn:metapathattention}
% X = \sum_{j=0}^{P-1} \gamma^{\rho_j} \cdot \mathbf{X}[\mathcal{N}_{\rho_j}].
% \end{equation}
% Hence, we have obtained the final aggregation result, which involves interaction information in both node-, path-, and metapath-level. 

\begin{figure}[t]
	\centering
	\vspace{-2mm}
	\includegraphics[width=0.9\textwidth]{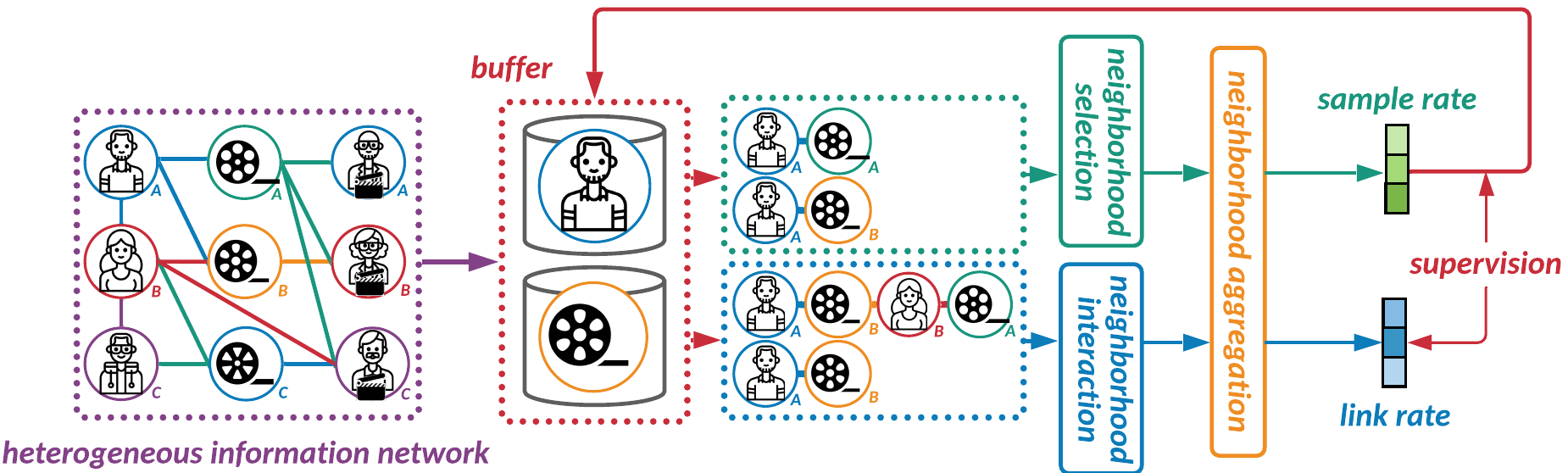}
	\vspace{-2mm}
	\caption{
		An illustrated example of the combination of neighborhood-based selection (NS) and interaction (NI) modules. We first sample and store metapath-guided neighborhoods of source and target nodes ($u_A$ and $m_B$ in this case) in the buffer. 
		Due to the large amount of high-order metapath-guided neighborhoods, we first use low-order neighborhoods to generate sample rates according to Eq.~(\ref{eqn:pathbeta}) for selection.
		We next sample high-order neighborhoods according to the sample rates and generate the final prediction of link rate according to Eq.~(\ref{eqn:pathattention}).
		Both NS and NI modules are under supervision according to Eq.~(\ref{eqn:loss}). 
	}
	\label{fig:comb}
	\vspace{-4mm}
\end{figure}

\subsection{Optimization Objective}
The final prediction result $\hat{Y}$ can be derived from final embedding $Z$ via a nonlinear projection (e.g.,~MLP).
The loss function of our model is a log loss:
\begin{equation}
\label{eqn:loss}
\mathcal{L}(Y, \hat{Y}) = \sum_{i, j \in \mathcal{Y}^+ \bigcup \mathcal{Y}^-} (y_{ij} \log \hat{y}_{ij} + (1- y_{ij}) \log(1-\hat{y}_{ij}))
\end{equation}
where $y_{ij}$ is the label of the instance (i.e.~source node $i$ and target node $j$), and  $\mathcal{Y}^+$, $\mathcal{Y}^-$ denote the positive instances set and the negative instances set, respectively.

\subsection{Combination of Neighborhood-based Selection \& Interaction}
\label{subsec:comb}
Previous approaches \citep{qu2019end,liu2018interactive} often suffer from high computations, especially when working on interactions of high-order neighborhoods and metapaths of different lengths.
These high-order neighborhoods or cross semantic interactions could result in (i) size explosion of metapath-guided neighborhoods; (ii) noise in interaction results. 
Equipped with \textsc{Definition}~\ref{def:neighbor}, we consider to filter the sampled neighborhood guided by complex metapaths through the results of those guided by simple metapaths.
The key motivation behind this is that if low-order neighborhoods are useful in the current task, it is likely that high-order neighborhoods extended from these neighborhoods are useful.
Take Fig.~\ref{fig:instance} as an example, User-Movie-User-Movie (UMUM) can be extended from User-Movie (UM).
Hence, we can first investigate the impact of each UM neighborhood and filter UMUM neighborhoods based on the results of UM neighborhoods.

\begin{figure}[t]
	\centering
	\includegraphics[width=0.9\textwidth]{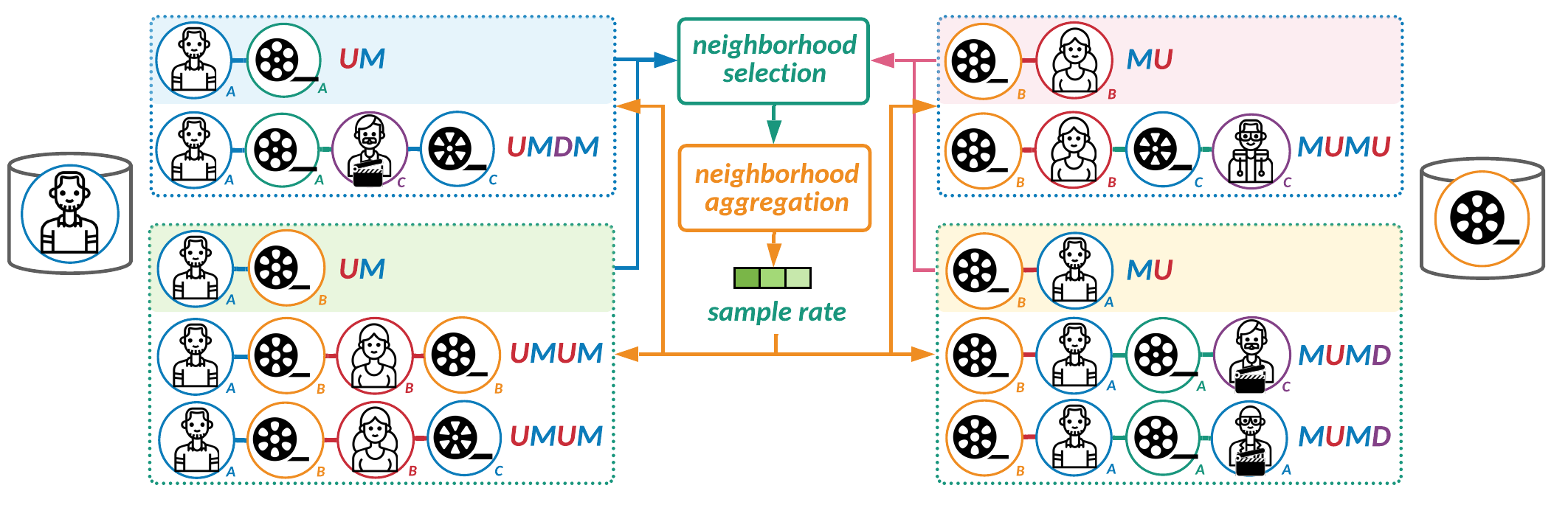}
	\vspace{-2mm}
	\caption{
		An illustrated example of the neighborhood-based selection (NS) module. We first group metapath-guided neighborhoods (of $u_\text{A}, m_\text{B}$ in this case) according to the inclusion relations of paths. 
		We next use low-order neighborhoods ($(u_\text{A}, m_\text{A})$ for path group 1, $(u_\text{A}, m_\text{B})$ for path group 2 in this case) as the input, and generate sample rates through the neighborhood-based selection module (Eq.~\ref{eqn:filter}) and the aggregation module (Eq.~\ref{eqn:pathbeta}).
		Then, we use the sample rates to sample all the paths in each path group according to its sample rate. 
	}
	\label{fig:select}
	\vspace{-2mm}
\end{figure}

As illustrated in Fig.~\ref{fig:comb}, when we study whether there exists a link between user $u_\text{A}$ and movie $m_\text{B}$ based on the given HIN, if we select UM and UMUM as the metapaths, we can sample metapath-guided neighborhoods for $u_\text{A}$ and $m_\text{B}$, and store them in the buffer.  
We then can examine the effect of each UM path and generate sample rates to capture the key paths and filter the noisy paths, which can be formulated as
% Inspired by convolutional neural network (CNN), we can regard shorter one of source and target paths as the filter and longer one as the input.
% One intuitive case behind this idea is to use source or target node (e.g.,~$u_A$) itself to filter the neighborhoods of the other (e.g., $m_B$).
% To achieve this, we can set the metapath length of filter side as 1.
% Suppose that we use source node to filter neighborhood of target node, which can be formulated as
% To achieve this, we set $\kappa=I_\text{min}$, $M = I_\text{max}+I_\text{min}-2I_\text{min}+1=I_\text{max}-I_\text{min}+1$ in Eq.~(\ref{eqn:crossinteract}), where $I_\text{min} = \min\{I_s, I_t\}$ and $I_\text{max} = \max\{I_s, I_t\}$.
% Assume that $I_s \leq I_t$, the filter operation in this case can be formulated as
\begin{equation}
\label{eqn:filter}
\mathbf{H}[\mathcal{N}^1_{\rho_s}(s), \mathcal{N}^1_{\rho_t}(t)]_{l, m} = \sum_{\substack{a, b\\ a+b ~ \text{mod} ~ M = m}} \mathbf{H}[\mathcal{N}^1_{\rho_s}(s)]_{l,a} \cdot \mathbf{H}[\mathcal{N}^1_{\rho_t}(t)]_{l,b},
\end{equation}
where $\mathcal{N}^1_{\rho_o}(o)$ denotes the 1-th neighborhood of $o$ guided by metapath $\rho_o$ (e.g., UM in the above case).
% The result of Eq.~(\ref{eqn:filter}) is the same to set source $n_s$ as the filter and use CNN on neighborhood of the target $n_t$.
It is straight forward to extend Eq.~(\ref{eqn:filter}) to use lower or higher-order neighborhoods as the filter.
While in this paper, we directly adopt Eq.~(\ref{eqn:filter}) as the filter.
% source $n_s$ or target $n_t$ together with its 1-hop neighborhood as the filter to maintain more useful information. 
% The formulation of this can be derived from Eq.~(\ref{eqn:filter}) through replacing $\mathcal{N}^0_{\rho_s}(s)$ with $\mathcal{N}^1_{\rho_s}(s)$.
Note that since Eq.~(\ref{eqn:filter}) can be regarded as a special case of the cross interaction operation defined in Eq.~(\ref{eqn:crossinteract}), it naturally can be accelerated by FFT formulated in Eq.~(\ref{eqn:crossinteraction}):
\begin{equation}
\label{eqn:filtering}
\mathbf{H}[\mathcal{N}^1_{\rho_s}(s), \mathcal{N}^1_{\rho_t}(t)] = \mathcal{F}^{-1}(\mathcal{F}(\mathbf{H}[\mathcal{N}^1_{\rho_s}(s)]) \cdot \mathcal{F}(\mathbf{H}[\mathcal{N}^1_{\rho_t}(t)])).
\end{equation}
% Eq.~(\ref{eqn:filter}) with $I_s\rangleI_t$ case by exchanging the position of $I_s$ and $I_t$.
% Note that filter operation defined in Eq.~(\ref{eqn:filter}) can be regard as the special case of cross interaction operation defined in Eq.~(\ref{eqn:crossinteract}), and thus can be accelerated by FFT formulated in Eq.~(\ref{eqn:crossinteraction}).
After the filter operation, we can next aggregate the information and obtain the attention vector for each path following Eqs.~(\ref{eqn:nodeh}), (\ref{eqn:nodealpha}), (\ref{eqn:nodeattention}), (\ref{eqn:pathbeta}), (\ref{eqn:pathattention}).
We treat the attention vector generated by Eq.~(\ref{eqn:pathbeta}) over the sampled paths as the sample rate on both low- and high-order neighborhoods (see Fig.~\ref{fig:comb}).

Note that this neighborhood-based selection procedure can closely incorporate with the neighborhood-based interaction module.
For instance, as shown in Fig.~\ref{fig:comb}, we evaluate the effects of two UM paths (e.g.,~($u_\text{A}, m_\text{A}$) and ($u_\text{A}, m_\text{B}$)) through the neighborhood-based selection (NS) module, and then the generated sample rate may push us to focus more on ($u_\text{A}, m_\text{B}$) and its related paths.
Hence, we update the inputs of the neighborhood-based interaction (NI) module as $(u_\text{A}, m_\text{B})$ and $(u_\text{A}, m_\text{B}, u_\text{B}, m_\text{C})$.
In order to update the NS module, we follow the learning procedure of GraphHINGE, where we supervise the whole model according to Eq.~(\ref{eqn:loss}).
The key difference between the training procedures of NI and NS modules is that input of NI module is the whole data, including both low- and high-order neighborhoods, while the input of NS module only contains low-order neighborhoods.
% Different from the output generated by Eq.~(\ref{eqn:pathattention}) of neighborhood aggregation operation in neighborhood selection procedure, the output in neighborhood interaction procedure is defined by Eq.~(\ref{eqn:metapathattention}), which is the final embedding over all paths under various metapath (combinations).

\begin{algorithm}[h]
	\caption{Neighborhood-based Selection (NS) Module}
	\label{algo:ns}
	\begin{algorithmic}[1]
		\Require
		HIN $\mathcal{G} = (\mathcal{V}, \mathcal{E})$; node feature $\{e_i, i\in \mathcal{V}\}$; metapath (combination) set $\{\rho_0, \rho_1, \cdots, \rho_{P-1}\}$; source node $s$ and target node $t$
		\Ensure
		filtered buffer $\mathcal{B}$ containing metapath-guided neighborhoods of $s$ and $t$
		% \vspace{1mm}
		\State Initialize all parameters and buffer $\mathcal{B}=\emptyset$.
		\State Sample metapath-guided neighborhoods according to Eq.~(\ref{eqn:neighbor}).
		\State Store neighborhoods in $\mathcal{B}$.
		\State Divide paths into $G$ groups $\{\rho^{0-0}, \rho^{0-1}, \ldots, \rho^{0-(L_0-1)}\}, \ldots, \{\rho^{(G-1)-0}, \rho^{(G-1)-1}, \ldots, \rho^{(G-1)-(L_{G-1}-1)}\}$ according to their inclusion relations.
		\label{line:group}
		\Repeat
		\For {each low-order path in groups $\rho^{0-0},\ldots, \rho^{g-0}, \ldots, \rho^{(G-1)-0}$} 
% 		\State Find metapath-guided neighborhoods of $n_s$, $n_t$: $\mathcal{N}_{\rho^k_s}(n_s)$, $\mathcal{N}_{\rho^k_t}(n_t)$ according to Eq.~(\ref{eqn:neighbor}).
% 		\label{line:generate}
		\State Obtain interaction result $\mathbf{H}[\mathcal{N}_{\rho^{g-0}}(s), \mathcal{N}_{\rho^{g-0}}(t)]$ according to Eq.~(\ref{eqn:filtering}).
		\State Calculate element-level embedding $z$ according to Eq.~(\ref{eqn:nodeattention}).
		\label{line:embedding}
		\State Calculate path-level attention vector $\beta$ (i.e.,~sample rate) according to Eq.~(\ref{eqn:pathbeta}).
% 		\label{line:samp}
% 		\State Filter neighborhoods guided by any metapath $\rho \in \{\rho^k_0, \rho^k_1, \dots, \rho^k_{P_k-1}\}$ in $\mathcal{B}$ according to sample rate.
		\label{line:filter}
		\State Obtain final prediction $Z$ according to Eq.~(\ref{eqn:pathattention}).
% 		\State Obtain final predication $\hat{Y}$ via MLP.
		\label{line:final}
		\State Calculate loss $\mathcal{L}(Y, \hat{Y})$ according to Eq.~(\ref{eqn:loss}), and Back propagation.
		\label{line:loss}
		\EndFor
		\Until convergence
		\State Select paths within each group according to their generated sample rates. 
		\label{line:select}
	\end{algorithmic}
\end{algorithm}

The overall procedure of the neighborhood-based selection (NS) module is given in Algorithm~\ref{algo:ns}.
Given an HIN $\mathcal{G}$, the task of NS module is to generate useful metapath-guided neighborhoods of source $s$ and target $t$.
As Algorithm~\ref{algo:ns} shows, we first divide sampled paths into $G$ groups (as illustrated in Fig.~\ref{fig:select}).
Let $\rho^{i-j}$ denote the $j$-th path of the $i$-th group in line~\ref{line:group}.
Especially, we use $\rho^{g-0}$\footnote{In this section, we store the low-order paths (e.g., UM and MU paths) on the top of each group, as illustrated in Fig.~\ref{fig:select}. Hence, $\mathbf{H}[\mathcal{N}_{\rho^{g-0}}(s), \mathcal{N}_{\rho^{g-0}}(t)]$ is equivalent to $\mathbf{H}[\mathcal{N}^1_{\rho_s}(s), \mathcal{N}^1_{\rho_t}(t)]$ in the Fig.~\ref{fig:select} case.} to denote the low-order path in the $g$-th group.
Then we push low-order paths into the neighborhood-based selection module and obtain sample rates to sample all paths over the group in line~\ref{line:filter}.
% , which allows filter operation within each group in line~\ref{line:filter}.
% For instance, as illustrated in Fig.~\ref{fig:comb}, we put metapaths UM and UMD in the same group.
% We then generate metapath-guided neighborhoods of both source and target nodes $n_s$ and $n_t$ in line~\ref{line:generate}.
% One should be noted that although interaction operation defined in Eq.~(\ref{eqn:filter}) allows input neighborhoods guided by any kind of metapath. 
% While practical, we often set one side (\emph{i.e.}, source or target) of metapath with sample formulation.
% For instance, we can set $\rho_s$ with UU or UM, and the other one $\rho_t$ fixed with UM.
% Next, we filter $\mathcal{N}_{\rho^k_t}(n_t)$ with $\mathcal{N}_{\rho^k_s}(n_s)$ through neighborhood interaction (see line~\ref{line:filter}) and aggregation (see line~\ref{line:aggregate}).
% We update the module with  in line~\ref{line:filter}.
% Take Fig.~\ref{fig:comb} as an example, one can adopt the sample rate of $(u_A, m_B)$ to sample/update both $(u_A, m_B)$ and $(u_A, m_B, d_B)$.
% In order to update the parameters of neighborhood interaction and aggregation modules, we calculate final prediction and loss function in line~\ref{line:final} and \ref{line:loss}, respectively.
% Note that the training procedures of neighborhood-based selection (NS) and neighborhood-based interaction (NI) along with cross neighborhood-based interaction (CNI) modules are pretty similar, where the parameters are updated according to the model performance.
Note that the goal of the NS module is to select appropriate high-order neighborhoods according to the performance of their corresponding low-order neighborhoods, while the goal of NI and CNI modules is to train a well-performed model for the prediction.

\subsection{Overall Algorithm}
\label{subsec:overall}
The learning algorithm of GraphHINGE is given in Algorithm~\ref{algo:framework}.  
In the HIN-based recommendation scenario, each source node $s$ corresponds to a user $u_s$, and each target node $t$ corresponds to an item $i_t$.
The task in this work can be either stated as link predictions on HINs or click-through predictions on HIN-based recommendations.
Also, one can easily extend the task into top-N recommendations.

As Algorithm~\ref{algo:framework} shows, we first obtain filtered buffer $\mathcal{B}$ in line~\ref{line:allbuffer}.
We then sample neighborhoods $\mathcal{N}_{\rho_{s}}(s)$, $\mathcal{N}_{\rho_{t}}(t)$ of $s$, $t$ from $\mathcal{B}$ in line~\ref{line:allsample}.
One should be noted that $\langle\rho_{s},\rho_{t}\rangle$ in line~\ref{line:allmetapath} means metapath combinations in the CNI module where the metapaths of source and target nodes are set with $\rho_{s}$ and $\rho_{t}$, respectively.  
Next, we calculate interactive information $\mathbf{H}[\mathcal{N}_{\rho_{s}}(s), \mathcal{N}_{\rho_{t}}(t)]$ through the cross interaction module described in Section~\ref{subsec:crossinteraction} in line~\ref{line:allinteract}.
To fuse the interaction results, we leverage the element-level attention mechanism to obtain representation $\mathbf{Z}[\mathcal{N}_{\rho_p}]$ in line~\ref{line:allnodeattention} .
We repeat the above procedure for different paths and obtain various embeddings $\mathbf{Z}[\mathcal{N}_{\rho_0}], \mathbf{Z}[\mathcal{N}_{\rho_1}], \cdots, \mathbf{Z}[\mathcal{N}_{\rho_{P-1}}]$.
To capture the key path in the current task, we employ the path-level attention mechanism to get embedding $Z$ in line~\ref{line:allpathattention}.
Finally, we obtain the final predication $\hat{Y}$ via MLP layers in line~\ref{line:allfinal} and update parameters through back propagation in line~\ref{line:allloss}.

\begin{algorithm}[h]
	\caption{GraphHINGE}
	\label{algo:framework}
	\begin{algorithmic}[1]
		\Require
		HIN $\mathcal{G} = (\mathcal{V}, \mathcal{E})$; node feature $\{e_i, i \in \mathcal{V}\}$; metapath (combination) set $\{\rho_0, \rho_1, \cdots, \rho_{P-1}\}$; source node $s$ and target node $t$
		\Ensure
		final link prediction $\hat{Y}$ between $s$ and $t$
		% \vspace{1mm}
		\State Initialize all parameters.
		\State \# (Obtain filtered buffer $\mathcal{B}$ according to Algorithm~\ref{algo:ns}).
		\Comment{for GraphHINGE$^+_{\text{SELECT}}$}
		\label{line:allbuffer}
		\Repeat 
		\For {each metapath (combination) $\rho_p$ ($\langle\rho_{s}, \rho_{t}\rangle$) in $\mathcal{B}$}
% 		\Comment{(for GraphHINGE$^+_\text{CROSS}$)}
		\label{line:allmetapath}
		\State Find metapath-guided neighborhoods of $s$, $t$: $\mathcal{N}_{\rho_{s}}(s)$, $\mathcal{N}_{\rho_{t}}(t)$ from $\mathcal{B}$.
		\label{line:allsample}
		\State Obtain interaction result $\mathbf{H}[\mathcal{N}_{\rho_{s}}(s), \mathcal{N}_{\rho_{t}}(t)]$ according to Eq.~(\ref{eqn:interaction}).
		\label{line:allinteract}
		\State \# (Obtain $\mathbf{H}[\mathcal{N}_{\rho_{s}}(s), \mathcal{N}_{\rho_{t}}(t)]$ according to Eq.~(\ref{eqn:crossinteraction}).) \Comment{for GraphHINGE$^+_\text{CROSS}$}
		\State Calculate element-level embedding $\mathbf{Z}[\mathcal{N}_{\rho_p}]$ according to Eq.~(\ref{eqn:nodeattention}).
		\label{line:allnodeattention}
% 		\State Calculate path-level embedding $x^{\rho_{p_s,p_t}}$ according to Eq.~(\ref{eqn:pathattention})  
% 		\label{line:allpathattention}
		\EndFor
		\State Fuse path-level embedding $Z$ according to Eq.~(\ref{eqn:pathattention}).
		\label{line:allpathattention}
		\State Obtain final predication $\hat{Y}$ via MLP.
		\label{line:allfinal}
		\State Calculate loss $\mathcal{L}(Y, \hat{Y})$ according to Eq.~(\ref{eqn:loss}), and Back propagation.
		\label{line:allloss}
		\Until convergence
	\end{algorithmic}
\end{algorithm}

\subsection{Model Analysis}
\label{subsec:analysis}

According to Algorithm~\ref{algo:framework}, we here give the analysis of the proposed GraphHINGE model as follows:

\minisection{Complexity}
The proposed CNI module is highly efficient and can be easily parallelized. 
We provide the model efficiency analysis for both the interaction module and the aggregation module. 
As stated in \cite{mathieu2013fast}, in the interaction module, the complexity of the FFT-based method is $\mathcal{O}(L_\rho I_\rho \log(I_\rho))$ where $L_\rho$ and $I_\rho$ denote the number of paths guided by $\rho$ and metapath length respectively.
As for the aggregation module, given a metapath $\rho$, the time complexity of the attention mechanism is $\mathcal{O}(V_\rho K + E_\rho K)$, where $V_\rho$ is the number of nodes, $E_\rho$ is the number of metapath-based node pairs, $K$ is the number of attention heads.
The attention can be computed individually across all nodes and metapaths.
The overall complexity is linear to the number of nodes and metapath-based node pairs. 
The proposed model can be easily parallelized, because the computation of the element- and path-level attentions can be parallelized across node pairs and metapaths, respectively.

\minisection{Interpretability}
The proposed model has potentially good interpretability for the learned interaction embedding through similarities and co-ratings among neighbor nodes.
Also, with the learned importance in node- and path-level, the proposed model can pay more attention to some meaningful interactions or metapaths for the specific task and give a more comprehensive description of a heterogeneous graph.
Based on the attention values, we can check which interactions or metapaths make the higher (or lower) contributions to our task, which is beneficial to analyze and explain our results.

\section{Experiment}
In this section, we present the details of the experiment setups and the corresponding results.
% \footnote{Reproducible code with instructions is available in the supplementary material. We are pushing process to integrate GraphHINGE into DGL \citep{wang2019deep} and make the code public.}. 
To illustrate the effectiveness of our proposed model, we compare it with some strong baselines on recommendation tasks, including click-through rate (CTR) predictions and top-N recommendations. 
We start with six research questions (RQ) to lead the experiments and the following discussions.
Furthermore, we have published our code for reproduction\footnote{Reproducible code based on Deep Graph Library (DGL) \citep{wang2019deep} with instructions is available at \url{https://github.com/Jinjiarui/GraphHINGE}.}. 

\begin{itemize}[topsep = 3pt,leftmargin =10pt]
	\item (\textbf{RQ1}) How does GraphHINGE compare to the existing state-of-the-art and other baselines in click-through rate prediction and top-N recommendation tasks?	
	\item (\textbf{RQ2}) What is the influence of different components in GraphHINGE? Are the proposed interaction and aggregation modules necessary for improving performance?
	\item (\textbf{RQ3}) What are the effects of the cross interaction module, neighborhood-based selection module, and other variants of GraphHINGE?
	\item (\textbf{RQ4}) How do various hyper-parameters, including the length and type of metapath-guided neighborhoods, the amount of training data, and the number of paths in each metapath-guided neighborhood, impact the model performance?
	\item (\textbf{RQ5}) What patterns does the proposed model capture for the final recommendation decision?
	\item (\textbf{RQ6}) What is the influence of the FFT technique? Does FFT accelerate the training procedure and gain high performance with low time cost?
% 	\item (\textbf{RQ7}) What is the convergence performance of GraphHINGE? Is the training process effective and stable?
% 	\item (\textbf{RQ7}) Is GraphHINGE feasible in the real-world practice?
\end{itemize}

\begin{table}[!h]
	\centering
	\caption{Statistics of the six datasets. The penult column reports the metapaths and the last column shows the sparsity in each dataset.}
	\vspace{-3mm}
	\label{tab:data}
	\resizebox{1.00\textwidth}{!}{
		\begin{tabular}{@{\extracolsep{4pt}}cccccccccccccc}
% 		\hline
        % \midrule
        \cmidrule{1-7}
        \cmidrule{8-14}
		Datasets & Relations (A-B) & A & B & A-B & Metapath & Sparsity & Datasets & Relations (A-B) & A & B & A-B & Metapath & Sparsity \\
% 		\hline
        \cmidrule{1-7}
        \cmidrule{8-14}
		\multirow{4}{*}{Amazon} 
		& User-Item & 6,170 & 2,753 & 195,791 & UIUI & 98.85\% & \multirow{4}{*}{Movielens} 
		& User-Movie & 943 & 1,682 & 100,000 & UMUM & 93.70\%\\
% 		\cline{2-6}
		{} & Item-View & 2,753 & 3,857 & 5,694 & UIVI & 99.95\% 
		& {} & Movie-Movie & 1,682 & 1,682 & 82,798 & UMMM & 97.07\%\\
% 		\cline{2-6}
		{} & Item-Brand & 2,753 & 334 & 2,753 & UIBI & 99.70\% 
		& {} & User-Occupation & 943 & 21 & 943 & UOUM & 95.24\%\\
% 		\cline{2-6}
		{} & Item-Category & 2,753 & 22 & 5,508 & UICI & 90.91\% 
		& {} & Movie-Genre & 1,682 & 18 & 2,861 & UMGM & 90.55\%\\ 
% 		\hline
        \cmidrule{1-7}
        \cmidrule{8-14}
% 		\cline{2-6}
% 		\cline{2-6}
% 		\cline{2-6}
% 		\cline{2-6}
% 		\hline
        % \midrule
		\multirow{4}{*}{Yelp} 
		& User-Business & 16,239 & 14,284 & 198,397 & UBUB & 99.91\% & 
		\multirow{4}{*}{DBLP} 
		& Paper-Author & 14,376 & 14,475 & 41,794 & PAPA & 99.97\% \\
% 		\cline{2-6}
		{} & User-User & 16,239 & 16,239 & 158,590 & UUUB & 99.93\% & 
		{} & Paper-Conference & 14,376 & 20 & 14,376 & PCPA & 95.00\% \\
% 		\cline{2-6}
		{} & Business-City & 14,284 & 47 & 14,267 & UBCB & 97.87\% & 
		{} & Paper-Type & 14,376 & 8,920 & 114,624 & PTPA & 99.91\%\\
% 		\cline{2-6}
		{} & Business-Category & 14,284 & 511 & 40,009 & UBAB & 99.45\% & 
		{} & Author-Label & 14,475 & 4 & 4,057 & PALA & 92.99\%\\ 
% 		\hline
        \cmidrule{1-7}
        \cmidrule{8-14}
		\multirow{2}{*}{Amazon Book} 
		& User-Book & 115,521 & 348,613 & 645,803 & UBUB & 99.99\% & \multirow{2}{*}{Movie-20M} 
		& User-Movie & 71,315 & 123,527 & 4,552,019 & UMUM & 99.95\%\\
% 		\cline{2-6}
		{} & Book-Attribute & 348,613 & 348,611 & 1,547,854 & UBA & 99.99\% & 
		{} & Movie-Attribute & 123,527 & 123,527 & 1,127,863 & UMA & 99.99\%\\
% 		\cline{2-6}
% 		\hline
% 		\midrule
% 		\cline{2-6}
% 		\hline
		\cmidrule{1-7}
        \cmidrule{8-14}
	\end{tabular}
	}
% 	\vspace{-3mm}
\end{table}

\setcounter{footnote}{0}
\subsection{Dataset Description}
We adopt six widely used datasets from different domains, namely \textbf{Amazon} e-commerce dataset\footnote{\url{http://jmcauley.ucsd.edu/data/amazon/}}, \textbf{Yelp} business dataset\footnote{\url{https://www.yelp.com/dataset/}}, \textbf{Movielens} movie dataset\footnote{\url{https://grouplens.org/datasets/movielens/}}, and \textbf{DBLP} academic dataset\footnote{\url{https://www.aminer.org/citation}}.
In order to evaluate whether GraphHINGE can scale to large scale datasets, we also compare our model with the baselines on \textbf{Amazon Book} dataset (which combines Amazon Book$^\ast$ and Freebase, and Amazon Book contains over 22.5 million ratings collected from 8 million users and 2.3 million items), and \textbf{Movie-20M} dataset (which combines Movielens-20M$^{\ddagger}$
% \footnote{\url{https://grouplens.org/datasets/movielens/}} 
and Freebase, and Movielens-20M contains ratings collected from the Movielens website).
We treat a rating as a relation record, indicating whether a user has rated an item.
Also, we provide the main statistics of the six datasets, which are summarized in Table~\ref{tab:data}.
The first row of each dataset corresponds to the numbers of users, items, and interactions, while the other rows correspond to the statistics of other relations.
We also report the selected metapaths for each dataset in the second last column of the table.

\subsection{Compared Methods}
We use eight baseline methods, which mainly can be concluded into three folds.
We list the brief descriptions of these strong baselines as follows:

The first class is heterogeneous and attributed graph embedding models containing TAHIN, HAN, HetGNN, and IPE.
\begin{itemize}[topsep = 3pt,leftmargin =10pt]
    \item \textbf{TAHIN}: \citet{bi2020heterogeneous} designed a cross domain model from both source and target domains and then employed three-level attention aggregations to get user and insurance product representations. 
	\item \textbf{HAN}: \citet{wang2019heterogeneous} introduced a hierarchical attention mechanism to capture node-level and semantics-level information.
	\item \textbf{HetGNN}:  \citet{zhang2019heterogeneous} introduced a unified framework to jointly consider heterogeneous structured information as well as heterogeneous contents information, adaptive to various HIN tasks. 
	\item \textbf{IPE}: \citet{liu2018interactive} proposed the interactive paths embedding to capture rich interaction information among metapaths.	
\end{itemize}
Notice that we do not compare GraphHINGE with several state-of-the-art methods on homogeneous graphs, since these approaches often perform poorly when extending to heterogeneous graphs setting.  
It is worth noting that TAHIN, HAN, HetGNN, IPE, are recently proposed, state-of-the-art models.

The second class is HIN-based recommendation models including NeuMF, LGRec, MCRec, NEM, GF.
\begin{itemize}[topsep = 3pt,leftmargin =10pt]
	\item \textbf{NeuMF}: \citet{he2017neural} introduced a generalized model consisting of a matrix factorization (MF) component and an MLP component.
	\item \textbf{LGRec}: \citet{hu2018local} proposed a unified model to explore and fuse local and global information for recommendation.
	\item \textbf{MCRec}: \citet{hu2018leveraging} leveraged rich metapath-based context to enhance the recommendation performance on HINs.
	\item \textbf{NEM}: \citet{yi2020heterogeneous} designed a network embedding mention recommendation model to recommend the right users in a message.
	\item \textbf{GF}: \citet{zhang2020graph} proposed a graph filtering recommendation methods on HINs.
% 	\item \textbf{KNI}: \citet{qu2019end} incorporated extra knowledge graphs and adopt graph neural network in neighborhood interaction model. 
\end{itemize}
These approaches mainly combine the expressive power of HIN and classical techniques to improve the performance of recommendation model.

The final class is non-neural-network-based models including ItemKNN and UserKNN, and classical models such as MF and DeepFM.
\begin{itemize}[topsep = 3pt,leftmargin =10pt]
    \item \textbf{DeepFM}: \citet{guo2017deepfm} proposed to combine the power of factorization machines (FM) for recommendation and deep learning for feature learning in a new architecture. 
    \item \textbf{MF}: \citet{koren2009matrix} designed to incorporate  implicit feedback, temporal effects and confidence levels with matrix factorization (MF).
	\item \textbf{ItemKNN}: A traditional collaborative filtering approach based on $k$-nearest-neighborhood (KNN) and item-item similarities \citep{sarwar2001item}.
	\item \textbf{UserKNN}: A neighborhood-based method using collaborative user-user similarities on $k$-nearest-neighborhood (KNN) \citep{wang2006unifying}.
\end{itemize}
As stated in \citep{dacrema2019we}, in several cases, non-neural-network-based models such as ItemKNN and UserKNN can obtain comparable results with those neural-network-based ones.
Also, we include MF and DeepFM here since these classical models often perform well even with few parameters \citep{rendle2020neural,koren2008factorization,koren2009matrix}.
For fair comparison, we involve the same metapaths when evaluating both our method and baselines on each dataset.
We provide the implementation details of both our model and baselines in Section~\ref{subsec:detail}.

\begin{figure}[b]
	\centering
	\vspace{-2mm}
	\includegraphics[width=0.7\textwidth]{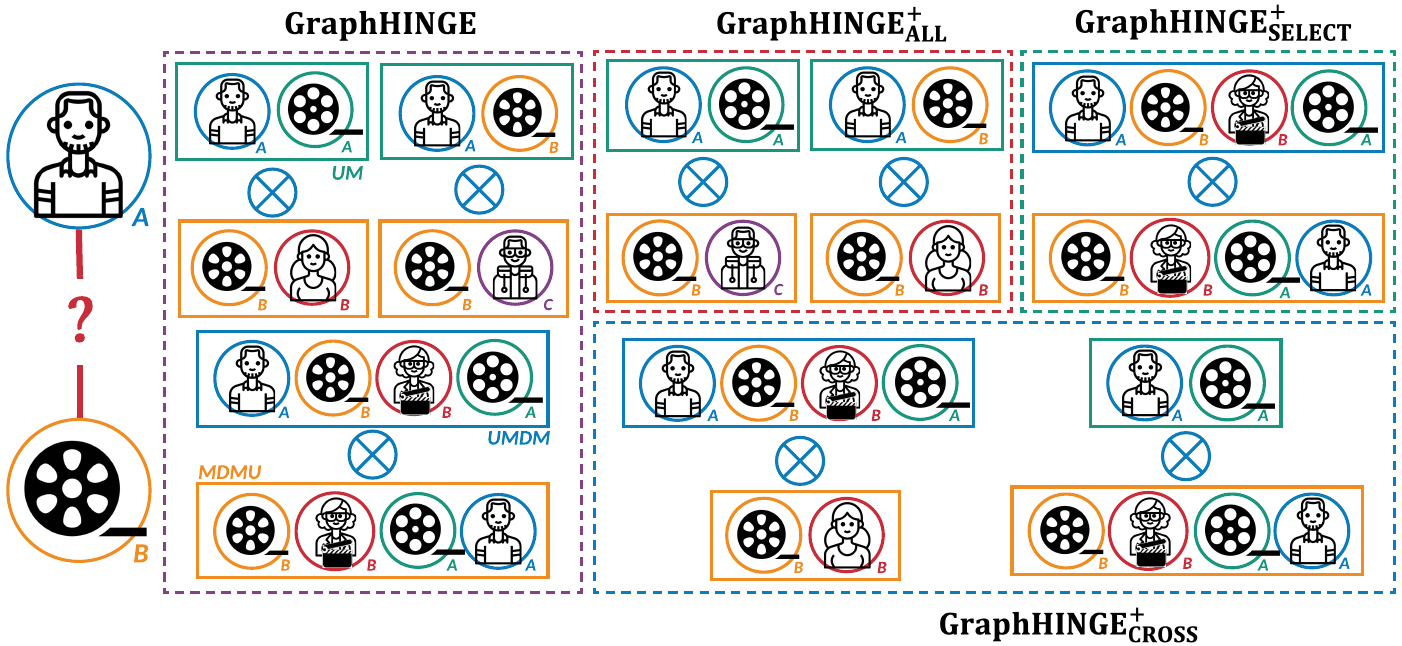}
	\vspace{-2mm}
	\caption{
		An illustrated example of GraphHINGE and its variants, namely GraphHINGE$^+_\text{ALL}$, GraphHINGE$^+_\text{CROSS}$ and GraphHINGE$^+_\text{SELECT}$.
		Note that GraphHINGE, which includes the  neighborhood-based interaction module and the  aggregation module, only conducts interaction operations between aligned paths guided by the same metapaths (e.g., $(u_\text{A},m_\text{A}) \odot (m_\text{B},u_\text{B})$, $(u_\text{A},m_\text{B}) \odot (m_\text{B},u_\text{C})$, $(u_\text{A},m_\text{B},d_\text{B},m_\text{A}) \odot (u_\text{B},d_\text{B},m_\text{A},u_\text{A})$). 
		Compared with GraphHINGE, GraphHINGE$^+_\text{ALL}$ operates interaction between all the paths, involving more interactive information (e.g., $(u_\text{A},m_\text{A}) \odot (m_\text{B}, u_\text{C})$, $(u_\text{A},m_\text{B}) \odot (m_\text{B}, u_\text{B})$); GraphHINGE$^+_\text{CROSS}$ allows interaction operations between paths guided by different metapaths (e.g., $(u_\text{A},m_\text{B},d_\text{B},m_\text{A}) \odot (m_\text{B},u_\text{B})$, $(u_\text{A},m_\text{A}) \odot (m_\text{B},d_\text{B},m_\text{A},u_\text{A})$). 
		GraphHINGE$^+_\text{SELECT}$ develops the neighborhood-based selection module to select appropriate high-order neighborhoods for GraphHINGE.
		Note that compared with GraphHINGE, GraphHINGE$^+_\text{ALL}$ and GraphHINGE$^+_\text{CROSS}$ involve additional information and GraphHINGE$^+_\text{SELECT}$ does not. 
	}
	\label{fig:model}
	\vspace{-2mm}
\end{figure}

\begin{table*}
	\centering
	\caption{The results of CTR prediction in terms of ACC, F1, LogLoss. Note: `*' indicates the statistically significant improvements over the best baseline, with $p$-value smaller than $10^{-6}$ in two-sided $t$-test.
	The best results for each metric are bold and the second best ones are underlined.}
	\label{tab:ctr}
	\vspace{-3mm}
	\resizebox{1.00\textwidth}{!}{
	\begin{tabular}{@{\extracolsep{4pt}}ccccccccccccc}
		\midrule
		\multirow{2}{*}{Model} & \multicolumn{3}{c}{Movielens} & \multicolumn{3}{c}{Amazon} & \multicolumn{3}{c}{Yelp} & \multicolumn{3}{c}{DBLP} \\ 
		\cmidrule{2-4}
		\cmidrule{5-7}
		\cmidrule{8-10}
		\cmidrule{11-13}
% 		\cline{2-9}
		& ACC & F1 & LogLoss & ACC & F1 & LogLoss & ACC & F1 & LogLoss & ACC & F1 & LogLoss\\
		\midrule
        TAHIN \citep{bi2020heterogeneous} 
        & 0.8811 & 0.8847 & 0.3029 
        & 0.8229 & 0.8267 & 0.4363 
        & 0.8904 & 0.8919 & 0.2666 
        & 0.9792 & 0.9793 & 0.1954\\
		HAN \citep{wang2019heterogeneous} 
		& 0.8713 & 0.8751 & 0.3210 
		& 0.8481 & 0.8490 & 0.4282 
		& 0.8894 & 0.8901 & 0.2697 
		& 0.9752 & 0.9751 & 0.2881\\ 
% 		\hline
		HetGNN \citep{zhang2019heterogeneous} 
		& 0.8510 & 0.8567 & 0.3528
		& 0.8313 & 0.8310 & 0.4402 
		& 0.8719 & 0.8843 & 0.2817 
		& 0.9512 & 0.9514 & 0.3125\\
% 		\hline
		IPE \citep{liu2018interactive} 
		& 0.8645 & 0.8693 & 0.3223 
		& 0.8430 & 0.8411 & 0.4366 
		& 0.8809 & 0.8876 & 0.2744 
		& 0.9570 & 0.9589 & 0.3099\\
% 		\hline
		\midrule
		NeuMF \citep{he2017neural} 
		& 0.8125 & 0.8421 & 0.4265 
		& 0.7510 & 0.7143 & 0.4907 
		& 0.7812 & 0.7742 & 0.4175 
		& 0.8125 & 0.8012 & 0.4103\\
% 		\hline
		LGRec \citep{hu2018local} 
		& 0.8144 & 0.8481 & 0.4290 
		& 0.7127 & 0.7100 & 0.5589 
		& 0.8308 & 0.8130 & 0.3642 
		& 0.8158 & 0.8172 & 0.3988\\ 
% 		\hline
		MCRec \citep{hu2018leveraging} 
		& 0.8125 & 0.8421 & 0.4234 
		& 0.7156 & 0.7114 & 0.5557 
		& 0.8113 & 0.8116 & 0.3302 
		& 0.8203 & 0.8215 & 0.3937\\
		NEM \citep{yi2020heterogeneous}
		& 0.8032 & 0.8322 & 0.4311
		& 0.7021 & 0.7012 & 0.5832
		& 0.7909 & 0.8087 & 0.3455
		& 0.8167 & 0.8110 & 0.4022\\
		GF \citep{zhang2020graph}
		& 0.8100 & 0.8277 & 0.4455
		& 0.7018 & 0.7103 & 0.5764
		& 0.8002 & 0.8045 & 0.3400
		& 0.8185 & 0.8155 & 0.4001\\
        \midrule
        MF \citep{koren2009matrix} 
		& 0.8098 & 0.8419 & 0.4433 
		& 0.7571 & 0.7135 & 0.4971 
		& 0.7841 & 0.7857 & 0.4151 
		& 0.7155 & 0.7381 & 0.5856\\
        DeepFM \citep{guo2017deepfm} 
		& 0.7284 & 0.7748 & 0.5371 
		& 0.6931 & 0.6893 & 0.5836 
		& 0.7162 & 0.7342 & 0.5122 
		& 0.7212 & 0.7897 & 0.5102\\
		ItemKNN \citep{sarwar2001item} 
		& 0.8066 & 0.8062 & 0.3890 
		& 0.6274 & 0.6355 & 0.6781 
		& 0.5512 & 0.5339 & 0.5914 
		& 0.7274 & 0.6940 & 0.5008\\
% 		\hline
		UserKNN \citep{wang2006unifying} 
		& 0.8061 & 0.8045 & 0.3920 
		& 0.6574 & 0.6471 & 0.6688 
		& 0.5974 & 0.5877 & 0.5577 
		& 0.7321 & 0.7319 & 0.4967\\
% 		\hline
        \midrule
% 		\hline
        GraphHINGE$_\text{CNN}$ 
        & 0.8482 & 0.8550 & 0.3549 
        & 0.8322 & 0.8317 & 0.4493 
        & 0.8751 & 0.8811 & 0.2233
        & 0.9582 & 0.9585 & 0.3051\\
% 		\hline
		GraphHINGE$_\text{GCN}$ 
		& 0.8727 & 0.8756 & 0.3120 
		& 0.8813 & 0.8815 & 0.3088 
		& 0.9014 & 0.9011 & 0.2213  
		& 0.9879 & 0.9842 & 0.1924\\
		\midrule
		GraphHINGE 
		& 0.8837$^*$ & 0.8854$^*$ & 0.2806$^*$
		& 0.8939$^*$ & 0.8912$^*$ & 0.2575$^*$
		& 0.9194$^*$ & 0.9106$^*$ & 0.2175$^*$
		& 0.9978$^*$ & 0.9979$^*$ & 0.0119$^*$\\
% 		\hline
        \midrule
        GraphHINGE$_\text{ALL}^+$ 
        & \underline{0.8857}$^*$ & \textbf{0.8868}$^*$ & \textbf{0.2749}$^*$
        & \underline{0.8960}$^*$ & \underline{0.8930}$^*$ & 0.2531$^*$
        & 0.9201$^*$ & 0.9109$^*$ & 0.2178$^*$
        & \underline{0.9983}$^*$ & \textbf{0.9982}$^*$ & \textbf{0.0009}$^*$\\
        GraphHINGE$_\text{CROSS}^+$ 
        & 0.8854$^*$ & \underline{0.8865}$^*$ & 0.2844$^*$
        & \textbf{0.8965}$^*$ & \textbf{0.8958}$^*$ & \textbf{0.2528}$^*$
        & \textbf{0.9225}$^*$ & \textbf{0.9131}$^*$ & \textbf{0.2064}$^*$
        & 0.9979$^*$ & 0.9980$^*$ & 0.0013$^*$\\
		GraphHINGE$_\text{SELECT}^+$ 
		& \textbf{0.8858}$^*$ & 0.8860$^*$ & \underline{0.2778}$^*$
		& 0.8945$^*$ & 0.8922$^*$ & \underline{0.2530}$^*$
		& \underline{0.9217}$^*$ & \underline{0.9110}$^*$ & \underline{0.2146}$^*$
		& \textbf{0.9984}$^*$ & \underline{0.9981}$^*$ & \underline{0.0010}$^*$\\
% 		\hline
        \midrule
	\end{tabular}
	}
% 	\vspace{-3mm}
\end{table*}

\begin{table*}
	\centering
	\caption{The results of top-N recommendation in terms of MAP, NDCG@3, NDCG@5. Note: `*' indicates the statistically significant improvements over the best baseline, with $p$-value smaller than $10^{-6}$ in two-sided $t$-test.
	The best results for each metric are bold and the second best ones are underlined.}
	\label{tab:topn}
	\vspace{-3mm}
	\resizebox{1.00\textwidth}{!}{
	\begin{tabular}{@{\extracolsep{4pt}}ccccccccccccc}
		\midrule
		\multirow{2}{*}{Model} & \multicolumn{3}{c}{Movielens} & \multicolumn{3}{c}{Amazon} & \multicolumn{3}{c}{Yelp} & \multicolumn{3}{c}{DBLP} \\ 
		\cmidrule{2-4}
		\cmidrule{5-7}
		\cmidrule{8-10}
		\cmidrule{11-13}
% 		\cline{2-9}
		& MAP & NDCG@3 & NDCG@5 & MAP & NDCG@3 & NDCG@5 & MAP & NDCG@3 & NDCG@5 & MAP & NDCG@3 & NDCG@5\\
		\midrule
		TAHIN \citep{bi2020heterogeneous} 
        & 0.6010 & 0.8908 & 0.8976 
        & 0.7029 & 0.9396 & 0.9312 
        & 0.7780 & 0.9523 & 0.9600 
        & 0.8261 & 0.9780 & 0.9734\\
		HAN \citep{zhang2019heterogeneous} 
		& 0.5930 & 0.8611 & 0.8723
		& 0.7111 & 0.9489 & 0.9370 
		& 0.7677 & 0.9509 & 0.9432 
		& 0.8238 & 0.9689 & 0.9695\\
		HetGNN \citep{liu2018interactive} 
		& 0.5906 & 0.8593 & 0.8699 
		& 0.7045 & 0.9340 & 0.9410 
		& 0.7459 & 0.9446 & 0.9409 
		& 0.8179 & 0.9502 & 0.9508\\
		IPE \citep{wang2019heterogeneous} 
		& 0.5910 & 0.8630 & 0.8725 
		& 0.7093 & 0.9389 & 0.9412 
		& 0.7503 & 0.9500 & 0.9406 
		& 0.8182 & 0.9509 & 0.9511\\ 
% 		\hline
		\midrule
		NeuMF \citep{he2017neural} 
		& 0.5816 & 0.8513 & 0.8675 
		& 0.6433 & 0.8906 & 0.8956 
		& 0.6501 & 0.8414 & 0.8403 
		& 0.7853 & 0.9029 & 0.9044\\
% 		\hline
		LGRec \citep{hu2018local} 
		& 0.5827 & 0.8504 & 0.8634 
		& 0.6127 & 0.8311 & 0.8435 
		& 0.7308 & 0.8930 & 0.8904 
		& 0.7821 & 0.9022 & 0.9042\\ 
% 		\hline
		MCRec \citep{hu2018leveraging} 
		& 0.5823 & 0.8565 & 0.8662 
		& 0.6242 & 0.8618 & 0.8756 
		& 0.7063 & 0.8717 & 0.8629 
		& 0.7946 & 0.9119 & 0.9140\\
		NEM \citep{yi2020heterogeneous}
		& 0.5712 & 0.8422 & 0.8499
		& 0.6087 & 0.8540 & 0.8579
		& 0.6978 & 0.8700 & 0.8599
		& 0.7812 & 0.9054 & 0.9023\\
		GF \citep{zhang2020graph}
		& 0.5801 & 0.8498 & 0.8510
		& 0.6123 & 0.8578 & 0.8665
		& 0.7001 & 0.8709 & 0.8602
		& 0.7899 & 0.9003 & 0.9104\\
% 		\hline
        % KNI \citep{qu2019end} 
        % & 0.9244 & 0.8522 & 0.0000 
        % & 0.9029 & 0.8196 & 0.0000 
        % & 0.9677 & 0.9109 & 0.0000 
        % & 0.8261 & 0.7411 & 0.0000\\
        \midrule
        MF \citep{koren2009matrix} 
		& 0.5197 & 0.8062 & 0.8090 
		& 0.6474 & 0.8955 & 0.8981 
		& 0.6512 & 0.8439 & 0.8414 
		& 0.7074 & 0.7940 & 0.8008\\
        DeepFM \citep{guo2017deepfm} 
		& 0.5002 & 0.7908 & 0.7982 
		& 0.5874 & 0.7855 & 0.7881 
		& 0.6514 & 0.8419 & 0.8397 
		& 0.7123 & 0.8076 & 0.8098\\
		ItemKNN \citep{sarwar2001item} 
		& 0.5161 & 0.8008 & 0.8015 
		& 0.6074 & 0.7876 & 0.7871 
		& 0.5409 & 0.6339 & 0.6845 
		& 0.7120 & 0.8012 & 0.8101\\
% 		\hline
		UserKNN \citep{wang2006unifying} 
		& 0.5153 & 0.8022 & 0.8037 
		& 0.6043 & 0.7871 & 0.7892 
		& 0.5506 & 0.6339 & 0.6903 
		& 0.7121 & 0.8023 & 0.8019\\
% 		\hline
        \midrule
% 		\hline
        GraphHINGE$_\text{CNN}$ 
        & 0.5749 & 0.8535 & 0.8740 
        & 0.7035 & 0.9488 & 0.9355 
        & 0.7432 & 0.9534 & 0.9424 
        & 0.8136 & 0.9516 & 0.9522\\
% 		\hline
		GraphHINGE$_\text{GCN}$ 
		& 0.5890 & 0.8730 & 0.8885 
		& 0.7302 & 0.9521 & 0.9414 
		& 0.7798 & 0.9578 & 0.9608  
		& 0.8260 & 0.9833 & 0.9850\\ 
		\midrule
		GraphHINGE 
		& 0.6066$^*$ & 0.9015$^*$ & 0.9074$^*$
		& 0.7530$^*$ & 0.9653$^*$ & 0.9525$^*$
		& 0.7810$^*$ & 0.9638$^*$ & 0.9622$^*$
		& 0.8293$^*$ & 0.9892$^*$ & 0.9888$^*$\\
% 		\hline
        \midrule
        GraphHINGE$_\text{ALL}^+$ 
        & \textbf{0.6098}$^*$ & 0.9037$^*$ & 0.9076$^*$
        & 0.7599$^*$ & 0.9644$^*$ & 0.9526$^*$
        & \textbf{0.7919}$^*$ & 0.9701$^*$ & 0.9704$^*$
        & \underline{0.8304}$^*$ & 0.9901$^*$ & \underline{0.9919}$^*$\\
        GraphHINGE$_\text{CROSS}^+$ 
        & 0.6068$^*$ & \textbf{0.9043}$^*$ & \textbf{0.9099}$^*$ 
        & \textbf{0.7769}$^*$ & \textbf{0.9670}$^*$ & \textbf{0.9538}$^*$
        & 0.7901$^*$ & \textbf{0.9747}$^*$ & \textbf{0.9714}$^*$
        & 0.8303$^*$ & \underline{0.9944}$^*$ & 0.9918$^*$\\
		GraphHINGE$_\text{SELECT}^+$ 
		& \underline{0.6070}$^*$ & \underline{0.9038}$^*$ & \underline{0.9085}$^*$
		& \underline{0.7755}$^*$ & \underline{0.9666}$^*$ & \underline{0.9532}$^*$
		& \underline{0.7909}$^*$ & \underline{0.9745}$^*$ & \underline{0.9706}$^*$
		& \textbf{0.8305}$^*$ & \textbf{0.9945}$^*$ & \textbf{0.9920}$^*$\\
% 		\hline
        \midrule
	\end{tabular}
	}
% 	\vspace{-3mm}
\end{table*}

\begin{table*}
	\centering
	\caption{The results of CTR prediction in term of ACC, F1, LogLoss, and top-N recommendation in terms of MAP, NDCG@3, NDCG@5. Note: `*' indicates the statistically significant improvements over the best baseline, with $p$-value smaller than $10^{-6}$ in two-sided $t$-test.
	The best results for each metric are bold and the second best ones are underlined.}
	\label{tab:big}
	\vspace{-3mm}
	\resizebox{1.00\textwidth}{!}{
	\begin{tabular}{@{\extracolsep{4pt}}ccccccccccccc}
		\midrule
		\multirow{2}{*}{Model} & \multicolumn{3}{c}{Amazon Book} & \multicolumn{3}{c}{Movie-20M} & \multicolumn{3}{c}{Amazon Book} & \multicolumn{3}{c}{Movie-20M} \\ 
		\cmidrule{2-4}
		\cmidrule{5-7}
		\cmidrule{8-10}
		\cmidrule{11-13}
% 		\cline{2-9}
		& ACC & F1 & LogLoss & ACC & F1 & LogLoss & MAP & NDCG@3 & NDCG@5 & MAP & NDCG@3 & NDCG@5\\
		\midrule
		TAHIN \citep{bi2020heterogeneous} 
        & 0.8349 & 0.8382 & 0.3789 
        & 0.9103 & 0.9094 & 0.2271 
        & 0.3911 & 0.7994 & 0.8667
        & 0.5578 & 0.8440 & 0.8512\\
		HAN \citep{wang2019heterogeneous} 
		& 0.8806 & 0.8822 & 0.2776 
		& 0.9129 & 0.9131 & 0.2285 
		& 0.3846 & 0.7688 & 0.8498 
		& 0.5444 & 0.8340 & 0.8461\\ 
% 		\hline
		HetGNN \citep{zhang2019heterogeneous} 
		& 0.7430 & 0.7412 & 0.4877
		& 0.8451 & 0.8457 & 0.3442 
		& 0.3800 & 0.7765 & 0.8566 
		& 0.5397 & 0.8276 & 0.8399\\
% 		\hline
		IPE \citep{liu2018interactive} 
		& 0.7692 & 0.7686 & 0.4504 
		& 0.8512 & 0.8510 & 0.3357 
		& 0.3817 & 0.7709 & 0.8524  
		& 0.5480 & 0.8346 & 0.8421\\
% 		\hline
		\midrule
		NeuMF \citep{he2017neural} 
		& 0.7377 & 0.7295 & 0.5490 
		& 0.8642 & 0.8645 & 0.3255 
		& 0.3856 & 0.7654 & 0.8463 
		& 0.5481 & 0.8355 & 0.8485\\
% 		\hline
		LGRec \citep{hu2018local} 
		& 0.7390 & 0.7398 & 0.5420 
		& 0.8650 & 0.8645 & 0.3233 
		& 0.3823 & 0.7704 & 0.8504 
		& 0.5455 & 0.8348 & 0.8467\\ 
% 		\hline
		MCRec \citep{hu2018leveraging} 
		& 0.7535 & 0.7451 & 0.4532 
		& 0.8669 & 0.8677 & 0.3155 
		& 0.3801 & 0.7848 & 0.8573 
		& 0.5435 & 0.8333 & 0.8456\\
		NEM \citep{yi2020heterogeneous}
		& 0.7211 & 0.7197 & 0.4912
		& 0.8501 & 0.8540 & 0.3367
		& 0.3755 & 0.7732 & 0.8411
		& 0.5311 & 0.8215 & 0.8367\\
		GF \citep{zhang2020graph}
		& 0.7210 & 0.7189 & 0.4921
		& 0.8510 & 0.8530 & 0.3365
		& 0.3765 & 0.7821 & 0.8399
		& 0.5324 & 0.8220 & 0.8298\\
% 		\hline
        % KNI \citep{qu2019end} 
        % & 0.9244 & 0.8522 & 0.0000 
        % & 0.9029 & 0.8196 & 0.0000 
        % & 0.9677 & 0.9109 & 0.0000 
        % & 0.8261 & 0.7411 & 0.0000\\
        \midrule
        MF \citep{koren2009matrix} 
		& 0.6391 & 0.6383 & 0.5890 
		& 0.8787 & 0.8777 & 0.3462 
		& 0.3512 & 0.7339 & 0.8214 
		& 0.4374 & 0.8240 & 0.8308\\
        DeepFM \citep{guo2017deepfm} 
		& 0.6066 & 0.6062 & 0.6890 
		& 0.7274 & 0.7355 & 0.5181 
		& 0.3034 & 0.7091 & 0.7890 
		& 0.4235 & 0.8160 & 0.8211\\
		ItemKNN \citep{sarwar2001item} 
		& 0.6181 & 0.6222 & 0.6990 
		& 0.7249 & 0.7301 & 0.5120 
		& 0.3508 & 0.7331 & 0.8244 
		& 0.4274 & 0.8126 & 0.8237\\
% 		\hline
		UserKNN \citep{wang2006unifying} 
		& 0.6234 & 0.6277 & 0.6680 
		& 0.7214 & 0.7209 & 0.5176 
		& 0.3511 & 0.7339 & 0.8156 
		& 0.4307 & 0.8190 & 0.8300\\
% 		\hline
        \midrule
% 		\hline
        GraphHINGE$_\text{CNN}$ 
        & 0.7833 & 0.7918 & 0.4493 
        & 0.8689 & 0.8711 & 0.3208 
        & 0.4383 & 0.8207 & 0.8700 
        & 0.5477 & 0.8330 & 0.8473\\
% 		\hline
		GraphHINGE$_\text{GCN}$ 
		& 0.8290 & 0.8376 & 0.3550 
		& 0.8680 & 0.8704 & 0.3267 
		& 0.4489 & 0.8310 & 0.8650  
		& 0.5408 & 0.8310 & 0.8426\\
		\midrule
		GraphHINGE 
		& 0.9619$^*$ & 0.9611$^*$ & 0.1173$^*$
		& \underline{0.9260}$^*$ & 0.9245$^*$ & \textbf{0.1931}$^*$
		& \underline{0.5105}$^*$ & 0.9341$^*$ & 0.9371$^*$
		& 0.5487$^*$ & 0.8430$^*$ & 0.8567$^*$\\
% 		\hline
        \midrule
        GraphHINGE$_\text{ALL}^+$ 
        & 0.9633$^*$ & 0.9624$^*$ & 0.1145$^*$
        & \textbf{0.9275}$^*$ & \textbf{0.9277}$^*$ & \underline{0.1115}$^*$
        & 0.5086$^*$ & \underline{0.9349}$^*$ & \textbf{0.9378}$^*$
        & 0.5500$^*$ & \textbf{0.8433}$^*$ & \underline{0.8568}$^*$\\
        GraphHINGE$_\text{CROSS}^+$ 
        & \textbf{0.9712}$^*$ & \textbf{0.9704}$^*$ & \textbf{0.0966}$^*$
        & 0.9224$^*$ & 0.9205$^*$ & 0.2019$^*$
        & \textbf{0.5164}$^*$ & \textbf{0.9354}$^*$ & \underline{0.9376}$^*$
        & \underline{0.5515}$^*$ & 0.8430$^*$ & 0.8567$^*$\\
		GraphHINGE$_\text{SELECT}^+$ 
		& \underline{0.9665}$^*$ & \underline{0.9640}$^*$ & \underline{0.1023}$^*$
		& 0.9255$^*$ & \underline{0.9273}$^*$ & 0.1995$^*$
		& 0.5090$^*$ & 0.9330$^*$ & \underline{0.9376}$^*$
		& \textbf{0.5568}$^*$ & \underline{0.8431}$^*$ & \textbf{0.8569}$^*$\\
% 		\hline
        \midrule
	\end{tabular}
	}
	\vspace{-3mm}
\end{table*}

\subsection{Evaluation Metric}
In this paper, we evaluate the models in two tasks: click-through rate prediction and top-N recommendation.
In click-through rate prediction task, we evaluate a model according to the performance on each data point, where we adopt ACC, LogLoss, and F1 score.
In top-N recommendation task, we take the position into consideration, where we adopt MAP at position 5 (denoted as MAP), NDCG at position 3 (denoted as NDCG@3), and NDCG at position 5 (denoted as NDCG@5).

\subsection{Model Variants Configuration}
\label{subsec:variant}
In order to investigate the impacts of different components in our
model, we set several variants of the GraphHINGE model as baselines.
\begin{itemize}[topsep = 3pt,leftmargin =10pt]
	\item \textbf{GraphHINGE}: is our proposed model as illustrated in Fig.~\ref{fig:overview}, which incorporates the neighborhood-based interaction (NI) module and the aggregation module, but excludes the neighborhood-based selection (NS) module.
    \item \textbf{GraphHINGE$_\text{CNN}$}: This variant does not consider the interpretability in the interaction module.
	That is, it directly fuses the neighborhood of each source and target node via Convolutional Neural Networks (CNN).
	This context embedding method is quite similar to the path instance embedding technique in \cite{hu2018leveraging}.
	The aggregation module is the same as GraphHINGE.
	This variant is designed to show the performance gain by our interaction module.
	\item \textbf{GraphHINGE$_\text{GCN}$}: This variant replaces the attention mechanism of GraphHINGE with a standard Graph Convolutional Network (GCN).
	That is, the element-level features are fed to GCN layers to get content embedding without fully considering different types of nodes and metapaths on HINs.
	The interaction module is the same as GraphHINGE.
	Hence, this variant is designed to show the performance gain by our aggregation module.
	\item \textbf{GraphHINGE$^-_\text{FFT}$}: This variant replaces FFT technique as introduced in Eq.~(\ref{eqn:interaction}) with the combination of product and sum operations as introduced in Eq.~(\ref{eqn:interact}).
	Therefore, this variant is proposed to show the efficiency of FFT technique.
\end{itemize}

In order to investigate the impact of different settings, cross neighborhood-based interaction (CNI) module, and neighborhood-based selection (NS) module, we set following variants of GraphHINGE model here.

\begin{itemize}[topsep = 3pt,leftmargin =10pt]
	\item \textbf{GraphHINGE$^+_\text{ALL}$}: This variant conducts interaction between all paths in source and target neighborhoods, while GraphHINGE only operates between aligned paths.
	Note that this setting improves performance but suffers from high time complexity (i.e.,~$\mathcal{O}(L^2_\rho I_\rho \log(I_\rho))$).
	Therefore, this variant can be regarded as a higher performance and complexity version of GraphHINGE.
	\item \textbf{GraphHINGE$^+_\text{CROSS}$}: This variant replaces the neighborhood-based interaction (NI) module with the cross neighborhood-based interaction (CNI) module.
	Hence, it allows the interaction operations between neighborhoods guided by different metapaths even in different lengths. 
	This variant is designed to show the performance of the CNI module.
	\item \textbf{GraphHINGE$^+_\text{SELECT}$}: This variant is the combination of GraphHINGE and neighborhood-based selection (NS) module, as introduced in Section~\ref{subsec:comb}.
	Therefore, this variant is proposed to illustrate the performance of our NS module.
\end{itemize}
We further illustrate the difference between GraphHINGE and its variants in Fig.~\ref{fig:model}.

\begin{figure}[b]
	\centering
% 	\vspace{-2mm}
	\includegraphics[width=1.0\textwidth]{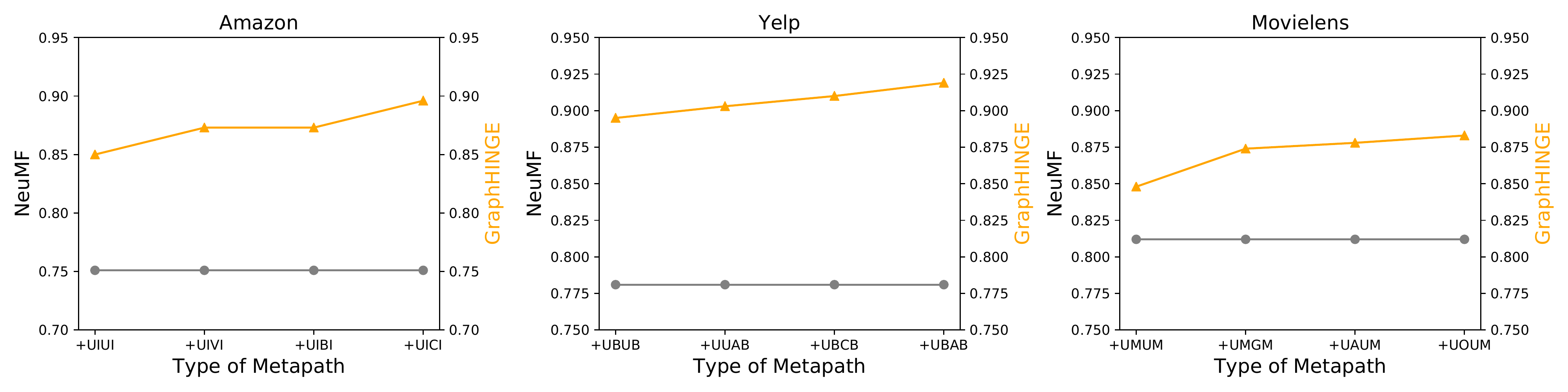}
	\vspace{-6mm}
	\caption{
		Performance change of GraphHINGE when gradually incorporating metapaths in terms of ACC.  
	}
	\label{fig:path}
% 	\vspace{-4mm}
\end{figure}

\begin{figure}[t]
	\centering
% 	\vspace{-2mm}
	\includegraphics[width=0.8\textwidth]{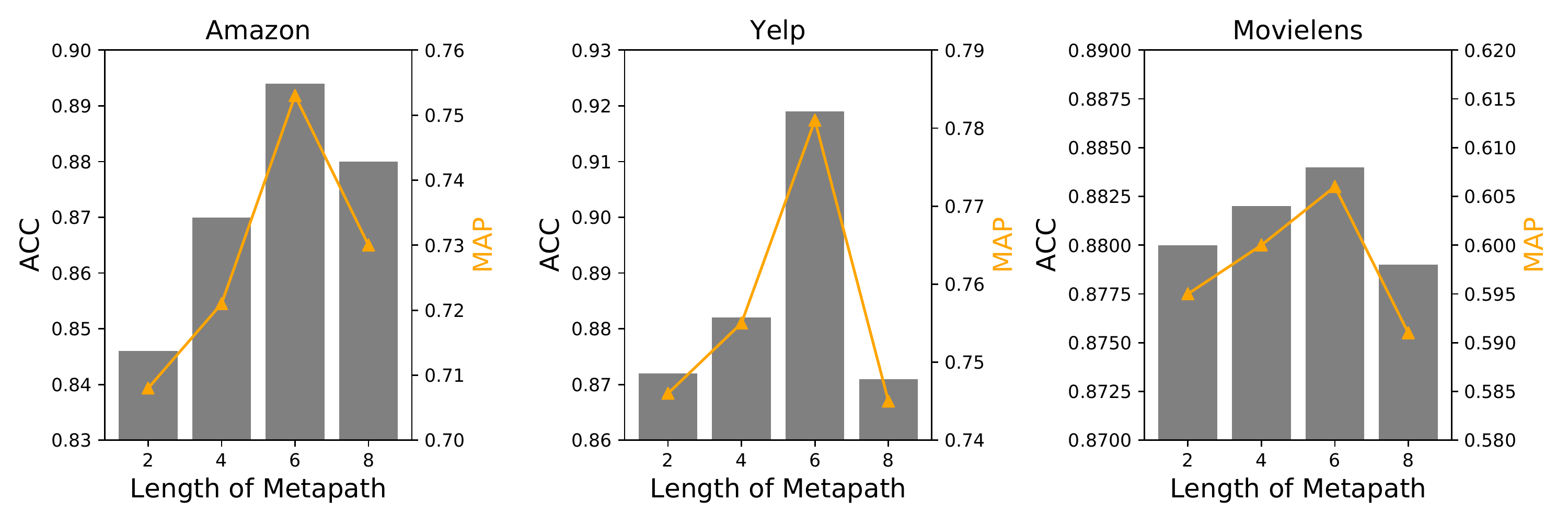}
% 	\vspace{-6mm}
	\caption{
		Performance change of GraphHINGE with different neighborhood lengths in terms of ACC and MAP.
	}
	\label{fig:neighbor}
% 	\vspace{-4mm}
\end{figure}

\subsection{Overall Performance (\textbf{RQ1})}
\label{subsec:result}
The comparison results of our proposed model and baselines on the six datasets are reported in Tables~\ref{tab:ctr}, \ref{tab:topn}, \ref{tab:big}.
The major findings from experimental results are summarized as follows:
\begin{itemize}[topsep = 3pt,leftmargin =10pt]
	\item Our model GraphHINGE is consistently better than all baselines on the six datasets.
	The results indicate the significance of capturing interactive information and the effectiveness of GraphHINGE on both CTR prediction and top-N recommendation, since GraphHINGE adopts a principled way to leverage interactive information to enhance the model performance on prediction tasks.
	\item Among the three kinds of baselines, the best of metapath-based methods (e.g., TAHIN, HAN) outperform graph-based or feature-based neural network methods (e.g., HetGNN, NeuMF) and non-neural network methods (e.g., ItemKNN, UserKNN) in most cases.
	An intuitive explanation is that those metapath-based methods can better capture the rich high-order structured information on HINs.
	It should be noted that our model GraphHINGE based on metapath-guided neighborhood is able to jointly consider low- and high-order neighborhood information and capture interactive patterns.
	\item Among HIN-based baselines, the recently proposed methods TAHIN and HAN gain better performance than the others.
	It is easy to notice that both of them try to distinguish different patterns hidden in context information among paths through attention mechanism.
	GraphHINGE outperforms those approaches.
	A possible reason is that simple aggregation of semantic messages on paths may lose some key information.
	It should be noted that GraphHINGE not only captures interaction information but also focus on key paths through the attention mechanism.
	\item Among all the methods, non-neural network approaches (e.g., ItemKNN, UserKNN) do not work well.
	One possible reason is that neural networks have good representation ability, which can largely improve the performances.
	Notice that DeepFM performs well on tubular data, but fails here.
	One probable reason is that for each node, its neighboring attributes on HIN are not suitable to be directly treated as its feature.
	This actually implies the necessity of developing interaction module on HIN.
	It also should be noted that MF is simple but works well on some datasets (e.g., Amazon, Yelp), but can not obtain stable performance on various datasets.
	It may indicate the need of adopting neural network based methods for more powerful generalization ability.
	\item Among all the datasets, GraphHINGE  outperforms state-of-the-art baseline models by a wide margin on those datasets (e.g., Amazon, Amazon Book) with high data sparsity.
	A possible explanation is that GraphHINGE leverages the interaction module to capture interactive information, which can mine more useful hidden relations than traditional methods. 
\end{itemize}

\subsection{Ablation Study (\textbf{RQ2})}
\label{subsec:ablation}
In order to investigate the contribution of each component to the final recommendation performance, we design two variants of GraphHINGE, namely GraphHINGE$_\text{CNN}$ and GraphHINGE$_\text{GCN}$ to study the neighborhood-based interaction (NI) and the aggregation modules, respectively.
The comparison results of our model and these variants on six datasets are shown in Tables~\ref{tab:ctr}, \ref{tab:topn}, \ref{tab:big}.
The major findings from experimental results are summarized as follows:
\begin{itemize}[topsep = 3pt,leftmargin =10pt]
    \item Result (GraphHINGE > GraphHINGE$_\text{CNN}$) indicates that our convolutional interaction strategy can better capture interaction information (i.e.,~similarities between the same type of nodes and ratings between different types of nodes) than simply employing CNN layers.
    \item Result (GraphHINGE > GraphHINGE$_\text{GCN}$) shows that the attention mechanism can better utilize the metapath-based interactive information, since the element- and path-level interactions may contribute differently to the final performance.
    Ignoring such influence may not be able to achieve optimal performance.
\end{itemize}

\subsection{Impacts of CNI \& NS Module (RQ3)}
In order to investigate the performances of the cross neighborhood-based interaction (CNI) and the selection (NS) modules, we develop two more variants of GraphHINGE, namely GraphHINGE$^+_\text{CROSS}$ for CNI module and GraphHINGE$^+_\text{SELECT}$ for NS module.
Also, we design GraphHINGE$^+_\text{ALL}$ to investigate the potential of the  interaction module.
The comparison results of our model and these variants on the six datasets are shown in Tables~\ref{tab:ctr}, \ref{tab:topn} and \ref{tab:big}.
The major findings from experimental results are summarized as follows:
\begin{itemize}[topsep = 3pt,leftmargin =10pt]
    \item Result (GraphHINGE$^+_\text{CROSS}$ > GraphHINGE) illustrates that in most cases, patterns in cross interaction information can help the model obtain more accurate predictions.
    Also, one can find that in some datasets (e.g., Movie-20M), GraphHINGE outperforms GraphHINGE$^+_\text{CROSS}$.
    One possible reason is that these cross interactive patterns obtained from different metapaths may include noise in some cases, which can impair the performance.
    \item Result (GraphHINGE$^+_\text{SELECT}$ > GraphHINGE) shows that the NS module can find valuable high-order neighborhoods to enhance the performance of GraphHINGE.
    The reason behind this is that almost in most HINs, high-order neighborhoods often are large-scale and full of noisy information.
    NS module is one of the solutions to find the key information and filter out the noise.
    \item Result (GraphHINGE$^+_\text{ALL}$ > GraphHINGE) illustrates that if we do not consider the time complexity, GraphHINGE can improve performance by involving more interaction information.
    We later show the comparison of time complexity in Section~\ref{subsec:complexity}.
\end{itemize}

\subsection{Impact of Metapath (\textbf{RQ4})}
\label{subsec:metapath}
In this section, we investigate the impact of different metapaths on the prediction performance.
To do this, we first include low-order metapaths (e.g., UI, UB, and UM) and then gradually incorporate metapaths into the proposed model. 
For ease of analysis, we include the NeuMF as the reference baseline. 
In Fig.~\ref{fig:path}, we can observe that the performance of GraphHINGE overall improves with the incorporation of more metapaths. 
Meanwhile, metapaths seem to have different effects on the prediction performance. 
Particularly, we can find that, when adding UMGM, GraphHINGE has a significant performance boost in the Movielens dataset.
A similar situation happens when adding UIVI in the Amazon dataset. 
These findings indicate that different metapaths contribute differently to the final result, consistent with previous observations in Section~\ref{subsec:ablation}.

\begin{figure}[t]
	\centering
% 	\vspace{-2mm}
	\includegraphics[width=0.8\textwidth]{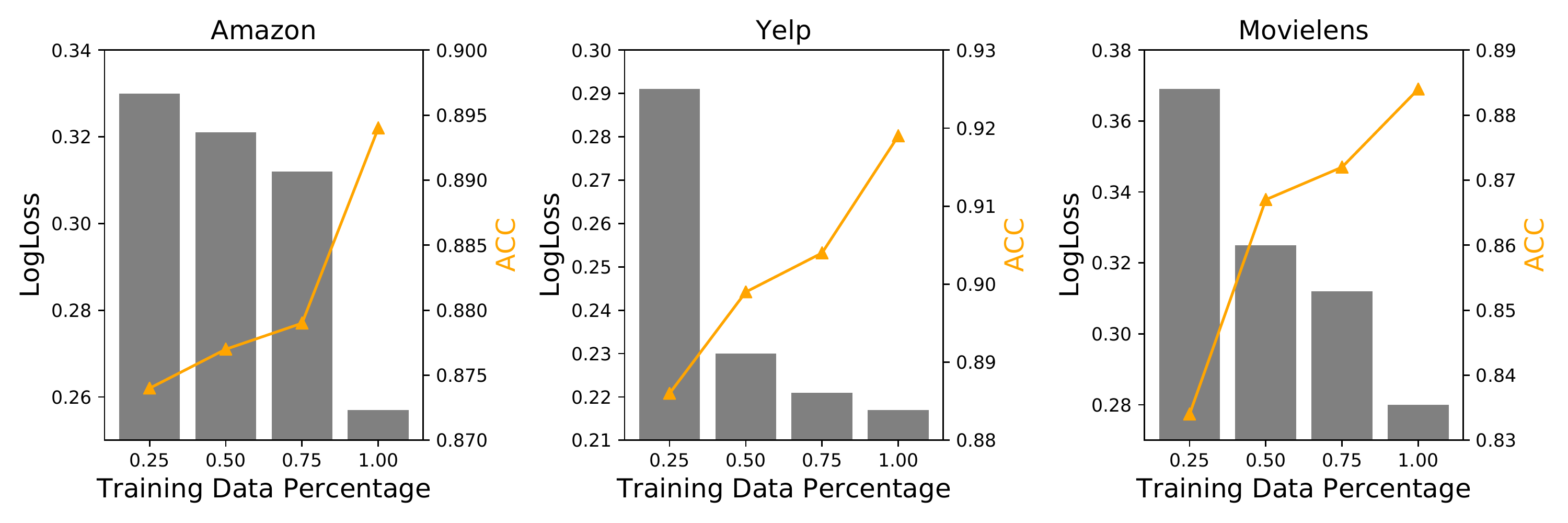}
% 	\vspace{-6mm}
	\caption{
		Performance change of GraphHINGE with different amount of training data in terms of ACC and LogLoss.
	}
	\label{fig:data}
% 	\vspace{-2mm}
\end{figure}

\begin{figure}[t]
	\centering
% 	\vspace{-2mm}
	\includegraphics[width=0.8\textwidth]{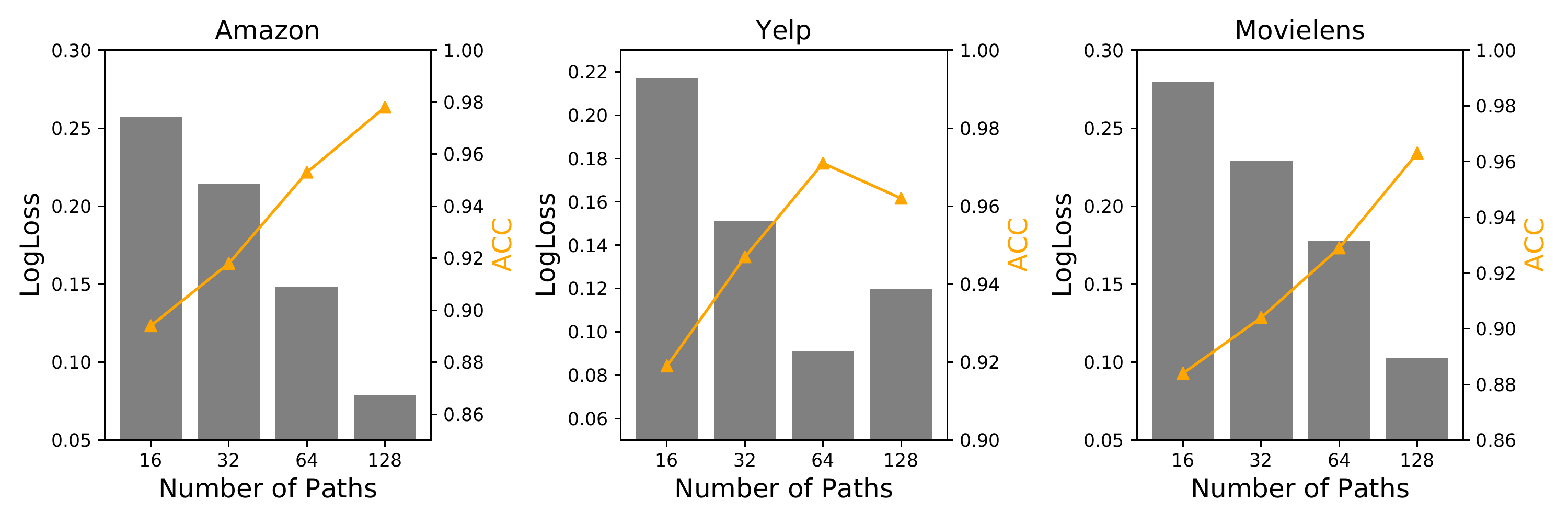}
% 	\vspace{-6mm}
	\caption{
		Performance change of GraphHINGE with different number of paths for each metapath in terms of ACC and LogLoss.
	}
	\label{fig:mount}
	\vspace{-2mm}
\end{figure}

\subsection{Impact of N-Hop Neighborhood (\textbf{RQ4})}
\label{subsec:neighbor}
In this section, we investigate the impact of different lengths of neighborhoods (i.e., n-hop neighborhood).
When changing the length of the neighborhood, we make other factors ( e.g.,~metapath type) fixed.
For example, when the metapath is UMUM, we study UM for length 2, UMUM for length 4, and UMUMUM for length 6.
We conduct similar procedures for the other metapaths and obtain the results in Fig.~\ref{fig:neighbor}.
We can observe that the performance of GraphHINGE first increases, next reaches the best performance and then decreases as the length of metapath increases.
The possible reason may be that as the length of the neighborhood increases, the metapath-guided neighborhood can contain more information.
When the length of the neighborhood is not very large, the information is mainly useful for the final performance.
However, when the length exceeds a certain value, the information includes noisy messages which harm the recommendation performance.
These findings indicate that different lengths of metapaths contribute differently to the final performance.
Also, it should be noted that our model is able to interact and aggregate neighbors in different lengths, as introduced in Section~ \ref{subsec:crossinteraction} and \ref{subsec:aggregation}.

\subsection{Impact of Amount of Data (\textbf{RQ4})}
In this section, we further evaluate the robustness of GraphHINGE with different amounts of training data.
We first randomly select a subset of training data (i.e.~25\% $\sim$ 100\%) to generate the new training data, and then use these data to train our methods.
For a fair comparison, we use the same data for evaluation across all experiments.
% For ease of analysis, we include the NeuMF as the reference baseline. 
As Fig.~\ref{fig:data} shows, when the amount of training data increases, the performance of GraphHINGE raises, especially during the period of the amount of training data expanding from 25\% to 50\% in Movielens dataset and 75\% to 100\% in Amazon dataset.

\begin{figure}[b]
	\centering
% 	\vspace{-2mm}
	\includegraphics[width=1.0\textwidth]{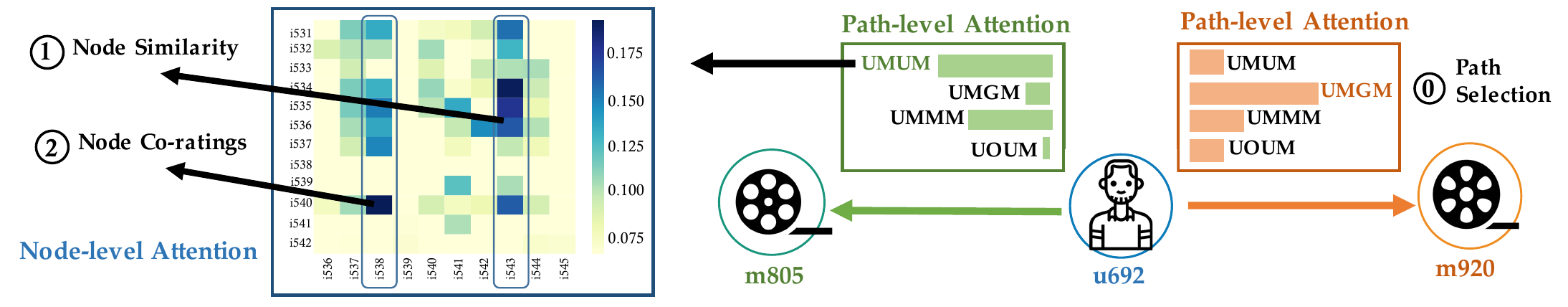}
	\vspace{-6mm}
	\caption{
		An illustrated example of the interpretability of interaction-specific attention distribution for GraphHINGE.
		The number denotes the logic flow of interpretation. 
	}
	\label{fig:case}
% 	\vspace{-2mm}
\end{figure}

\subsection{Impact of Number of Paths (\textbf{RQ4})}
\label{subsec:numberpath}
In this section, we investigate the performance change with different number of paths for each metapath.
As illustrated in Fig~\ref{fig:mount}, as the number of paths increases, the performance of GraphHINGE becomes better.
It is easy to understand since a larger number of paths always means more data for training.
There is a drop in the Yelp dataset when the number of paths increases from 64 to 128.
A possible reason is that the data might involve some noise.
However, one should note that a larger number of paths also means higher time complexity.

% \subsection{Impact of Batch Size (\textbf{RQ4})}
% In this section, we study the influence brought from different batch sizes.
% As shown in Fig.~\ref{fig:batch}, 

% \begin{figure}[h]
% 	\centering
% % 	\vspace{-2mm}
% 	\includegraphics[width=0.8\textwidth]{fig/batch.pdf}
% % 	\vspace{-6mm}
% 	\caption{
% 		Performance change of GraphHINGE with different amount of paths in each metapath in terms of ACC.
% 	}
% 	\label{fig:batch}
% 	\vspace{-2mm}
% \end{figure}

\subsection{Case Study (\textbf{RQ5})}
\label{subsec:case}
A major contribution of GraphHINGE is the incorporation of the element- and path-level attention mechanism, which takes the interaction relation into consideration in learning effective representations for recommendation.
Besides the performance effectiveness, another benefit of the attention mechanism is that it makes the recommendation result highly interpretable.
To see this, we select the user u692 in Movielens dataset as an illustrative example.
Two interaction records of this user have been used here, namely m805 and m920.
In Fig.~\ref{fig:case}, we can see that each user-movie pair corresponds to a unique attention distribution, summarizing the contributions of the metapaths. 
The relation between u692-m805 pair mainly relies on the metapaths UMUM and UMMM, while u692-m920 pair mainly relies on metapath UMGM (denoted by \circled{0} in Fig.~\ref{fig:case}). 
By inspecting into the dataset, it is found that at least five first-order neighbors of u692 have watched m805, which explains why user-oriented metapaths UMUM plays the key role in the first pair. 
As for the second pair, we find that the genre of m805 is g3, which is the favorite movie genre of u692. 
This explains why genre-oriented metapath UMGM plays
the key role in the second interaction. 
Our path-level attention is able to produce path-specific attention distributions.

If we wonder more specific reasons, we can have a look at the  element-level attention.
Here, we plot the interactive attention values in Fig.~\ref{fig:case}.
We can observe that the attention value of the similarity for the same type elements is very high (denoted by \circled{1} in Fig.~\ref{fig:case}).
By inspecting into the dataset, we can find that there are metapaths connected between u692 and m920.
In other words, some parts of the metapath-guided neighborhood of u692 and m920 overlap, which causes high similarity.
Also, the co-ratings between two neighborhoods play another key role (denoted by \circled{2} in Fig.~\ref{fig:case}).
It is natural, since in these neighborhoods, many users are fans of g3, and movies are in the type of g3, which causes the co-ratings among them to become really high. 
According to the analysis above, we can see that the distributions of attention weights are indeed very skew, indicating some interactions and metapaths are more important than the others.

\begin{table*}
	\centering
	\caption{The results of inference time of different models on Movielens and Amazon datasets.
	We visualize the data as shown in Fig.~\ref{fig:time}.
	For each model, we evaluate it with 10 runs and report the average time here.
% 	The number of paths (denoted as $\#\text{path}$) is set as 16 unless otherwise stated.
}
	\label{tab:time}
	\vspace{-3mm}
	\resizebox{1.00\textwidth}{!}{
	\begin{tabular}{@{\extracolsep{4pt}}cccccccccc}
		\midrule
		\multicolumn{2}{c}{Model on Movielens} & HAN \citep{wang2019heterogeneous} & TAHIN \citep{bi2020heterogeneous} & $\text{GraphHINGE}^-_\text{FFT}$ & GraphHINGE & $\text{GraphHINGE}_\text{Conv}$ & $\text{GraphHINGE}^+_\text{ALL}$ & GraphHINGE & $\text{GraphHINGE}_\text{Conv}$\\ 
% 		\cmidrule{1-2}
% 		\cmidrule{3-8}
% 		\cmidrule{9-10}
		\multicolumn{2}{c}{Number of path} & (Whole Graph) & (Whole Graph) & (16) & (16) & (16) & (16) & (32) & (32)\\ 
		\cmidrule{1-2}
		\cmidrule{3-8}
		\cmidrule{9-10}
		\multirow{2}{*}{Time (s)} 
		& Training
		& 14.3812 & 31.5352 & 51.6120 & 25.8209 & 26.5538 & 261.2001 & 31.8346 & 32.3579\\ 
		& Inference 
		& 2.9573 & 6.6152 & 7.2971 & 3.4670 & 3.5096 & 31.3405 & 4.4262 & 3.4689\\
		\cmidrule{1-2}
		\cmidrule{3-8}
		\cmidrule{9-10}
		Number of epochs
		& Best Performance
		& 85 & 13 & 10 & 12 & 13 & 9 & 13 & 9\\ 
% 		& Early Stop 
% 		& 110 & 38 & 35 & 37 & 38 & 34 & 38 & 34\\
        \midrule
        \multicolumn{2}{c}{Model on Amazon} & HAN \citep{wang2019heterogeneous} & TAHIN \citep{bi2020heterogeneous} & $\text{GraphHINGE}^-_\text{FFT}$ & GraphHINGE & $\text{GraphHINGE}_\text{Conv}$ & $\text{GraphHINGE}^+_\text{ALL}$ & GraphHINGE & $\text{GraphHINGE}_\text{Conv}$\\ 
% 		\cmidrule{1-2}
% 		\cmidrule{3-8}
% 		\cmidrule{9-10}
		\multicolumn{2}{c}{Number of path} & (Whole Graph) & (Whole Graph) & (16) & (16) & (16) & (16) & (32) & (32)\\ 
		\cmidrule{1-2}
		\cmidrule{3-8}
		\cmidrule{9-10}
		\multirow{2}{*}{Time (s)} 
		& Training
		& 114.6312 & 149.4831 & 148.8860 & 75.8216 & 75.7214 & 753.6511 & 150.0904 & 152.0212\\ 
		& Inference 
		& 15.7382 & 23.4506 & 25.4387 & 10.3340 & 9.9237 & 90.2116 & 11.2200 & 9.8833\\
		\cmidrule{1-2}
		\cmidrule{3-8}
		\cmidrule{9-10}
		Number of epochs
		& Best Performance
		& 87 & 18 & 9 & 7 & 6 & 5 & 8 & 5\\ 
% 		& Early Stop 
% 		& 112 & 43 & 34 & 32 & 31 & 30 & 33 & 30\\
		\midrule
	\end{tabular}
	}
	\vspace{-3mm}
\end{table*}

\begin{figure}[t]
	\centering
	\vspace{-2mm}
	\includegraphics[width=1.0\textwidth]{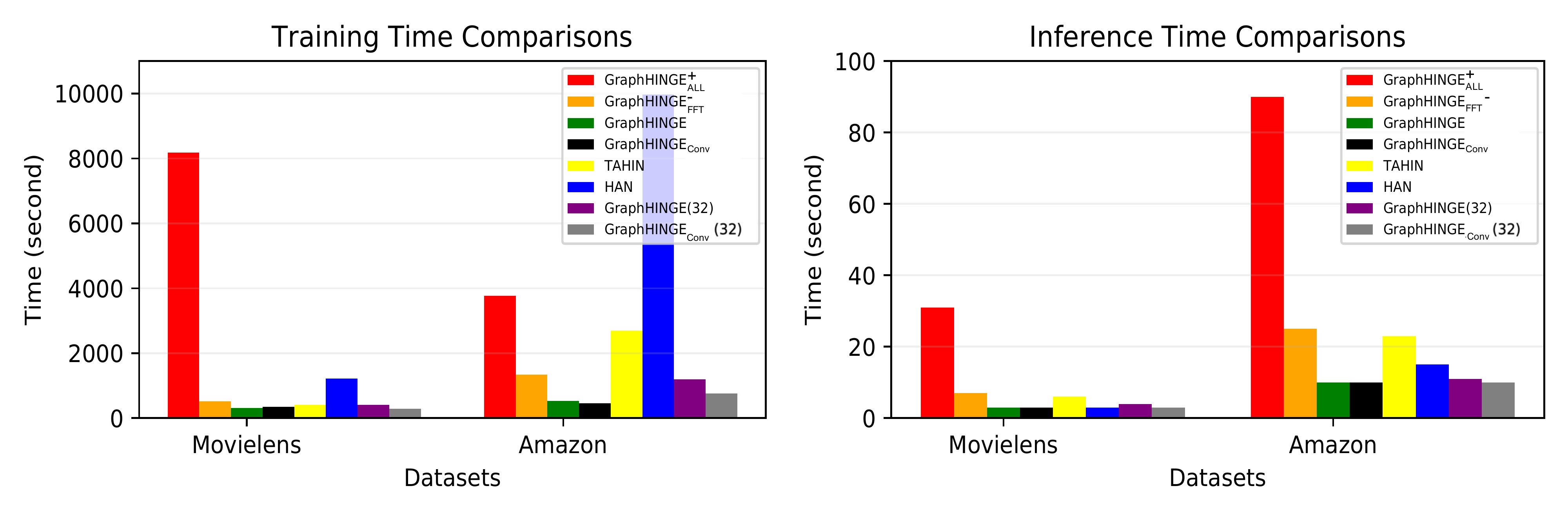}
	\vspace{-6mm}
	\caption{
		Time comparisons of GraphHINGE, its variants, and baselines. Training time (Left) and inference time (Right) comparisons on Movielens and Amazon datasets.
		See Table~\ref{tab:time} with tabular format.
	}
	\label{fig:time}
	\vspace{-2mm}
\end{figure}

\subsection{Complexity Study (\textbf{RQ6})}
\label{subsec:complexity}
In this section, we investigate whether the FFT technique can truly accelerate the convolution operation through comparisons between $\text{GraphHINGE}$ and $\text{GraphHINGE}^-_\text{FFT}$.
For ease of analysis, we also include recent proposed state-of-the-art heterogeneous graph embedding baseline models namely HAN and TAHIN according to the result in Table~\ref{tab:ctr}, \ref{tab:topn} and \ref{tab:big}.
We evaluate all the models on Movielens dataset and report the time of training and inference on one bath data, and the number of epochs to gain the model with best performance on training and inference in Table~\ref{tab:time}.
For convenience, we also demonstrate the time of learning the model with best performance in training and inference in Fig.~\ref{fig:time}.
Specifically, the training time reported in the left part of Fig.~\ref{fig:time} is the time to train the model with the best performance, calculated by multiplying training time of each epoch by the number of epochs shown in Table~\ref{tab:time}.
We also evaluate the inference time of one epoch data, and report the results in the right part of Fig.~\ref{fig:time}.
From Fig.~\ref{fig:time}, one can see that GraphHINGE is more efficient than HAN and TAHIN in terms of both training and inference. 
One possible explanation is that the input of HAN and TAHIN contains the whole graph, which is much more complex than sampled paths.
Also, as shown in Fig.~\ref{fig:time} and Table~\ref{tab:time}, we also demonstrate the performance gain of the FFT technique.
We further study the time complexity of $\text{GraphHINGE}^+_\text{ALL}$.
One can easily see that the result is consistent with the previous analysis in Section~\ref{subsec:variant}.
Notice that the convolution operation in GraphHINGE also can be implemented directly with Application Programming Interface (API)\footnote{\url{https://pytorch.org/docs/stable/nn.functional.html}} in Pytorch \citep{paszke2019pytorch}.
Hence, we also include this implementation (denoted as GraphHINGE$_\text{Conv}$.) in the experiment.
One should note that this implementation has involved several system optimization techniques that may also include FFT.
Notice that GraphHINGE is based on sampling, interaction and aggregation.
Hence, as stated in Section~\ref{subsec:variant}, the time complexity of GraphHINGE can be largely influenced by the number of paths.
In the experiments, we mainly adopt $16$ paths for each metapath.
Therefore, we here further investigate the time complexity of implying GraphHINGE with $32$ paths for each metapath.
If we increase the number of paths for each metapath, then the performance of GraphHINGE also can be further improved, as illustrated in Section~\ref{subsec:numberpath}.

\begin{wrapfigure}{r}{0.6\textwidth} 
	\centering
	\vspace{-2mm}
	\includegraphics[width=0.6\textwidth]{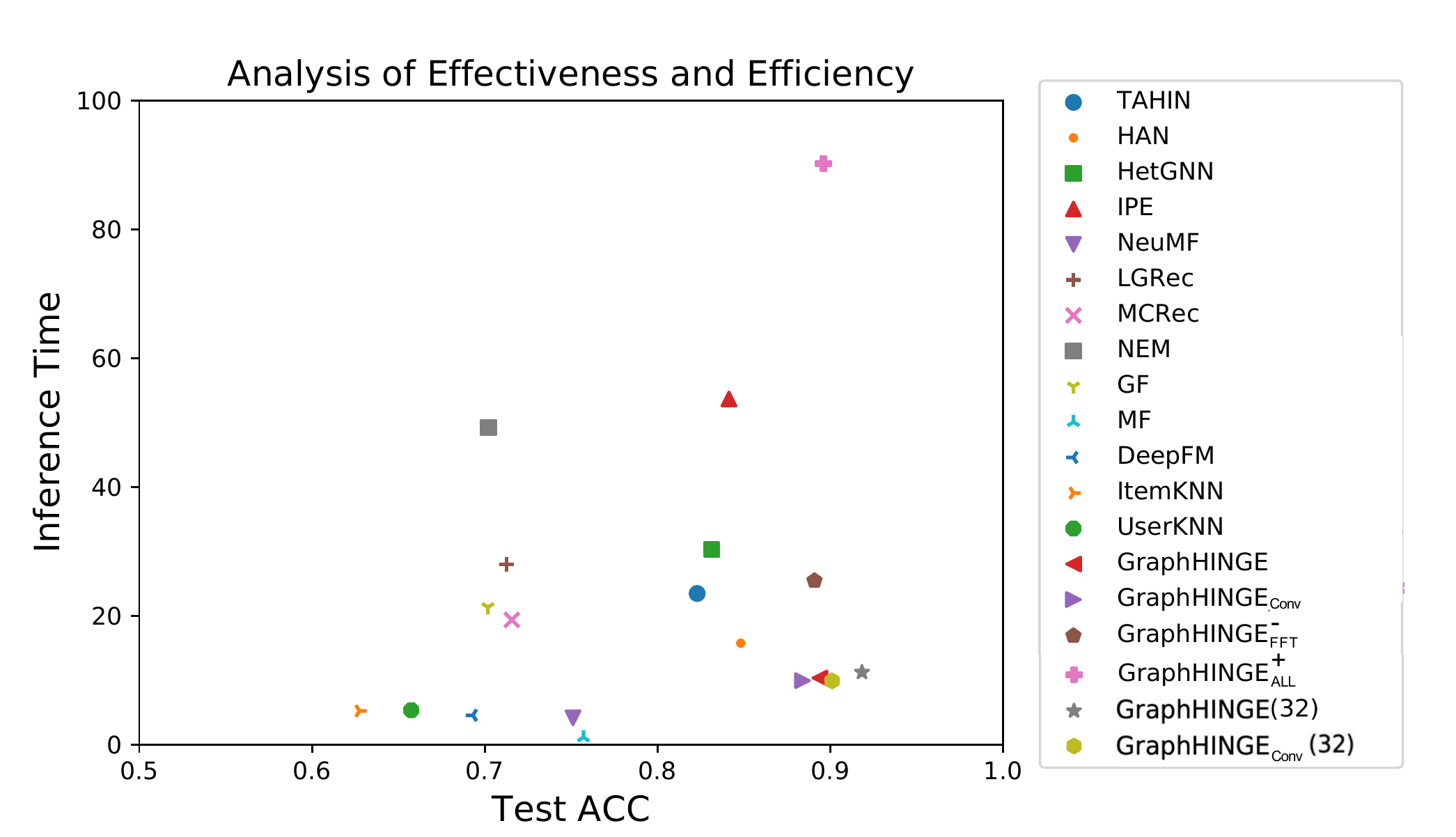}
	\vspace{-6mm}
	\caption{
		Analysis of the trade-off between effectiveness and efficiency of all the models in term of ACC and inference time on Amazon dataset. We report the average inference time of each method, evaluated on the whole dataset (i.e., one epoch) after 10 runs.
	}
	\label{fig:analysis}
	\vspace{-2mm}
\end{wrapfigure}

In order to find the connections between the results of effectiveness and efficiency, we also study the performance and inference time of all the baselines on Amazon dataset.
As illustrated in Fig.~\ref{fig:analysis}, one can see that "simple" baselines such as NeuMF is much faster than "complex" baselines such as HAN, TAHIN, and any variant of GraphHINGE, but are not able to obtain the good performance.
And methods including HAN, TAHIN, especially $\text{GraphHINGE}^+_\text{ALL}$ act well but are time-consuming.
We conclude that $\text{GraphHINGE}$ and $\text{GraphHINGE}_\text{Conv}$ located at the right and bottom part of the figure would be good choices to trade off between effectiveness and efficiency.
The detailed implementation settings including hardware setting of our model and these baselines can be found in Section~\ref{subsec:detail}.

% \subsection{Learning Process (\textbf{RQ7})}
% In Fig.~\ref{fig:curve}, we provide the learning curves of GraphHINGE when training on different datasets.
% From these figures, we find that GraphHINGE converges effectively.

% \begin{figure}[h]
% 	\centering
% % 	\vspace{-2mm}
% 	\includegraphics[width=1.0\textwidth]{fig/curve.pdf}
% 	\vspace{-6mm}
% 	\caption{
% 		Learning curves of GraphHINGE over different datasets.
% 	}
% 	\label{fig:curve}
% 	\vspace{-2mm}
% \end{figure}

\subsection{Implementation Details}
\label{subsec:detail}
\minisection{Data Processing}
Each dataset is processed as follows.
Since the relations between users and items are originally in rating format, we convert ratings to binary feedbacks: ratings with 4-5 stars are converted to positive feedbacks (denoted as ``1''), and other ratings are converted to negative feedbacks (denoted as ``0'').
After the datasets are processed, we split each dataset into training/validation/test sets at a ratio of 6:2:2.

\minisection{Hyper-parameter Setting}
The embedding dimension of GraphHINGE is 128.
As stated in Section~\ref{subsec:sample}, we perform the metapath-guided random walk method with restart. The number of attention heads is 3.
% The size of sampled neighbors set of each node is 128.
The temperatures are set to $0.2$.
GraphHINGE samples 16 paths for each metapath in metapath-guided neighborhood while $\text{GraphHINGE}^+_\text{SELECT}$ first samples 128 paths and then adopt neighborhood-based selection (NS) module to learn to select 16 paths from sampled 128 candidate paths for each metapath in metapath-guided neighborhoods.
The length of the metapath-guided neighborhoods\footnote{For simplification, we use ``length of (the metapath-guided) neighborhoods'' to denote ``length of paths in the metapath-guided neighborhoods''.} equals to the length of the metapath.
We investigate the impacts of different lengths of neighborhoods in Section~\ref{subsec:neighbor}.
The patience value of selecting the model with the best performance is set as 25 for all the experiments.

\minisection{Model Configuration}
For fair comparisons, we implement the proposed method and baselines as following:
(i) In the data processing procedure, we sample and store paths guided by metapaths in a unified framework.
For instance, when sampling data from Movielens following UMUM metapath, we can obtain 16 paths connecting $u_0$ and $m_0$ (denoted as $\rho(u_0\rightarrow m_0)$), as well as 16 paths in metapath-guided neighborhoods of $u_0$ (denoted as $\mathcal{N}(u_0)$) and 16 paths in metapath-guided neighborhoods of $m_0$ (denoted as $\mathcal{N}(m_0)$).
That is, the first node of both path $\rho(u_0\rightarrow m_0)$ and metapath-guided neighborhood $\mathcal{N}(u_0)$ is sampled from the same source node $u_0$.
The difference is that the path $\rho(u_0\rightarrow m_0)$ must end with target node $m_0$ while the neighborhood $\mathcal{N}(u_0)$ can end with any node in the type of movie.
(ii) For non-neural-network based methods, we first transform the graph data into the tabular data where the neighborhood nodes are regarded as the feature of central node.
For graph based models, such as HAN and TAHIN, we directly includes all the neighbors following the metapath. 
For metapath-based baselines such as MCRec and IPE, we feed models with the same paths with GraphHINGE.
% For graph-based baselines, such as HAN \citep{wang2019heterogeneous} and HetGNN \citep{zhang2019heterogeneous}, we generate n-hop neighbor nodes from sampled paths.
% For neighborhood interaction based baselines such as KNI \citep{qu2019end}, we follow the description in the paper, where high-order neighborhood information is integrated by graph neural network. 
All the methods involving metapath share the same ones with GraphHINGE.
We report the metapath used in each dataset in Table~\ref{tab:data}.
(iii) The embedding dimensions of all baselines are set to $128$ (same as GraphHINGE).

\minisection{Hardware Setting}
The models are trained under the same hardware settings with an Amazon EC2 p3.8$\times$large instance\footnote{Detailed setting of AWS E2 instance can be found at \url{https://aws.amazon.com/ec2/instance-types/?nc1=h_ls}.}, where the GPU processor is NVIDIA Tesla V100 processor and the CPU processor is Intel Xeon E5-2686 v4 (Broadwell) processor.

\subsection{Deployment Feasibility Analysis}
In this section, we mainly discuss the feasibility of the industrial deployment of GraphHINGE framework. 
First, it is not difficult to switch the current model pipeline to GraphHINGE, because the main change brought from GraphHINGE is how to utilize graph structure.
Previous methods such as GraphSAGE \citep{hamilton2017inductive} developed sample and aggregation frameworks, where neighbors are first sampled and then aggregated to obtain the final prediction.
Interactive patterns need to be built to update the model pipeline to GraphHINGE, while the whole HIN-based recommendation remains almost the same but adding an additional interaction module between sampling and aggregation modules.
As Fig.~\ref{fig:overview} shows, the sampling and aggregation modules have no substantial difference from that of traditional solutions.
Efficiency is another essential concern in industrial applications.
We analyze the time complexity of GraphHINGE that contains interaction and aggregation modules.
As stated in Section~\ref{subsec:complexity}, the time complexity of the interaction model is $O(L_\rho I_\rho \log(I_\rho))$ and that of the aggregation model is $O(V_\rho K + E_\rho K)$.
Note that the overall complexity is linear to the number of nodes and metapath-based node pairs.
From the perspective of metapath-guided neighborhood construction, we propose a novel neighborhood-based selection module to update the buffer (as illustrated in Fig.~\ref{fig:comb}), where we first adopt a neighborhood sampling strategy to obtain the raw neighborhood paths, and then use the performance of low-order neighborhood to distinguish useful high-order neighborhood and filter out the noise.

% Moreover, we compare the actual inference time between GraphHINGE and previous interaction baselines in experiments.
% The average inference time of the models are illustrated in Fig.~\ref{fig:time}.
% The time is calculated by dividing the overall time (only the time that contains the forward computations and behavior searching) on test dataset with the number of prediction targets.
% From the figure, we could find that the absolute value of GraphHINGE's inference time on three datasets is less than 1ms which is efficient enough for online serving \citep{wang2016display}.

\section{Conclusion and Future Work}
In this paper, we introduce the problem of ``early summarization'' and propose a novel framework named GraphHINGE to address this issue.
We first introduce the definition of metapath-guided neighborhoods to preserve the heterogeneity of HINs.
Then, we elaborately design an interaction module to capture the similarity of each source and target node pair through their neighborhoods, where we propose the neighborhood-based interaction (NI) module for neighborhoods guided by the same metapaths and the cross neighborhood-based interaction (CNI) module for neighborhoods guided by different metapaths.
To fuse the rich semantic information, we propose the element- and path-level attention mechanism to capture the key interactions from element and path levels respectively.
Extensive experimental results have demonstrated the superiority of our model in effectiveness, interpretability, and efficiency.

Currently, our approach is able to capture interactive information only in the structured (graph) side effectively.
However, there is rich semantic information on both the structured (graph) side and the non-structured (node) side.
In the future, a promising direction is extending the  neighborhood-based interaction and aggregation modules to capture key messages from both sides and adapt to more general scenarios.

\begin{acks}
The corresponding author Weinan Zhang thanks the support of Shanghai Municipal Science and Technology Major Project (No. 2021SHZDZX0102) and National Natural Science Foundation of China (Grant No. 62076161, 61772333, 61632017). 
We would also like to thank Wu Wen Jun Honorary Doctoral Scholarship from AI Institute, Shanghai Jiao Tong University.
\end{acks}

%%
%% The next two lines define the bibliography style to be used, and
%% the bibliography file.
\clearpage
\bibliographystyle{ACM-Reference-Format}
\bibliography{nirec}
\clearpage
\setcounter{section}{0}
\renewcommand{\thedefinition}{R\arabic{section}}%

\noindent
Dear TOIS Editors,
~\\

We would like to submit the enclosed manuscript entitled `Learning Interaction Models of Structured Neighborhood on Heterogeneous Information Network’, which we wish to be considered for publication in TOIS Special Issue on Graph Technologies for User Modeling and Recommendation.

In the Heterogeneous Information Network (HIN) based recommendation scenario, we investigate the `early summarization’ issue that is not well addressed in the previous works. In this paper, we propose GraphHINGE (Heterogeneous INteract and aggreGatE), which captures and aggregates the interactive patterns between each pair of nodes' structured neighborhoods. 
We propose and formulate interaction modules in a convolutional way and learn efficiently with fast Fourier transform. Extensive experiments on 6 widely used datasets from various domains illustrate the effectiveness of our model compared with 8 strong baselines on both click-through rate prediction and top-N recommendation tasks. 
This work would shed light on building efficient interaction models on HINs for downstream tasks such as recommender systems.

Please note that an earlier and shorter presentation of this work was accepted for publication by the 26th ACM SIGKDD International Conference on Knowledge Discovery and Data Mining (KDD 2020), namely `An Efficient Neighborhood-based Interaction Model for Recommendation on Heterogeneous Graph'. And the conference version is attached as a supporting document.

We list the differences (compared to the conference paper) as follows:
\begin{itemize}[topsep = 3pt,leftmargin =10pt]
\item We analyze the ``early summarization'' issue in the HIN-based recommendation. And we propose a general framework to capture and aggregate interactive patterns among the structured neighborhoods of source and target nodes.
\item We model the structured neighborhoods guided by the metapaths and build the Cross Neighborhood-based interaction (CNI) module for neighborhoods guided by different metapaths even in different lengths. CNI is designed to generalize the neighborhood-based interaction (NI) module in the previous conference version, and to enhance the prediction performance with rich cross semantic information.
\item Our previous work is limited by the random neighborhood sampling, while in this paper, we further propose a Neighborhood-based Selection (NS) module to capture the key paths and filter out the noise.
\item We adopt more baseline models. Besides, we conduct new experiments on more datasets, including two large-scale HIN datasets on both click-through rate prediction and top-N recommendation tasks.
\end{itemize}
We confirm that at least 50\% of this article is new material. 
And all of the authors are aware of the submission and agree to the review by TOIS. 
We have checked all guest editors and confirm that there is no editor with conflicts of interest: they neither work in our institutes nor have any collaborative publications with us within the recent 3 years.

Thanks for taking care of this submission.

~\\
\noindent
Manuscript authors:

\noindent
Jiarui Jin, Kounianhua Du, Weinan Zhang, Jiarui Qin, Yuchen Fang, Yong Yu, Zheng Zhang and Alex Smola.

\noindent
November 14, 2020

\end{document}